\pdfoutput=1
\documentclass[10pt]{article}
\usepackage[left=.75in,right=.75in,top=.75in,bottom=.75in]{geometry}
\usepackage[utf8]{inputenc}
\usepackage{amssymb,amsmath,latexsym,amsthm,amsfonts,mathtools}

\usepackage{mathrsfs}
\usepackage{multirow}
\usepackage{graphicx}
\usepackage{color}
\usepackage{textcomp,siunitx}
\usepackage{float}
\usepackage{subcaption}
\usepackage{booktabs}
\usepackage{enumitem}

\usepackage[hyperref]{xcolor}
\definecolor{ao(english)}{rgb}{0.0, 0.5, 0.0}
\usepackage{hyperref}
\hypersetup{colorlinks, breaklinks, citecolor=blue, linkcolor=ao(english), urlcolor=blue}

\usepackage[nameinlink,noabbrev,sort&compress]{cleveref}
\crefname{assumption}{Assumption}{Assumptions}
\usepackage{setspace}
\usepackage{cite}

\theoremstyle{plain}
\newtheorem{theorem}{Theorem}
\newtheorem{lemma}{Lemma}
\newtheorem{corollary}{Corollary}

\newtheorem{assumption}{Assumption}

\theoremstyle{remark}
\newtheorem{remark}{Remark}

\theoremstyle{definition}
\newtheorem{definition}{Definition}

\usepackage{newtxmath}

\usepackage{anyfontsize}
\usepackage{accents}

\usepackage{scalerel}
\usepackage{tikz}
\usetikzlibrary{automata, shapes, arrows, calc, arrows.meta, fit, positioning}
\usetikzlibrary{svg.path}

\definecolor{orcidlogocol}{HTML}{A6CE39}
\tikzset{
	orcidlogo/.pic={
		\fill[orcidlogocol] svg{M256,128c0,70.7-57.3,128-128,128C57.3,256,0,198.7,0,128C0,57.3,57.3,0,128,0C198.7,0,256,57.3,256,128z};
		\fill[white] svg{M86.3,186.2H70.9V79.1h15.4v48.4V186.2z}
		svg{M108.9,79.1h41.6c39.6,0,57,28.3,57,53.6c0,27.5-21.5,53.6-56.8,53.6h-41.8V79.1z M124.3,172.4h24.5c34.9,0,42.9-26.5,42.9-39.7c0-21.5-13.7-39.7-43.7-39.7h-23.7V172.4z}
		svg{M88.7,56.8c0,5.5-4.5,10.1-10.1,10.1c-5.6,0-10.1-4.6-10.1-10.1c0-5.6,4.5-10.1,10.1-10.1C84.2,46.7,88.7,51.3,88.7,56.8z};
	}
}

\newcommand\orcidicon[1]{\href{https://orcid.org/#1}{\mbox{\scalerel*{
				\begin{tikzpicture}[yscale=-1,transform shape]
					\pic{orcidlogo};
				\end{tikzpicture}
			}{|}}}}

\title{Trajectory Tracking for Unmanned Aerial Vehicles in 3D Spaces under Motion Constraints}

\author{Saurabh~Kumar\textsuperscript{\orcidicon{0000-0002-8344-5966}}
	\thanks{Corresponding author.\newline Saurabh Kumar and Shashi Ranjan Kumar are with the Intelligent Systems \& Control Lab, Department of Aerospace Engineering, Indian Institute of Technology Bombay, Powai-- 400076, Mumbai, India. Abhinav Sinha is with the Guidance, Autonomy, Learning, and Control for Intelligent Systems (GALACxIS) Lab, Department of Aerospace Engineering and Engineering Mechanics, University of Cincinnati, Cincinnati, OH, 45221,  USA
 e-mails: saurabh.k@aero.iitb.ac.in, srk@aero.iitb.ac.in, abhinav.sinha@uc.edu}	
    \and Shashi~Ranjan~Kumar\textsuperscript{\orcidicon{0000-0001-6446-7281}}
	\and Abhinav~Sinha\textsuperscript{\orcidicon{0000-0001-6419-2353}}
    }

\date{}

\begin{document}

\maketitle
\doublespacing

\begin{abstract}
This article presents a three-dimensional nonlinear trajectory tracking control strategy for unmanned aerial vehicles (UAVs) in the presence of spatial constraints. As opposed to many existing control strategies, which do not consider spatial constraints, the proposed strategy considers spatial constraints on each degree of freedom movement of the UAV. Such consideration makes the design appealing for many practical applications, such as pipeline inspection, boundary tracking, etc. The proposed design accounts for the limited information about the inertia matrix, thereby affirming its inherent robustness against unmodeled dynamics and other imperfections. We rigorously show that the UAV will converge to its desired path by maintaining bounded position, orientation, and linear and angular speeds. Finally, we demonstrate the effectiveness of the proposed strategy through various numerical simulations.
\medskip
	
\end{abstract}

\section{Introduction}\label{sec:intro}
In recent years, unmanned aerial vehicles, especially quadrotors, have been extensively used for many military applications, e.g., \cite{doi:10.2514/1.G006608,doi:10.2514/1.G004626,doi:10.2514/1.G003773,doi:10.2514/1.G007964,doi:10.1016/j.ast.2024.109225}. As UAVs become pervasive, their deployment in civil applications (e.g., search and rescue, pipeline and railway track inspection, agriculture spraying, environmental boundary tracking, package delivery, overhead transmission line surveillance, patrolling, and so forth) is also expected to grow. As such, efficient motion planning becomes critical. In this paper, we address a class of constrained motion control problems in which the UAV may require, for example, (\romannumeral 1) to pass through a narrow space (e.g., passing through a tunnel or emergency exit during any critical situation), (\romannumeral 2) drop food and essentials at a precise location during a disaster, (\romannumeral 3) perform surveillance of railway track or overhead line with a camera attached where it cannot move independently in order to keep the railway track or transmission line within the field-of-view of the camera. Such applications necessitate the quadrotor to operate under physical constraints that impose limitations on its movement. To elucidate the complexity of the problem at hand, let us consider a hypothetical scenario in which a quadrotor is navigating through a pipeline system, where its deviation from a predefined flight path is restricted to a mere few meters in both the horizontal and vertical planes. Furthermore, the quadrotor's orientation must remain within carefully defined boundaries to prevent collision with the surrounding walls, imposing additional limitations on its motion. Controlling the quadrotor in such confined spaces with predetermined precision under motion constraints is an interesting and challenging task, which is our main focus.

 There have been several techniques proposed for trajectory tracking using a quadrotor, albeit limitations on its complete motion have been less explored. For example, in \cite{1389776}, the authors designed proportional-integral-derivative (PID) controllers, and the performance was compared with a linear quadratic controller. It was shown that the PID performed better in the presence of minor perturbations in the attitude angles. A linear quadratic regulator (LQR) based method for multiple quadrotors was discussed in \cite{doi:10.1007/s10846-012-9708-3}. In \cite{doi:10.1016/j.isatra.2018.11.015}, the authors designed an $H_{\infty}$ controller for the attitude tracking of the quadrotor. Note that these controllers were designed by linearizing the quadrotor dynamics around an operating point. These control methods were simple in design and showed satisfactory performance around the operating conditions at the expense of a reduced operating regime. 

To circumvent the limitations of the linear controllers, several nonlinear controllers were also proposed for the trajectory tracking problem. For example, feedback linearization-based controllers were designed in \cite{761745,6425945}, whereas a nested saturation function-based design was discussed in \cite{doi:10.2514/1.27882}. The authors in \cite{doi:10.1016/j.jfranklin.2018.01.039,doi:10.2514/1.43768,doi:10.1007/s10846-009-9331-0} discussed backstepping-based approaches for a quadrotor. A procedure to design a hierarchical controller for a quadrotor by utilizing its structural properties was also presented in \cite{doi:10.2514/1.43768}. Various trajectory-tracking techniques were also proposed for a quadrotor using sliding mode control, see for example, \cite{doi:10.1016/j.ast.2019.105306,doi:0.1016/j.ast.2021.106616,9836191,10093309}. The authors in \cite{doi:10.2514/1.G004710} presented a geometric control based on feedback linearization for attitude and position control of the quadrotor. An augmented $\mathcal{L}_1$ adaptive controller was discussed in \cite{6978900}. In this approach, two nonlinear feed-forward compensators with the proportional derivative controller were used to cancel out the known nonlinearities, and then the $\mathcal{L}_1$ controllers were designed to mitigate the effect of model uncertainties and external disturbances. Using a convex multi-objective filter optimization method, the authors in \cite{doi:10.2514/1.G000566} discussed an optimal $\mathcal{L}_1$ adaptive controller. A robust tuning method for geometric attitude controller of multirotor UAVs was discussed in \cite{doi:10.2514/1.G004457}. The controller gains were obtained by linearizing the dynamics around an operating point together with structured $H_{\infty}$ controller synthesis. In the absence of complete information about the UAV's model, an adaptive incremental nonlinear dynamic inversion (INDI) for the attitude dynamics of UAV was presented in \cite{doi:10.2514/1.G001490}. Typically, the INDI approach for attitude control requires the control effectiveness model together with the estimate of angular accelerations. The authors in \cite{doi:10.2514/1.G001490} used an adaptation rule to estimate the control effectiveness matrix, allowing the method to remain effective when it is changing over the flight envelope. The authors in \cite{doi:10.2514/1.G001439} discussed linear matrix inequality (LMI) based adaptive control for the quadrotor, while a geodesic method for a tethered quadrotor was presented in \cite{doi:10.2514/1.G004356}.

It is worth noting that most of the above-mentioned control strategies were only concerned with tracking a predefined trajectory either by considering all six degrees of freedom or with the attitude dynamics only. In practice, inherent motion limitations cannot be ignored while navigating a quadrotor with a predetermined accuracy in a restricted space. A controller addressing the motion constraints on a UAV will be more suitable for several real-time applications such as pipeline, railway track, overhead transmission line inspections, etc. Some attempts were also made earlier to tackle the constrained trajectory tracking problem of the quadrotor, see for example, \cite{doi:10.1016/j.ast.2020.105935,doi:10.1016/j.ast.2021.107063,doi:10.1017/S0263574723000735,9836037}. However, the works in \cite{doi:10.1016/j.ast.2020.105935,doi:10.1016/j.ast.2021.107063} only considered the constraints on the position of the quadrotor. In essence, the constraints were on the three degrees of freedom movement, whereas in \cite{doi:10.1017/S0263574723000735,9836037}, symmetrical constraints were considered. The efficacy of such methods may be limited in the applications that we are considering in this paper. Suppose a quadrotor equipped with a camera is performing an overhead line or railway track inspection, where its yaw angle must lie within a certain limit (perhaps a narrow zone) in order to keep the overhead line or railway track inside the field of view of the camera. At the same time, the position of the quadrotor also becomes constrained as part of a safe standoff. Hence, it is pragmatic to design a nonlinear controller that is capable of accounting for motion constraints on all six degrees of freedom of the quadrotor, thereby enabling it to accomplish such tasks effectively. In light of the above-mentioned discussions, we articulate the main contributions below.


We propose a nonlinear control strategy that adeptly guides the quadrotor UAV along its desired trajectory in a three-dimensional (3D) space, exhibiting a high degree of efficacy across a wide range of operating conditions. The controller's robustness is demonstrated by its ability to maintain the desired performance despite limited knowledge of the inertia matrix, which may arise from the precision of identification methods or the accuracy of experimental data. This inherent robustness enables the controller to mitigate the effects of unmodeled dynamics and other imperfections, thereby ensuring a reliable and stable response.

Furthermore, our proposed strategy guarantees that the UAV tracks its desired trajectory while respecting its spatial constraints, thereby facilitating its deployment in a diverse array of applications, including pipeline inspections, precision landing, transmission and railway line surveillance, and other tasks that demand precise motion control under spatial trajectory constraints. It is worth noting that our approach is distinct from previous work in that it explicitly considers constraints on all six degrees of freedom of the quadrotor, thereby enabling trajectory tracking in 3D spaces with precision and flexibility. Consequently, the proposed controller is expected to outperform earlier designs that operate under partial motion constraints.

In addition, we conduct a rigorous analysis of the theoretical guarantees underlying our controller design, providing a comprehensive examination of the boundedness of the quadrotor's position, velocity, and orientation, as well as their respective rates. This detailed analysis provides a thorough understanding of the controller's behavior and performance, thereby establishing a foundation for its reliable implementation in real-world scenarios.

The remainder of this paper is organized as follows. The dynamics of the quadrotor and problem statement, along with some preliminaries, are discussed in \Cref{sec:problem}. The proposed control scheme will be derived and its stability analysis will be performed in \Cref{sec:main}, which will be followed by performance validation using the numerical simulations in \Cref{sec:simulation}. Finally, \Cref{sec:conclusion} presents concluding remarks while indicating some of the future directions for investigation.
	
\section{Background and Problem Statement}\label{sec:problem}
\begin{figure}[!ht]
		\centering
		\includegraphics[width=0.45\linewidth]{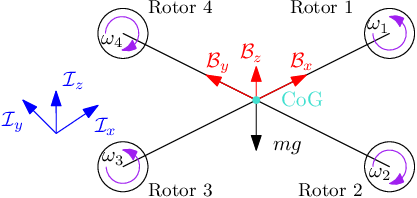}
		\caption{Schematic representation of a quadrotor.}
		\label{fig:fbd}
	\end{figure}

	In this section, we first discuss the kinematics and dynamics of a quadrotor UAV. We then formulate the main problem addressed in this paper, followed by some preliminaries that aid in the subsequent design. We consider a quadrotor UAV with a \emph{plus} configuration and coplanar rotors, as shown in \Cref{fig:fbd}. Without loss of generality, we assume that rotors labeled $1$ and $3$ spin in the anticlockwise direction while those labeled $2$ and $4$ spin in the clockwise direction to counterbalance the reaction moments they generate. This quadrotor belongs to a class of under-actuated systems since it cannot apply an instantaneous force in the direction perpendicular to its rotor axis. In other words, it cannot move in the lateral direction without rolling or pitching by some angle, thus making it a nonholonomic vehicle (which imposes turn rate constraints naturally).
\subsection{Kinematics and Dynamics of Quadrotor}
\begin{figure}[!ht]
     \centering
     \includegraphics[width=0.85\linewidth]{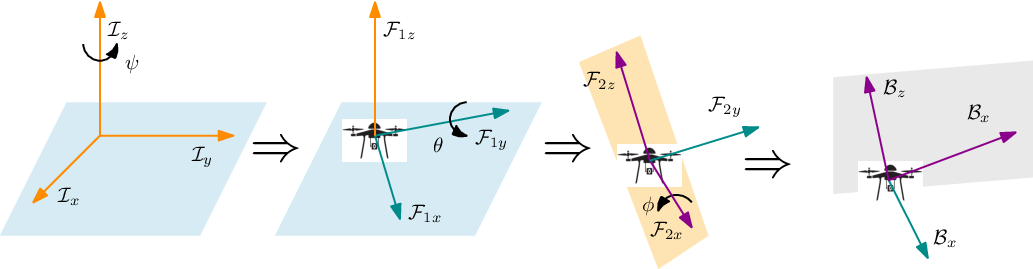}
     \caption{Geometrical representation of coordinate frames.}
     \label{fig:coordinates_frames}
 \end{figure}

	We consider four frames of reference as shown in \Cref{fig:coordinates_frames} to derive the kinematics and dynamics of the quadrotor UAV. The inertial frame ($\mathcal{I}$) is defined by a set of three mutually orthogonal axes, $\mathcal{I}_x$, $\mathcal{I}_y$, and $\mathcal{I}_z$, with $\mathcal{I}_z$ pointing upwards. There are two intermediate frames, namely, vehicle frame 1 ($\mathcal{F}_1$, obtained by first rotating the inertial frame through an angle $\psi$ about the $\mathcal{I}_z$ axis) and vehicle frame 2 ($\mathcal{F}_2$, obtained by rotating $\mathcal{F}_1$ by an angle $\theta$ about the $y$--axis of the rotated frame). Finally, the body frame, $\mathcal{B}$, is obtained by rotating $\mathcal{F}_2$ through an angle $\phi$ about its $x$--axis. Therefore, the order of rotation from the inertial to the body-fixed frame is $zy'x''$. Consideration of $\mathcal{I}$ as an inertial frame of reference essentially implies that Newton's laws are valid in this frame, whereas $\mathcal{B}$ is a body-fixed frame that is constrained to move with the UAV. The origin of $\mathcal{B}$ is $O_b$, which is an invariant point that belongs to the quadrotor's structure and is at its center of gravity (CoG). The $b_x$--axis points towards rotor 1, whereas the $b_y$--axis lies in the same plane but perpendicular to the $b_x$--axis and points towards rotor 4. The $b_z$--axis points upward following the right-hand rule and is perpendicular to both $b_x$ and $b_y$--axes, as depicted in \Cref{fig:fbd}. Thus, the \emph{East-North-Up} standard is adopted.
	
	We represent the quadrotor's attitude using \emph{Tait-Bryan angles}. Accordingly, the roll, pitch, and yaw angles are chosen to be around the $x$, $y$, and $z$--axes in the corresponding frames, respectively. These roll, pitch, and yaw angles are commonly referred to as the \textit{Euler angles}. We have defined different frames of reference to characterize the equations of motion of the quadrotor. Thus, we now obtain the relationship among these frames using coordinate transformation. The transformation matrix for the given sequence of rotation $(zy'x'')$ from $\mathcal{I}$ to $\mathcal{B}$ is given by
	\begin{align}
		\nonumber\mathcal{R}(\phi,\theta,\psi)=&~\mathcal{R}(x'',\phi)\mathcal{R}(y',\theta)\mathcal{R}(z,\psi)
		=~\begin{bmatrix}
			1 & 0 & 0 \\
			0 & \cos{\phi} & \sin{\phi} \\
			0 & -\sin{\phi} & \cos{\phi}
		\end{bmatrix}
		\begin{bmatrix}
			\cos{\theta} & 0 & -\sin{\theta} \\
			0 & 1 & 0 \\
			\sin{\theta} & 0 & \cos{\theta}
		\end{bmatrix}
		\begin{bmatrix}
			\cos{\psi} & \sin{\psi} & 0 \\
			-\sin{\psi} & \cos{\psi} & 0 \\
			0 & 0 & 1
		\end{bmatrix}\\
		=&~\begin{bmatrix}
			\cos{\theta} \cos{\psi} &\cos{\theta} \sin{\psi}  & -\sin{\theta} \\
			\sin{\phi} \sin{\theta} \cos{\psi}-\cos{\phi} \sin{\psi} & \sin{\phi} \sin{\theta} \sin{\psi}+\cos{\phi} \sin{\psi} & \sin{\phi} \cos{\theta} \\
			\cos{\phi} \sin{\theta} \cos{\psi}+\sin{\phi} \sin{\psi} & \cos{\phi} \sin{\theta} \sin{\psi}-\sin{\phi} \cos{\psi} & \cos{\phi} \cos{\theta}
		\end{bmatrix}, \label{eq:rotationmatrix}
	\end{align}
	where $\mathcal{R}\in \mathcal{SO}(3) \left (\mathcal{SO}(3)\in\mathbb{R}^{3\times3}\,|\,\mathcal{R}^\top\mathcal{R}=I_{3\times3} \;\mathrm{and}\; \det(\mathcal{R})=1\right)$. 
 
 Now, assume $\mathbf{V}(t)=\left[u(t)\quad v(t)\quad w(t)\right]^\top\in\mathbb{R}^3$ and $\mathbf{\Omega}(t)=\left[p(t)\quad q(t) \quad r(t)\right]^\top\in\mathbb{R}^3$ be the linear and angular velocity vectors, respectively, of the CoG of the quadrotor expressed in $\mathcal{B}$. The quadrotor's CoG position vector in the inertial frame is represented by $\mathbf{P}(t)=\left[x(t)\quad y(t)\quad z(t)\right]^\top\in\mathbb{R}^3$, whereas its orientation in the corresponding frame is denoted by $\mathbf{\Theta}(t)=\left[\phi(t)\quad \theta(t)\quad \psi(t)\right]^\top\in\mathbb{S}^3$ denotes the orientation vector, where $\mathbb{S}^3$ is a \textit{torus} of dimension $3$. Hence, the velocity of the quadrotor expressed in $\mathcal{B}$ can be transformed into $\mathcal{I}$ as
	\begin{equation}\label{eq:rhodot}
		\dot{\mathbf{P}}(t)=\left[\dot{x}(t)\quad \dot{y}(t)\quad \dot {z}(t)\right]^\top=\mathcal{R}^\top\mathbf{V}(t)=\mathcal{R}^\top\left[u(t)\quad v(t) \quad w(t)\right]^\top.
	\end{equation}
	On expanding \eqref{eq:rhodot}, one may obtain the linear 
   velocities expressed in the inertial frame as
	\begin{subequations}\label{eq:linearvelocities}
		\begin{align}
			\dot{x}(t)=&~   (\cos \theta \cos \psi) u+(\cos \psi  \sin \theta  \sin \phi-\cos \phi  \sin \psi)  v
			+(\cos \psi  \sin \theta  \cos \phi+\sin \psi  \sin \phi)  w,\\
			\dot{y}(t)=&~  (\cos \theta \sin \psi) u+(\sin \psi  \sin \theta  \sin \phi+\cos \psi  \cos \phi)  v 
			+(\sin \psi  \sin \theta  \cos \phi-\sin \phi  \cos \psi) w,\\
			\dot{z}(t)=&~(-\sin \theta)  u+(\sin \phi  \cos \theta)  v+(\cos \phi\cos \theta  )  w.
		\end{align}
	\end{subequations}
	The body-fixed angular velocity vector, $\mathbf{\Omega}$, and the Euler rate vector, $\dot{\mathbf{\Theta}}$, are related through a transformation matrix $\mathcal{T}(\Theta)$ as
	\begin{equation}\label{eq:angularrate}
		\dot{\mathbf{\Theta}}=\mathcal{T}(\mathbf{\Theta})\mathbf{\Omega}\implies
		\left[\dot{\phi} \quad \dot{\theta} \quad \dot {\psi}\right]^\top=\mathcal{T}(\mathbf{\Theta})\left[p\quad q \quad r\right]^\top,~ \mathrm{where}~\mathcal{T}(\mathbf{\Theta})=\begin{bmatrix}
			1 & \tan{\theta}  \sin{\phi} & \tan{\theta}  \cos{\phi} \\
			0 & \cos{\phi} & -\sin{\phi} \\
			0 & \sec \theta  \sin \phi & \sec \theta  \cos \phi
		\end{bmatrix}.
	\end{equation}
	On expanding \eqref{eq:angularrate}, we obtain the relation between Euler rates and the angular velocities in $\mathcal{B}$ as
	\begin{subequations}\label{eq:angtoeuler}
		\begin{align}
			\dot{\phi}=&~p+q  \tan \theta  \sin \phi+r  \tan \theta  \cos \phi,\\
			\dot{\theta}=&~q  \cos \phi-r  \sin \phi,\\
			\dot{\psi}=&~q  \sec \theta  \sin \phi+r  \sec \theta  \cos \phi.
		\end{align}
	\end{subequations}
	Equation \eqref{eq:linearvelocities} together with \eqref{eq:angtoeuler} represent the kinematic equations of the quadrotor. 
 \begin{remark}
     Note that $\mathbf{\Omega}(t)$ cannot be integrated directly to obtain the actual angular coordinates due to the fact that $\int_0^t \mathbf{\Omega}(\tau)\;\mathbf{d\tau}$ does not have any physical interpretation. However, the vector $\mathbf{\Theta}(t)$ does represent the proper generalized coordinates.
 \end{remark}
  We now state a few commonly used assumptions for deriving the dynamics of the quadrotor \cite{doi:10.2514/1.59869}.
	\begin{assumption}\label{assum:eulerangles}
		The quadrotor structure is symmetrical in the lateral plane. The roll and pitch angles are assumed to be bounded as $-\dfrac{\pi}{2}<\phi<\dfrac{\pi}{2}$ and $-\dfrac{\pi}{2}<\theta<\dfrac{\pi}{2}$, respectively.
       The thrust and reaction moment generated by the quadrotor varies with the square of the angular speed of rotors. Moreover, the rotors are assumed to be rigid. Thus, no blade flapping is considered in the design.
	\end{assumption}
 \Cref{assum:eulerangles} implies that the quadrotor will not perform an acrobatic maneuver in which it has to undergo an inverted flight. This assumption remains true for almost every application of the quadrotor considered in this paper, such as pipeline inspection, surveillance, etc., and a majority of other practical applications. Further, let $u_T$, $u_\phi$, $u_\theta$, and $u_\psi$ denote the net thrust, roll moment, pitch moment, and yaw moment generated by the quadrotor, respectively. The relation between these quantities can be obtained as
	\begin{equation}\label{eq:omegatoinputs}
		\begin{bmatrix}
			u_T\\
			u_\phi\\
			u_\theta\\
			u_\psi
		\end{bmatrix}
		=
		\begin{bmatrix}
			C_T &  C_T & C_T & C_T \\
			0 & -dC_T & 0 & dC_T \\
			-dC_T & 0 & dC_T & 0 \\
			-C_Q & C_Q & -C_Q & C_Q 
		\end{bmatrix}
		\begin{bmatrix}
			\omega_1^2\\
			\omega_2^2\\
			\omega_3^2\\
			\omega_4^2
		\end{bmatrix},
	\end{equation}
	where $\omega_\mathrm{i}$ denotes the angular speed of the $i$\textsuperscript{th} rotor, and $C_T$ and $C_Q$ are thrust and moment coefficients, respectively. The term $d$ denotes the distance of each rotor from the center of mass of the quadrotor. It is a reasonable assumption to make that the thrust and moments exerted on the quadrotor's body are proportional to the squares of the angular velocity, given that the intensity of aerodynamic forces and moments is known to increase at a rate no faster than the square of the velocity.
 \begin{remark}
     The relationship \eqref{eq:omegatoinputs} is bijective, and one can obtain $C_T$ and $C_Q$ by the static thrust tests. This further allows to designate control inputs, $u_T$, $u_\phi$, $u_\theta$, and $u_\psi$ in lieu of the individual rotor controllers.
 \end{remark}
  The individual rotor speed can be obtained by allocating the control inputs through \eqref{eq:omegatoinputs}. Based on {Newton-Euler} formulation, we can write the translation dynamics of the quadrotor as
	\begin{equation}\label{eq:transdyn1} 
     m\ddot{\mathbf{P}}=\mathbf{F}_{g}+\mathcal{R}^\top u_T-\mathbf{F}_{a},
	\end{equation}
	where $\mathbf{F}_{g}$ represents the gravitational force and $\mathbf{F}_{a}$ is the aerodynamic drag force acting on the quadrotor's CoG. The gravitational and aerodynamic drag force is obtained as $\mathbf{F}_{g}=-mg\left[0\quad 0\quad 1\right]^\top$ and $\mathbf{F}_{a}=\mathrm{diag}\left[K_x\quad K_y \quad K_z\right]\dot{\mathbf{P}}$, where $K_{\ell}$ for $\ell=x$, $y$, and $z$ are drag coefficients and diag represents a diagonal matrix of appropriate dimension. 
    \begin{assumption}
        We assume a linear drag force, consistent with Stokes' law, which is a simple assumption yet well justified in the context of quadrotor UAVs with symmetrical structures and small rotors, as discussed in \Cref{assum:eulerangles} and supported by existing literature, e.g., \cite{doi:10.2514/1.59869}.
    \end{assumption}
On substituting the value of $\mathbf{F}_{g}$ and $\mathbf{F}_{a}$ in \eqref{eq:transdyn1} and using \eqref{eq:omegatoinputs}, we obtain
	\begin{equation*}
		m\ddot{\mathbf{P}}=\begin{bmatrix}0 \\ 0\\ -mg \end{bmatrix}+\mathcal{R}^\top\begin{bmatrix}0 \\ 0\\ u_{T} \end{bmatrix}-\begin{bmatrix}
			K_x & 0 & 0 \\
			0 & K_y & 0 \\
			0 & 0 & K_z 
		\end{bmatrix}\dot{\mathbf{P}}\implies
		\ddot{\mathbf{P}}=\begin{bmatrix}0 \\ 0\\ -g \end{bmatrix}+\mathcal{R}^\top\begin{bmatrix}0 \\ 0\\ {u_{T}}/{m} \end{bmatrix}-\dfrac{1}{m}\begin{bmatrix}
			K_x & 0 & 0 \\
			0 & K_y & 0 \\
			0 & 0 & K_z 
		\end{bmatrix}\dot{\mathbf{P}},
	\end{equation*}
	which, alternately, can be written as
 \begin{subequations}
		\begin{align}
			\ddot x &= \frac{u_T}{m}( \cos\phi \sin\theta \cos\psi + \sin\phi \sin\psi )-\dfrac{K_x\dot x}{m},  \label{eq:ddotx}\\
			\ddot y &= \frac{u_T}{m}( \cos\phi \sin\theta \sin\psi - \sin\phi \cos\psi )-\dfrac{K_y\dot y}{m}, 
			\label{eq:ddoty}\\
			\ddot z &= \frac{u_T}{m}\left(  \cos\phi\cos\theta  \right) -g-\dfrac{K_z\dot z}{m}
			\label{eq:ddotz}
		\end{align}
	\end{subequations}
 in the inertial frame.
 
	In addition, for the quadrotor UAV rotational dynamics, we have
	\begin{equation}\label{eq:rotdyn}
		\mathcal{J}\dot{\mathbf{\Omega}}=-\mathbf{\Omega}^\times (\mathcal{J}{ \mathbf{\Omega}})+\tau_{\mathcal{B}},
	\end{equation}
	where $m$ is the mass of quadrotor, $``\times"$ denotes the cross product, $\mathcal{J}  \in \mathbb{R}^{3\times3}$ represents the inertia matrix, $\tau_{\mathcal{B}} \in \mathbb{R}^{3}$ is the vector of torques acting on the CoG of the quadrotor, expressed in the body-fixed frame, $\mathcal{B}$. Due to the symmetrical structure of the quadrotor, the inertia matrix reduces to a diagonal matrix. Thus, $\mathcal{J}=\mathrm{diag}[J_{\rm xx}\quad  J_{\rm yy}\quad  J_{\rm zz}]$. 
    \begin{remark}
        the performance of the control design usually relies on the knowledge of the principal moments of inertia, $J_{\rm xx}$, $J_{\rm yy}$, and $J_{\rm zz}$, which can be determined to some accuracy via experiments or some computer-aided design. Most of the works in the literature assume that the inertia matrix is precisely known, which may not be true in practice due to limited manufacturing and experimental accuracy.
    \end{remark}
    Different from such considerations and to enhance the generalizability, we consider that we only have partial information about the inertia matrix to bring the proposed design closer to practice. We decompose the partially known inertia matrix into a known and an unknown part as $\mathcal{J}=\mathcal{J}_{\textsubscript{0}}+\mathcal{J}_\Delta$, where $\mathcal{J}_{\textsubscript{0}} \in\mathbb{R}^{3\times3}$ is the nominal part which is known and  $\mathcal{J}_\Delta\in\mathbb{R}^{3\times3}$ is the uncertain part, leading to the following assumption on the boundedness of the uncertainty.
	\begin{assumption}\label{assum:inertia}
		The uncertain part of inertia matrix $\mathcal{J}_\Delta$ is upper bounded by $\mu$, that is, $\|\mathcal{J}_\Delta\|\leq \mu$, where $\mu\in \mathbb{R}_+$.
	\end{assumption}
	Realizing that $\mathcal{J}=\mathcal{J}_{\textsubscript{0}}+\mathcal{J}_\Delta$, one can  write \eqref{eq:rotdyn} as
	\begin{equation*}
  \left(\mathcal{J}_{\textsubscript{0}}+\mathcal{J}_\Delta\right)\dot{\mathbf{\Omega}}=-\mathbf{\Omega}^\times \left(\left(\mathcal{J}_{\textsubscript{0}}+\mathcal{J}_\Delta\right){ \Omega}\right)+\tau_{\mathcal{B}} \implies
 \mathcal{J}_{\textsubscript{0}}\dot{\mathbf{\Omega}}=-\mathbf{\Omega}^\times (\mathcal{J}_{\textsubscript{0}}{ \mathbf{\Omega}})-\mathbf{\Omega}^\times(\mathcal{J}_\Delta\mathbf{\Omega})-\mathcal{J}_\Delta\dot{\mathbf{\Omega}}+\tau_{\mathcal{B}},
	\end{equation*}
	resulting in
	\begin{equation}\label{eq:rotdyn1}
		\dot{\mathbf{\Omega}}=\mathcal{J}^{-1}_{\textsubscript{0}}\left[-\mathbf{\Omega}^\times (\mathcal{J}{ \mathbf{\Omega}})+\tau_{\mathcal{B}}\right]+\mathbf{h},
	\end{equation}
	 where $\mathbf{h}= [{h}_\phi\quad{h}_\theta\quad{h}_\psi]^\top=\mathcal{J}^{-1}_0\left[-\mathbf{\Omega}^\times(\mathcal{J}_\Delta\mathbf{\Omega})-\mathcal{J}_\Delta{\dot{\mathbf{\Omega}}}\right]$ denotes a vector of lumped uncertainties.  Note that under \Cref{assum:inertia}, the lumped uncertainty $\mathbf{d}$ satisfies the condition $\|\mathbf{h}\|\leq \gamma$, where $\gamma\in\mathbb{R}_+$ is known constant that depends on $\mu$. One can obtain the net moment acting on the quadrotor's CoG, expressed in $\mathcal{B}$ as
	\begin{align}\label{eq:tau}
		\tau_{\mathcal{B}}=\begin{bmatrix} d(F_{4}-F_{2}) \\ d(F_{3}-F_{1}) \\ C_Q(F_{2}+F_4-F_{1}-F_{3}) \end{bmatrix} - {J}_{r}\left({\mathbf{\Omega}}^\times\begin{bmatrix}0 \\0 \\1\end{bmatrix}\right)\omega_{\mathrm{r}}
		=\begin{bmatrix} u_{\phi} \\ u_{\theta} \\ u_{\psi} \end{bmatrix} - {J}_{\rm r}\begin{bmatrix}q \\-p \\0 \end{bmatrix}\omega_{\mathrm{r}},
	\end{align}
	where $F_{\mathrm{i}}=C_T\omega_{\mathrm{i}}^2~\forall\,i=1,2,3,4$ denotes the net thrust generated by $\mathrm{i}^{\textsuperscript{th}}$ rotor, $\omega_{\mathrm{r}}$ represents the relative speed between opposite rotating rotor pairs which can be determined from
 \begin{equation}
     \omega_{\mathrm{r}}=\left(\omega_{1}-\omega_{2}+\omega_{3}-\omega_{4}\right),
 \end{equation}
and ${J}_{\mathrm{r}}$ represents the inertia of the rotor blades. In \eqref{eq:tau}, the first term corresponds to the moments generated by the actuators, whereas the second term refers to the gyroscopic moment. 

In practical applications, adhering to strict safety requirements limits the attitude angles, which are always kept small during the flight. Thus it follows from \eqref{eq:angtoeuler} that $[\dot\phi\quad \dot\theta \quad \dot\psi]^\top\approxeq [p \quad q \quad r]^\top$. Now, using this relation together with \eqref{eq:tau}, one can write the attitude dynamics of the quadrotor from \eqref{eq:rotdyn1} as
\begin{align*}
    \ddot{\phi} =&~ \xi_1\dot \theta \dot \psi -\xi_2\Omega_r\dot \theta + \dfrac{u_{\phi}}{J_{\rm xx}}  + h_{\phi},\\
     \ddot{\theta} =&~ \xi_3\dot \phi \dot \psi +\xi_4\Omega_r \dot \phi + \dfrac{u_{\theta}}{J_{\rm yy}}  + h_{\theta},\\
     \ddot{\psi} =&~ \xi_5\dot \phi \dot\theta + \dfrac{u_{\psi}}{J_{\rm zz}}  + h_{\psi},
\end{align*}
which can be written as
 \begin{equation}\label{eq:attitudedynamics}
    \ddot{\Theta}_{k}= \mathcal{F}_{k}+\mathcal{G}_{k} u_k+h_{k},
 \end{equation}
 where $k=\phi, \theta, \psi$. Additionally, $\mathcal{F}_\phi=\xi_1\dot \theta \dot \psi -\xi_2\Omega_r\dot \theta$, $\mathcal{F}_\theta=\xi_3\dot \phi \dot \psi +\xi_4\Omega_r \dot \phi$, and $\mathcal{F}_\psi=\xi_5\dot \phi \dot\theta$, $\mathcal{G}_\phi=1/J_{\rm xx}$, $\mathcal{G}_\theta=1/J_{\rm yy}$, and $\mathcal{G}_\psi=1/J_{\rm zz}$. The terms $\xi_1=\frac{J_{\rm yy}-J_{\rm zz}}{J_{\rm xx}}$, $\xi_2=\frac{J_{\rm r}}{J_{\rm xx}}$, $\xi_3=\frac{J_{\rm zz}-J_{\rm xx}}{J_{\rm yy}}$, $\xi_4=\frac{J_{\rm r}}{J_{\rm yy}}$, and $\xi_5=\frac{J_{\rm xx}-J_{\rm yy}}{J_{\rm zz}}$.
\begin{remark}
    The relations \eqref{eq:attitudedynamics} can be thought of as a second-order nonlinear system with state variables $\left[\phi, \dot{\phi}, \theta, \dot{\theta}, \psi, \dot{\psi}\right]^\top$, and control inputs $\left[u_{\phi},u_{\theta},u_{\psi}\right]^\top$, subject to bounded uncertainty, $\left[h_{\phi},h_{\theta},h_{\psi}\right]^\top$.
\end{remark}
\subsection{Problem Statement}
\begin{figure}[ht!]
    \centering
    \includegraphics[width=0.5\linewidth]{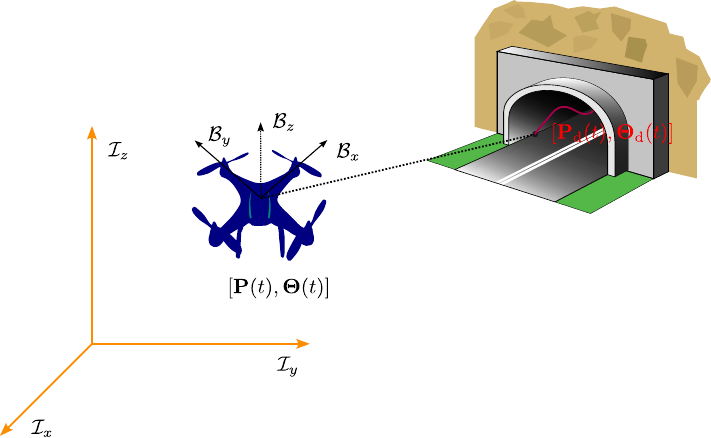}
    \caption{Sample illustration of the problem: UAV inspecting a tunnel.}
    \label{fig:controlproblem}
\end{figure}

Consider a quadrotor UAV flying in three dimensions with spatial constraints on its motion, as depicted in \Cref{fig:controlproblem}. Let us define $\mathbf{P}_\mathrm{d}(t)=\left[x_\mathrm{d}(t)\quad y_\mathrm{d}(t) \quad z_\mathrm{d}(t)\right]^\top$ and $\mathbf{\Theta}_\mathrm{d}(t)= \left[\phi_\mathrm{d}(t)\quad \theta_\mathrm{d}(t)\quad \psi_\mathrm{d}(t)\right]^\top$ as the vectors of smooth time-varying desired position and orientation trajectories of the quadrotor, respectively, such that, the elements of $\mathbf{P}_\mathrm{d}(t)$ and $\mathbf{\Theta}_\mathrm{d}(t)$ are at least thrice differentiable. Similarly, $\mathbf{P}(t)$ and $\mathbf{\Theta}(t)$ denote the instantaneous position and orientation of the quadrotor, respectively. From hereafter, we denote the elements of the position vector $\mathbf{P}(t)=\left[x(t)\quad y(t) \quad z(t) \right]^\top$ as $\mathcal{P}_\ell$ and $\mathbf{P}_\mathrm{d}=\left[x_\mathrm{d}(t)\quad y_\mathrm{d}(t) \quad z_\mathrm{d}(t)\right]^\top$ as $\mathcal{P}_{d_\ell}$, for $\ell=\{x,y,z\}$. Quite similarly, elements of $\mathbf{\Theta}(t)$ are $\Theta_k$ and those of $\mathbf{\Theta}_\mathrm{d}(t)$ are $\Theta_{d_k}$ for $k=\{\phi,\theta,\psi\}$.

 The goal of this paper is to devise a robust nonlinear control strategy that enables the quadrotor UAV to track the desired 3D time-varying trajectories, $\mathbf{P}_\mathrm{d}(t)$ and $\mathbf{\Theta}_d(t)$, without violating the spatial constraints imposed on it. In other words, $\mathcal{P}_\ell <\mathcal{L}_\ell$ and $\Theta_k<\mathcal{Q}_k$ for some positive constants $\mathcal{L}_\ell$ and $\mathcal{Q}_k$. It is worth noting that our control strategy is developed within a nonlinear framework, which enables the proposed design to maintain its validity over a broader range of operating conditions. Next, we present some preliminaries that aid the design.
 \begin{assumption}\label{assum:li}
 For any given positive value of $\mathcal{L}_\ell$, there exist positive constants $\underline{\mathcal{Y}_{0_\ell}}$, $\overline{\mathcal{Y}_{0_\ell}}$, $\mathcal{A}_{0_\ell}$, $\mathcal{Y}_{1_\ell}$, $\mathcal{Y}_{2_\ell}$ that satisfy the inequality $\max\{\underline{\mathcal{Y}_{0_\ell}}$, $\overline{\mathcal{Y}_{0_\ell}}\}\leq \mathcal{A}_{0_\ell} <\mathcal{L}_\ell$. Furthermore, these constants ensure that the elements of $\mathbf{P}_\mathrm{d}(t)$ and their time derivatives are bounded as follows: $-\underline{\mathcal{Y}_{0_\ell}}\leq|\mathcal{P}_{\mathrm{d}_\ell}(t)|\leq \overline{\mathcal{Y}_{0_\ell}}$, $|\dot{\mathcal{P}}_{\mathrm{d}_\ell}(t)|\leq \mathcal{Y}_{1_\ell}$, $|\ddot{\mathcal{P}}_{\mathrm{d}_\ell}(t)|\leq \mathcal{Y}_{2_\ell}$, $\forall\, t \geq 0$.

\end{assumption}	
\begin{assumption}\label{assum:qi}
For any given positive value of $\mathcal{Q}_{k}$, there exist positive constants $\underline{\mathcal{X}_{0_k}}$, $\overline{\mathcal{X}_{0_k}}$, $\mathcal{B}_{0_k}$, $\mathcal{X}_{1_k}$, $\mathcal{X}_{2_k}$ that satisfy the inequality $\max\{\underline{\mathcal{X}_{0_k}}$, $\overline{\mathcal{X}_{0_k}}\}\leq \mathcal{B}_{0_k} <\mathcal{Q}_k$. Moreover, these constants ensure that the elements of $\mathbf{\Theta}_\mathrm{d}(t)$ and their time derivatives are bounded as follows: $\underline{\mathcal{X}_{\mathrm{0i}}}\leq |\Theta_{\mathrm{d}_k}(t)|\leq \overline{\mathcal{X}_{0_k}}$, $| \dot{\Theta}_{\mathrm{d}_k}(t)| \leq \mathcal{X}_{1_k}$, $|\ddot{\Theta}_{\mathrm{d}_k}(t)| \leq \mathcal{X}_{2_k}$, $\forall \,t \geq 0$.
\end{assumption}
\Cref{assum:li,assum:qi} essentially mean that the desired trajectories and their first and second-time derivatives are bounded, which holds true in most of the practical applications. Mathematically, it points to the fact that $\mathbf{P}_{d_\ell}$ and $\mathbf{\Theta}_{d_k}$ belong to $\mathbb{C}^{2}$, where $\mathbb{C}$ denotes the class of differentiable functions. In order to guarantee that the quadrotor UAV never violates the given spatial constraints, we employ a Barrier Lyapunov function (BLF) defined below. Thereafter, in the subsequent lemma, a sufficient condition is provided for the Lyapunov function candidate so that the states of the nonlinear system (the dynamics of the quadrotor UAV) remain in a prespecified open set.
    \begin{definition}
		(\cite{doi:10.1016/j.automatica.2008.11.017}) A BLF is a scalar function $V(x)$, defined with respect to the system $\dot{x} = f(x)$ on an open region $D$ containing the origin that is continuous, positive definite, has continuous first-order partial derivatives at every point of $D$, has the property $V(x)\to\infty$ as $x$ approaches the boundary of $D$, and satisfies $V\left(x\left(t\right)\right) \leq b \; \forall \; t \geq 0$ along the solution of $\dot x = f\left(x\right)$ for $x\left(0\right) \in D$.
	\end{definition}
	\begin{lemma}\label{lem:lemma1}(\cite{doi:10.1016/j.automatica.2008.11.017})
        Consider the system 
        \begin{equation}
            \dot{\eta} = 	\varpi (t, \eta)
        \end{equation}
		where $\eta:=\left[w, z_{1}\right]^{\top} \in \mathcal{N}$, and $	\varpi: \mathbb{R}_+\times \mathcal{N} \rightarrow \mathbb{R}^{l+1}$ is piecewise continuous in $t$ and locally Lipschitz in $z$, uniform in $t$, on $\mathbb{R}_{+} \times \mathcal{N}$. For any positive constants $k_{a_{1}}, k_{b_{1}}$, let $z_{1}:=\left\{z_{1} \in \mathbb{R}:-k_{a_{1}}<z_{1}<k_{b_{1}}\right\} \subset \mathbb{R}$ and $\mathcal{N}:=\mathbb{R}^{l} \times z_{1} \subset \mathbb{R}^{l+1}$ be open sets.  Suppose that there exist functions $U: \mathbb{R}^{l} \rightarrow \mathbb{R}$, and $V_{1}: L_{1} \rightarrow \mathbb{R}_{+}$ continuously differentiable and positive definite in their respective domains, such that
		\begin{align*}
			V_{1}\left(z_{1}\right) &\rightarrow \infty \text { as } z_{1} \rightarrow-k_{a_{1}} \text { or } z_{1} \rightarrow k_{b_{1}} ,\\
			\mathcal{X}_{1}(\|w\|) &\leq U(w) \leq \mathcal{X}_{2}(\|w\|),
		\end{align*}
		where $\mathcal{X}_{1}$ and $\mathcal{X}_{2}$ are class $K_{\infty}$ functions. Let $V(\eta):=V_{1}\left(z_{1}\right)+U(w)$, and $z_{1}(0)$ belong to the set $z_{1} \in\left(-k_{a_{1}}, k_{b_{1}}\right)$. If the inequality $$\dot{V}=\frac{\partial V}{\partial \eta} 	\varpi \leq 0$$
		holds then $z_{1}(t)$ remains in the open set $z_{1} \in\left(-k_{a_{1}}, k_{b_{1}}\right) \forall t \in[0, \infty)$.
	\end{lemma}
\section{Main Results}\label{sec:main}
\begin{figure}[!ht]
		\centering
		\includegraphics[width=0.65\linewidth]{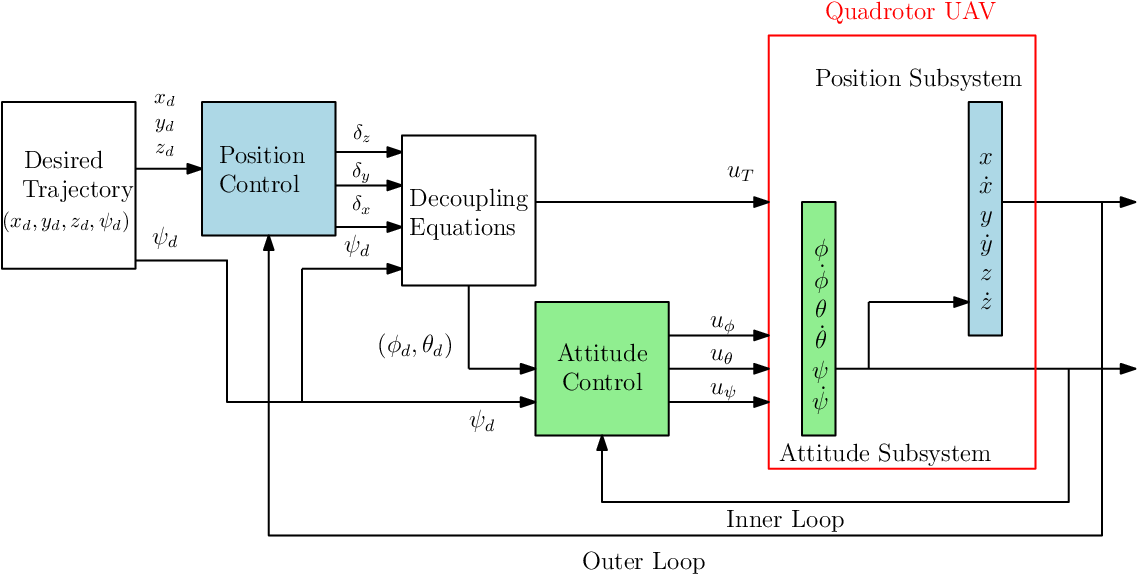}
		\caption{The proposed control architecture for the quadrotor UAV.}
        \label{fig:cloop}
	\end{figure}

In this section, we develop position and attitude controllers to steer the quadrotor on the desired trajectory while adhering to the spatial constraints imposed on its motion.  The results in \cite{doi:10.2514/1.43768} demonstrate that the dynamics of a rotorcraft (quadrotor UAV) can be decoupled into two cascaded subsystems, where the outer loop is characterized by slow dynamics, and the inner loop is governed by fast dynamics. This allows us to develop a hierarchical control architecture for the UAV wherein the inner loop consists of its rotational dynamics, whereas the outer loop consists of the translational dynamics, as shown in \Cref{fig:cloop}. The desired trajectories, $x_d$, $y_d$, $z_d$, and $\psi_d$, are application-specific and provided through a high-level guidance system. The desired trajectories ($x_d$, $y_d$, and $z_d$) are then fed to the positioning subsystem, which subsequently generates the net thrust and the desired roll and pitch angle for the UAV. In essence, the outer loop positioning subsystem serves two purposes: (i) to generate the net thrust command for the UAV to track the given trajectory and (ii) to generate the desired roll and pitch angle for the inner loop. Meanwhile, the inner loop subsystem takes the desired roll and pitch angle from the outer loop and the desired yaw angle or high-level guidance strategy and generates the roll, pitch, and yaw moments for the UAV to track its desired orientation. We now present detailed controller designs for the position and the attitude subsystems in the following subsections.
\subsection{Position Control}\label{subsec:position_control}
\begin{figure}[!ht]
		\centering
		\includegraphics[width=0.85\linewidth]{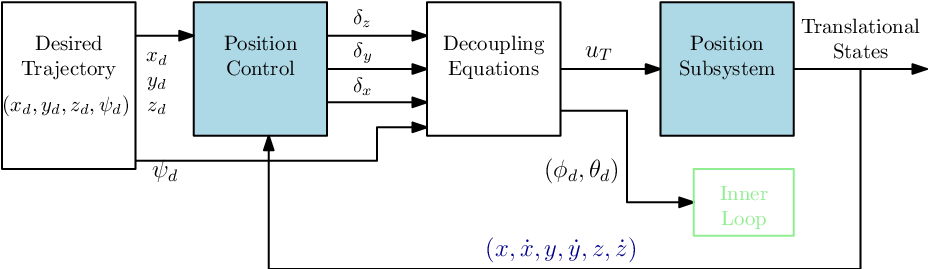}
		\caption{Illustration of the outer loop design.}
        \label{fig:outer_loop}
	\end{figure}

The goal of the proposed trajectory tracking control for a quadrotor is to guide it along a predetermined path while respecting spatial constraints imposed on its motion. To achieve this, we must develop control inputs that steer the UAV toward its desired trajectory in the 3D space in the presence of the given spatial constraints. A closed-loop block diagram representation of the outer loop position control is depicted in \Cref{fig:outer_loop}. Based on the mission requirements, the desired position trajectory is fed into the position control subsystem, and using the UAV's current position information, the net thrust and the desired roll and pitch angles are generated. To this end, let us define $\mathbf{E}_\mathrm{P}=\mathbf{P}-\mathbf{P}_{\mathrm{d}}$ and refer to the elements of $\mathbf{E}_\mathrm{P}$ as $\left[\gamma_x\quad\gamma_y\quad\gamma_z\right]^\top$. To be consistent with our previous notations for elementwise referencing, it is understood that $\gamma_\ell$ is an element of $\mathbf{E}_\mathrm{P}$ for all $\ell=\{x,y,z\}$. Owing to the fact that the position subsystem is underactuated, we introduce virtual control inputs, which effectively transform it into a fully actuated system. To this end, we define a virtual control input vector $\mathbf{\Delta} \in \mathbb{R}^{3}$ as
\begin{equation}\label{eq:Delta}
    \mathbf{\Delta}=\mathscr{F}(u_T,\phi_{d},\theta_{d},\psi_{d})=\dfrac{\mathcal{R}^{\top}(\phi_{d},\theta_{d},\psi_{d}) u_T \mathcal{I}_{z}}{m} - g\mathcal{I}_{z},
\end{equation}
where $\mathscr{F}(.):\mathbb{R}^{3} \to \mathbb{R}^{3}$ is a continuous invertible function. Physically, the vector $m\mathbf{\Delta}$ represents the desired force vector whose magnitude denotes the net thrust $u_T$, and its orientation is given by the Euler angles $(\phi,\theta,\psi)$. Thus, the angles in \eqref{eq:Delta} correspond to the desired roll, pitch, and yaw angles. The expression in \eqref{eq:Delta} can be written in the elementwise form as
	\begin{subequations}\label{eq:deltas}
		\begin{align}
			\delta_x=&~\frac{u_T}{m}( \cos{\phi_{d}} \sin{\theta_{d}} \cos\psi_{d} + \sin\phi_{d} \sin\psi_{d} ),\label{eq:deltaxdef}\\
			\delta_y=&~\frac{u_T}{m}( \cos\phi_{d} \sin\theta_{d} \sin\psi_{d} - \sin\phi_{d} \cos\psi_{d} ),\label{eq:deltaydef}\\
			\delta_z=&~\frac{u_T}{m}\left(  \cos\phi_{d} \cos\theta_{d}  \right) -g. \label{eq:deltazdef}
		\end{align}
	\end{subequations}
One may also think of these virtual control inputs as the component of the acceleration vector required for the translational dynamics (positioning subsystem) to track a predefined 3D trajectory while adhering to the spatial constraints. In other words, the same can be thought of as the force vector components (after multiplying with the mass of the quadrotor) needed to track the desired position trajectory.
 
Using \eqref{eq:deltas}, the translational dynamics of the quadrotor, given by \eqref{eq:ddotx} -- \eqref{eq:ddotz}, can be written as
	\begin{equation}\label{eq:transdynnew}
		\ddot{\mathcal{P}}_{\ell}=\delta_\ell-\frac{K_\ell\dot{\mathcal{P}}_\ell}{m}.
	\end{equation}
	On subsequent time differentiation of position error vector, $\mathbf{E}_\mathrm{P}$, one may obtain
	\begin{equation}\label{eq:gammapiddot}
		\ddot{\gamma}_{\ell}=\ddot{\mathcal{P}}-\ddot{\mathcal{P}}_{d_\ell}=\delta_\ell-\frac{K_\ell\dot{\mathcal{P}_{\ell}}}{m}-\ddot{\mathcal{P}}_{d_\ell}.
	\end{equation}
It may be noted from \eqref{eq:gammapiddot} that the error dynamics \eqref{eq:gammapiddot} has a relative degree of two with respect to the virtual control input $\delta_\ell$. 

We next endeavor to design the virtual control input, $\delta_\ell$, that steer the quadrotor towards its desired position trajectory $\mathbf{P}_\mathrm{d}(t)$, without violating the spatial constraints. In essence, our aim is to drive the position error vector to zero while respecting its bound. This is discussed in the next theorem.
	\begin{theorem}\label{thm:deltai}
		Consider the UAV's translational dynamics, given in \eqref{eq:transdynnew}, and the position error dynamics as \eqref{eq:gammapiddot}. If the UAV's virtual control input for the position subsystem is designed as
		\begin{align}
\nonumber			\delta_\ell=&~\dfrac{K_\ell\mathcal{P}_{\ell}}{m}+\ddot{\mathcal{P}}_{d_\ell}-\gamma_{\ell}\left[\dfrac{1-q(\gamma_{\ell})}{\underline{\gamma_{\ell}}^2-\gamma_{\ell}^2}
			+\dfrac{q(\gamma_{\ell})}{\overline{\gamma_{\ell}}^2-\gamma_{\ell}^2}\right]-\mathcal{K}_{\ell}\gamma_{\ell}^2 \dot{\gamma}_{\ell} \left[3q(\gamma_{\ell})\left(\overline{\gamma_{\ell}}^2-\underline{\gamma_{\ell}}^2\right)+\left(3\underline{\gamma_{\ell}}^2-5\gamma_{\ell}^2\right)\right]\\
   &~-\mathcal{M}_{\ell}\left[\dot{\gamma}_{\ell}+\left\{q(\gamma_{\ell})\left(\overline{\gamma_{\ell}}^2-\gamma_{\ell}^2\right)+\left(1-q(\gamma_{\ell})\right)\left(\underline{\gamma_{\ell}}^2-\gamma_{\ell}^2\right)\right\}\mathcal{K}_{\ell}\gamma_{\ell}^3\right],\label{eq:deltai}
		\end{align}
where $\mathcal{K}_{\ell}$, $\mathcal{M}_{\ell}$ are positive constants, with the error bounds $\underline{\gamma_{\ell}}$ and $\overline{\gamma_{\ell}}$, chosen as, $\underline{\gamma_{\ell}}=\mathcal{L}_{\ell}-\underline{\mathcal{Y}_{0_\ell}}$ and $\overline{\gamma_{\ell}}=\mathcal{L}_{\ell}-\overline{\mathcal{Y}_{0_\ell}}$, 
with $q(\gamma_{\ell})$, as
\begin{equation}
			q(\gamma_{\ell})=\begin{cases}
				1; \quad \gamma_{\ell} \in (0,\infty),\\
				0;  \quad \gamma_{\ell}\in (-\infty,0],
	       \end{cases}
		\end{equation}
  then the position errors remain bounded within  $\underline{\gamma_{\ell}}<\gamma_{\ell}<\overline{\gamma_{\ell}}$, $\forall$ $t\geq0.$ 
	\end{theorem}   
	\begin{proof}
		Consider an asymmetric barrier Lyapunov function candidate as
		\begin{align}\label{eq:pos_lyapunov}
			\mathcal{W}_{\ell}=&~\dfrac{1-q(\gamma_{\ell})}{2}\ln\left[\dfrac{\underline{\gamma_{\ell}}^2}{\underline{\gamma_{\ell}}^2-\gamma_{\ell}^2}\right]+\dfrac{q(\gamma_{\ell})}{2}\ln\left[\dfrac{\overline{\gamma_{\ell}}^2}{\overline{\gamma_{\ell}}^2-\gamma_{\ell}^2}\right].
		\end{align}
		We drop the argument of $q(\gamma_{\ell})$ for brevity, and it will be understood that $q$ is a function of $\gamma_{\ell}$. It can be observed that the asymmetric barrier Lyapunov candidate, given in \eqref{eq:pos_lyapunov}, is a convex combination of two barrier Lyapunov functions based on the function $q$. We define $\gamma_{a_\ell}:=\gamma_{\ell}/\underline{\gamma_{\ell}},\gamma_{b_\ell}=\gamma_{\ell}/\overline{\gamma_{\ell}}$, and the term $\hat{\gamma}_{\ell}=(1-q)\gamma_{a_\ell}+q\gamma_{b_\ell}$,
		which is essentially a convex combination of $\gamma_{a_\ell}$ and $\gamma_{b_\ell}$. Then, one may simplify the Lyapunov function candidate to
		\begin{equation}\label{eq:pos_wcompactblf}
		     \mathcal{W}_{\ell}=\dfrac{1-q}{2}\ln\left[\dfrac{1}{1-\gamma_{a_\ell}^2}\right]+\dfrac{q}{2}\ln\left[\dfrac{1}{1-\gamma_{b_\ell}^2}\right]=-\dfrac{1-q}{2}\ln\left[1-\gamma_{a_\ell}^2\right]-\dfrac{q}{2}\ln\left[1-\gamma_{b_\ell}^2\right].
		\end{equation}
		The expression in \eqref{eq:pos_wcompactblf} can be written in compact form using the relation $\hat{\gamma}_{\ell}=(1-q)\gamma_{a_\ell}+q\gamma_{b_\ell}$ as
		\begin{equation}\label{eq:pos_compactBLF}
			\mathcal{W}_{\ell}= \dfrac{q}{2}\ln\left[\dfrac{1}{1- \hat{\gamma}_{\ell}}\right],
		\end{equation}
		which is positive definite and continuously differentiable in the set $|\hat{\gamma}_{\ell}|<1$, thus it is a valid Lyapunov function candidate. On differentiating \eqref{eq:pos_lyapunov} with respect to time, one may obtain
		\begin{equation*}
		   \dot{\mathcal{W}}_{\ell}=\dfrac{1-q}{2}\left[\dfrac{\underline{\gamma_{\ell}}^2-\gamma_{\ell}^2}{\underline{\gamma_{\ell}}^2}\right]\left[\dfrac{-\left(\underline{\gamma_{\ell}}^2\right)^2}{\left(\underline{\gamma_{\ell}}^2-\gamma_{\ell}^2\right)^2}\right]\left(\dfrac{-2\gamma_{\ell}\dot \gamma_{\ell}}{\underline{\gamma_{\ell}}^2}\right)
			+
			\dfrac{q}{2}\left[\dfrac{\overline{\gamma_{\ell}}^2-\gamma_{\ell}^2}{\overline{\gamma_{\ell}}^2}\right]\left[\dfrac{-\left(\overline{\gamma_{\ell}}^2\right)^2}{\left(\overline{\gamma_{\ell}}^2-\gamma_{\ell}^2\right)^2}\right]\left(\dfrac{-2\gamma_{\ell}\dot\gamma_{\ell}}{\overline{\gamma_{\ell}}^2}\right),
		\end{equation*}
		which, on some algebraic simplifications, results in
		\begin{equation}\label{eq:wdot1}
			\dot{\mathcal{W}}_{\ell}=\dfrac{1-q}{2}\left[\dfrac{2\gamma_{\ell}\dot \gamma_{\ell}}{\underline{\gamma_{\ell}}^2-\gamma_{\ell}^2}\right]
			+
			\dfrac{q}{2}\left[\dfrac{2\gamma_{\ell}\dot \gamma_{\ell}}{\overline{\gamma_{\ell}}^2-\gamma_{\ell}^2}\right].
		\end{equation}
		On rearranging the terms of \eqref{eq:wdot1}, we get
		\begin{align}\label{eq:wdot2}
			\dot{\mathcal{W}}_{\ell}=&~\left[\dfrac{1-q}{\underline{\gamma_{\ell}}^2-\gamma_{\ell}^2}
			+
			\dfrac{q}{\overline{\gamma_{\ell}}^2-\gamma_{\ell}^2}\right]  \gamma_{\ell}\dot \gamma_{\ell}.
		\end{align}
		Now, assume a variable $\zeta_\ell:=\dot \gamma_{\ell}-\beta_\ell$ for $\ell=x,y,z$, where $\beta_\ell$ is the stabilizing function to be designed. From this relation, we have $\dot \gamma_{\ell}=\zeta_\ell+\beta_\ell$, which upon substituting in \eqref{eq:wdot2}, leads to
		\begin{equation}\label{eq:wdot3}
			\dot{\mathcal{W}}_{\ell}=\left[\dfrac{1-q}{\underline{\gamma_{\ell}}^2-\gamma_{\ell}^2}
			+
			\dfrac{q}{\overline{\gamma_{\ell}}^2-\gamma_{\ell}^2}\right]  \gamma_{\ell}\left(\zeta_i+\beta_\ell\right).
		\end{equation}
		If we choose the stabilizing function $\beta_\ell$, with $\mathcal{K}_{\ell}>0$, such that
		\begin{equation}\label{eq:betapi}
		  \beta_\ell=  -\dfrac{\left(\underline{\gamma_{\ell}}^2-\gamma_{\ell}^2\right)\left(\overline{\gamma_{\ell}}^2-\gamma_{\ell}^2\right)}{\left[\left(1-q\right)\left(\overline{\gamma_{\ell}}^2-\gamma_{\ell}^2\right)+q\left(\underline{\gamma_{\ell}}^2-\gamma_{\ell}^2\right)\right]}\mathcal{K}_{\ell}\gamma_{\ell}^3
			=-\left[q\left(\overline{\gamma_{\ell}}^2-\gamma_{\ell}^2\right)+\left(1-q\right)\left(\underline{\gamma_{\ell}}^2-\gamma_{\ell}^2\right)\right]\mathcal{K}_{\ell}\gamma_{\ell}^3,
		\end{equation}
		then the derivative of the Lyapunov function candidate given in \eqref{eq:wdot3} becomes
		\begin{equation*}
		    \dot{\mathcal{W}}_{\ell}=\gamma_{\ell}\left[\dfrac{1-q}{\underline{\gamma_{\ell}}^2-\gamma_{\ell}^2}
			+
			\dfrac{q}{\overline{\gamma_{\ell}}^2-\gamma_{\ell}^2}\right]  \left[\zeta_i-\dfrac{\left(\underline{\gamma_{\ell}}^2-\gamma_{\ell}^2\right)\left(\overline{\gamma_{\ell}}^2-\gamma_{\ell}^2\right)\mathcal{K}_{\ell}\gamma_{\ell}^3}{\left\{\left(1-q\right)\left(\overline{\gamma_{\ell}}^2-\gamma_{\ell}^2\right)+q\left(\underline{\gamma_{\ell}}^2-\gamma_{\ell}^2\right)\right\}}\right],
		\end{equation*}
		which is equivalent to
		\begin{equation}\label{eq:wdot4}
		 \dot{\mathcal{W}}_{\ell}=-\mathcal{K}_{\ell}\gamma_{\ell}^4\left[\dfrac{1-q}{\underline{\gamma_{\ell}}^2-\gamma_{\ell}^2}
			+\dfrac{q}{\overline{\gamma_{\ell}}^2-\gamma_{\ell}^2}\right]  \dfrac{\left(\underline{\gamma_{\ell}}^2-\gamma_{\ell}^2\right)\left(\overline{\gamma_{\ell}}^2-\gamma_{\ell}^2\right)}{\left[\left(1-q\right)\left(\overline{\gamma_{\ell}}^2-\gamma_{\ell}^2\right)+q\left(\underline{\gamma_{\ell}}^2-\gamma_{\ell}^2\right)\right]}+\gamma_{\ell}\zeta_i\left[\dfrac{1-q}{\underline{\gamma_{\ell}}^2-\gamma_{\ell}^2}
			+\dfrac{q}{\overline{\gamma_{\ell}}^2-\gamma_{\ell}^2}\right].
		\end{equation}
		After some algebraic simplifications, one may also express \eqref{eq:wdot4} as
		\begin{equation}\label{eq:wdot5}
			\dot{\mathcal{W}}_{\ell}=-\mathcal{K}_{\ell}\gamma_{\ell}^4+    \gamma_{\ell}\zeta_i\left[\dfrac{1-q}{\underline{\gamma_{\ell}}^2-\gamma_{\ell}^2}
			+\dfrac{q}{\overline{\gamma_{\ell}}^2-\gamma_{\ell}^2}\right].
		\end{equation}
		We now consider another Lyapunov candidate, $\mathcal{E}_{\ell}$, by augmenting $\mathcal{W}_{\ell}$ with a quadratic Lyapunov function as
		\begin{equation}\label{eq:wiplusone1}
			\mathcal{E}_{\ell}=\mathcal{W}_{\ell}+\dfrac{1}{2}\zeta_\ell^2
			=
			\dfrac{1-q}{2}\ln\left[\dfrac{\underline{\gamma_{\ell}}^2}{\underline{\gamma_{\ell}}^2-\gamma_{\ell}^2}\right]+\dfrac{q}{2}\ln\left[\dfrac{\overline{\gamma_{\ell}}^2}{\overline{\gamma_{\ell}}^2-\gamma_{\ell}^2}\right]+\dfrac{1}{2}\zeta_i^2.
		\end{equation}
		The time derivative of the Lyapunov candidate, 
		$\mathcal{E}_{\ell}$, can be obtained using \eqref{eq:wdot5} as
		\begin{equation}\label{eq:wiplusone2}
            \dot{\mathcal{E}}_{\ell}=\dot{\mathcal{W}}_{\ell}+\zeta_i\dot \zeta_i=-\mathcal{K}_\ell\gamma_{\ell}^4+    \gamma_{\ell}\zeta_{\ell}\left[\dfrac{1-q}{\underline{\gamma_{\ell}}^2-\gamma_{\ell}^2}+
			+\dfrac{q}{\overline{\gamma_{\ell}}^2-\gamma_{\ell}^2}\right]+\zeta_i\left(\ddot \gamma_{\ell}-\dot \beta_i\right).
		\end{equation}
		One may obtain the derivative of the stabilizing function, $\beta_\ell$, by differentiating  \eqref{eq:betapi} with respect to time, as
		\begin{equation*}
\dot{\beta}_\ell= -3\mathcal{K}_\ell\gamma_{\ell}^2 \dot{\gamma}_{\ell} \left[q\left(\overline{\gamma_{\ell}}^2-\gamma_{\ell}^2\right)+\left(1-q\right)\left(\underline{\gamma_{\ell}}^2-\gamma_{\ell}^2\right)\right]-\mathcal{K}_\ell\gamma_{\ell}^3\left[ q\left(-2\gamma_{\ell} \dot{\gamma}_{\ell}\right)+\left(1-q\right)\left( -2\gamma_{\ell} \dot{\gamma}_{\ell}\right)\right],
		\end{equation*}
		which, on some algebraic simplification, becomes
		\begin{equation}\label{eq:betapidot1}
			\dot \beta_\ell=-3\mathcal{K}_\ell\gamma_{\ell}^2 \dot \gamma_{\ell} \left[q\left(\overline{\gamma_{\ell}}^2-\gamma_{\ell}^2\right)+\left(1-q\right)\left(\underline{\gamma_{\ell}}^2-\gamma_{\ell}^2\right)\right]+2\mathcal{K}_\ell\gamma_{\ell}^4 \dot \gamma_{\ell}.
		\end{equation}
		On rearranging the terms of \eqref{eq:betapidot1}, we get 
		\begin{equation}\label{eq:betapidot}
			    \dot \beta_\ell= -\mathcal{K}_\ell\gamma_{\ell}^2 \dot \gamma_{\ell} \left[3q\left(\overline{\gamma_{\ell}}^2-\gamma_{\ell}^2\right)+3\left(1-q\right)\left(\underline{\gamma_{\ell}}^2-\gamma_{\ell}^2\right)\right]-2\gamma_{\ell}^2
			= -\mathcal{K}\gamma_{\ell}^2 \dot \gamma_{\ell} \left[3q\left(\overline{\gamma_{\ell}}^2-\underline{\gamma_{\ell}}^2\right)+\left(3\underline{\gamma_{\ell}}^2-5\gamma_{\ell}^2\right)\right].
		\end{equation}
		Further, upon substituting the value of $\dot\beta_\ell$ and $\ddot \gamma_{\ell}$ from \eqref{eq:betapidot} and \eqref{eq:gammapiddot} in \eqref{eq:wiplusone2}, one may obtain
		\begin{equation}\label{eq:wiplusone3}
		  \dot{\mathcal{E}}_{\ell}=-\mathcal{K}_\ell\gamma_{\ell}^4+    \gamma_{\ell}\zeta_\ell\left[\dfrac{1-q}{\underline{\gamma_{\ell}}^2-\gamma_{\ell}^2}
			+\dfrac{q}{\overline{\gamma_{\ell}}^2-\gamma_{\ell}^2}\right]+\zeta_\ell\left[\delta_\ell-\frac{K_\ell\dot{\mathcal{P}}_{\ell}}{m}-\ddot{\mathcal{P}}_{d_\ell}+\mathcal{K}\gamma_{\ell}^2 \dot \gamma_{\ell} \left\{3q\left(\overline{\gamma_{\ell}}^2-\underline{\gamma_{\ell}}^2\right)+\left(3\underline{\gamma_{\ell}}^2-5\gamma_{\ell}^2\right)\right\}\right],
		\end{equation}
		where $\delta_{\ell}$ given in \eqref{eq:deltai}. After some algebraic manipulations, the expression in \eqref{eq:wiplusone3} can be simplified to
		\begin{equation}
			\dot{\mathcal{E}}_{\ell}=-\mathcal{K}\gamma_{\ell}^4-\zeta_\ell\mathcal{M}_{\ell}\left[\gamma_{\ell}+\left\{q\left(\overline{\gamma_{\ell}}^2-\gamma_{\ell}^2\right)+\left(1-q\right)\left(\underline{\gamma_{\ell}}^2-\gamma_{\ell}^2\right)\right\}\mathcal{K}_{\ell}\gamma_{\ell}^3\right],
         \end{equation}
  which can be further simplified, using the relation $\zeta_\ell=\gamma_{\ell}-\beta_\ell$ and \eqref{eq:betapi}, to
  \begin{equation}\label{eq:wiplusonefinal}
     \dot{\mathcal{E}}_{\ell}=-\mathcal{K}_{\ell}\gamma_{\ell}^4-\mathcal{M}_{\ell}\zeta_\ell^2\leq 0.
  \end{equation}
		Thus, the considered system,\eqref{eq:gammapiddot}, is stable in the Lyapunov sense. It can be further deduced, using the results of \Cref{lem:lemma1}, that the position error never violates its bound, that is, $-\underline{\gamma_{\ell}}<\gamma_{\ell}(t)<\overline{\gamma_{\ell}}\,\, \forall\,\, t>0 $, provided that the initial errors satisfy $-\underline{\gamma_{\ell}}<\gamma_{\ell}(0)<\overline{\gamma_{\ell}}$. 
\end{proof}
\Cref{thm:deltai} infers that if the UAV's initial position lies in a given set, then its instantaneous position will always remain inside the same set throughout its movement. While we have analytically demonstrated in \Cref{thm:deltai} that the quadrotor position remains within a specified set, the subsequent theorem takes this result a step further by establishing that the position converges asymptotically to its desired value, with the added guarantee of bounded position and velocity signals. 
 \begin{theorem}\label{thm:position_bound}
     Consider the closed-loop systems \eqref{eq:transdynnew} and \eqref{eq:deltai}. Provided the UAV's initial position is such that $\gamma_\ell(0)\in\left(\underline{\gamma_{\ell}},\overline{\gamma_{\ell}}\right)$ in line with \Cref{assum:li}, we can guarantee that the UAV's position remains bounded for all time $t\geq0$. Furthermore, the position remains confined to the set $\mathcal{S}_{\ell}\in  \left(-\mathcal{L}_{\ell},\mathcal{L}_{\ell}\right)$, which is defined as the region $\Lambda_\ell\coloneqq \left\{\mathcal{P}_\ell\;|\;\mathcal{P}_\ell\in \left(-\underline{\mathcal{H}_{\ell}}-\underline{\mathcal{Y}_{\mathrm{0}\ell}},\overline{\mathcal{H}_{\ell}}+\overline{\mathcal{Y}_{\mathrm{0}\ell}}\right)\right\}\subset \mathcal{S}_\ell$, where $\underline{\mathcal{H}_{\ell}}=\underline{\gamma_{\ell}}\sqrt{1-e^{-2\dot{\mathcal{E}}_{\ell}(0)}}$ and $\overline{\mathcal{H}_{\ell}}=\overline{\gamma_{\ell}}\sqrt{1-e^{-2\dot{\mathcal{E}}_{\ell}(0)}}$.
 \end{theorem}
 \begin{proof}
      It is immediate from \eqref{eq:wiplusonefinal} that $\dot{\mathcal{E}}_{\ell}(t)<\dot{\mathcal{E}}_{\ell}(0).$ It follows from \Cref{thm:deltai} that the position error remains within $-\underline{\gamma_{\ell}}<\gamma_{\ell}(t)<\overline{\gamma_{\ell}}$ $\forall$ $t>0$. Together these two conditions imply
      \begin{equation}\label{eq:wiplus1dot01}
     \dot{\mathcal{E}}_{\ell}(0)\geq     
            \begin{cases}
              \dfrac{1}{2}\ln\left(\dfrac{\overline{\gamma_{\ell}}^2}{\overline{\gamma_{\ell}}^2-\gamma_{\ell}^2(t)}\right);& \gamma_{\ell}\in(0,\overline{\gamma_{\ell}}),\\
              \dfrac{1}{2} \ln\left(\dfrac{\underline{\gamma_{\ell}}^2}{\underline{\gamma_{\ell}}^2-\gamma_{\ell}^2(t)}\right);& \gamma_{\ell}\in (-\underline{\gamma_{\ell}},0].
          \end{cases}
      \end{equation}
  Taking exponential on both sides of \eqref{eq:wiplus1dot01} and after some algebraic simplification, one may arrive at
  \begin{equation}\label{eq:wiplus1dot02}
     e^{2\dot{\mathcal{E}}_{\ell}(0)}\geq
     \begin{cases}
              \dfrac{\overline{\gamma_{\ell}}^2}{\overline{\gamma_{\ell}}^2-\gamma_{\ell}^2(t)};& \gamma_{\ell}\in(0,\overline{\gamma_{\ell}}),\\
            \dfrac{\underline{\gamma_{\ell}}^2}{\underline{\gamma_{\ell}}^2-\gamma_{\ell}^2(t)};& \gamma_{\ell}\in (-\underline{\gamma_{\ell}},0].
          \end{cases}
  \end{equation}
 Cross-multiplying and rearranging the terms of \eqref{eq:wiplus1dot02} leads us to arrive at
  \begin{equation*}
     e^{2\dot{\mathcal{E}}_{\ell}(0)}\gamma_{\ell}^2(t)\leq
     \begin{cases}
     \overline{\gamma_{\ell}}^2\left(e^{2\dot{\mathcal{E}}_{\ell}(0)}-1\right);& \gamma_{\ell}\in(0,\overline{\gamma_{\ell}}),\\
    \underline{\gamma_{\ell}}^2\left(e^{2\dot{\mathcal{E}}_{\ell}(0)}-1\right)  ;& \gamma_{\ell}\in (-\underline{\gamma_{\ell}},0],
          \end{cases}
  \end{equation*}
  which, on some algebraic simplification, results in
  \begin{equation}\label{eq:wiplus1dot03}
     \gamma_{\ell}(t)\leq
     \begin{cases}
     \overline{\gamma_{\ell}}\sqrt{1-e^{-2\dot{\mathcal{E}}_{\ell}(0)}};& \gamma_{\ell}\in(0,\overline{\gamma_{\ell}}),\\
    \underline{\gamma_{\ell}}\sqrt{1-e^{-2\dot{\mathcal{E}}_{\ell}(0)}}  ;& \gamma_{\ell}\in (-\underline{\gamma_{\ell}},0].
          \end{cases}
  \end{equation}
  It is immediate from \eqref{eq:wiplus1dot03} that $\gamma_{\ell}(t)$ is bounded by $-\underline{\gamma_{\ell}}\sqrt{1-e^{-2\dot{\mathcal{E}}_{\ell}(0)}}:=-\underline{\mathcal{H}_{\ell}}\leq \gamma_{\ell}(t)\leq \overline{\gamma_{\ell}}\sqrt{1-e^{-2\dot{\mathcal{E}}_{\ell}(0)}}:=\overline{\mathcal{H}_{\ell}}$. We also have $\mathcal{P}_{\ell}(t)=\gamma_{\ell}(t)+\mathcal{P}_{d_\ell}(t)$ from error definition and $-\underline{\mathcal{Y}_{0_\ell}}\leq\mathcal{P}_{d_\ell}(t)\leq \overline{\mathcal{Y}_{0_\ell}}$ from \Cref{assum:li}, which together imply
  \begin{equation*}
      -\underline{\mathcal{H}_{\ell}}-\underline{\mathcal{Y}_{0_\ell}}\leq\mathcal{P}_{\ell}(t)\leq\overline{\mathcal{H}_{\ell}}+\overline{\mathcal{Y}_{0_\ell}}.
  \end{equation*}
  Also, $\underline{\mathcal{H}_{\ell}}<\underline{\gamma_{\ell}}$ and $\overline{\mathcal{H}_{\ell}}<\overline{\gamma_{\ell}}$, implies  $\underline{\mathcal{H}_{\ell}}+\underline{\mathcal{Y}_{0_\ell}}<\underline{\gamma_{\ell}}+\underline{\mathcal{Y}_{0_\ell}}=\mathcal{L}_{\ell}$ and $\overline{\mathcal{H}_{\ell}}+\overline{\mathcal{Y}_{0_\ell}}<\overline{\gamma_{\ell}}+\underline{\mathcal{Y}_{0_\ell}}=\mathcal{L}_{\ell}$. Thus, it can be concluded that quadrotor's position remains in the set $\mathcal{S}_{\ell}$, that is, $\mathcal{P}_{\ell}(t)\in \mathcal{S}_{\ell}\,\forall\,t\geq0$.
\end{proof}
\begin{corollary}
        If the UAV's initial position remains in the set given by $\gamma_\ell(0)\in\left(\underline{\gamma_{\ell}},\overline{\gamma_{\ell}}\right)$ in line with \Cref{assum:li}, then, the UAV's velocity remains bounded for all time $t\geq0$.  Furthermore, the error $\gamma_\ell$ converges asymptotically to zero. This essentially implies that the UAV ultimately converges to its desired position trajectory and remains on it for all future times without violating its spatial constraints.
\end{corollary}
\begin{proof}
One may write the position of quadrotor as $\mathcal{P}_{\ell}(t)=\gamma_{\ell}(t)+\mathcal{P}_{\mathrm{d}\ell}(t)$ from the error definition given in \Cref{subsec:position_control}.
As a direct consequence of \Cref{thm:deltai}, we can conclude that $\gamma_{\ell}(t)$ remains bounded. Furthermore, combining this result with the boundedness of $\mathcal{P}_{\mathrm{d}\ell}(t)$, we can infer that $\mathcal{P}_{\ell}(t)$ also remains bounded.
 
 One may express $\dot{\mathcal{P}}_{\ell}$ using the relations $\zeta_{\ell}=\dot{\gamma}_{\ell}-\beta_{\ell}$ and $\gamma_{\ell}=\mathcal{P}_{\ell}-\mathcal{P}_{d_\ell}$ as
\begin{equation}\label{eq:dot_p_boundedness}
    \dot{\mathcal{P}}_{\ell}= \zeta_{\ell}+\beta_{\ell}+\dot{\mathcal{P}}_{\mathrm{d}\ell}.
\end{equation}
Using the fact that, $\dfrac{1}{2}\zeta_{\ell}^2(t)\leq \mathcal{E}_{\ell}(0),$ one can readily verify that $|\zeta_{\ell}(t)|\leq\sqrt{2\mathcal{E}_{\ell}(0)}$, and thus  $\zeta_{\ell}$ is bounded. On the other hand, $\beta_{\ell}$ given in \eqref{eq:betapi} is only a function of $\gamma_{\ell}$, and thus the boundedness of $\gamma_{\ell}$ implies that $\beta_{\ell}$ remains bounded.
Combining this inference with \Cref{assum:li} guarantees the boundedness of the UAV's velocity.

On differentiating \eqref{eq:wiplusonefinal} with respect to time, one may obtain
\begin{equation}\label{eq:ddotwplus1}
\ddot{\mathcal{E}}_{\ell}=-4\mathcal{K}_{\ell}\gamma_{\ell}^3\dot{\gamma}_{\ell}-2\mathcal{M}_{\ell}\zeta_\ell\dot{\zeta}_\ell.
\end{equation}
The right-hand side of the expression given in \eqref{eq:ddotwplus1} is a function of $\gamma_{\ell}$, $\dot{\gamma}_{\ell}$, $\zeta_{\ell}$, and $\dot{\zeta}_{\ell}$, which are bounded. This implies that $\ddot{\mathcal{E}}_{\ell}$ is bounded. This essentially points to the fact that $\dot{\mathcal{E}}_{\ell}$ is uniformly continuous. Using Barbalet's lemma, it can be concluded that the error $\gamma_{\ell}(t)\to 0$ as $t\to \infty$. Consequently, the UAV converges to its desired trajectory asymptotically without violating the spatial constraints.
This completes the proof.
\end{proof}

As stated earlier, the translational subsystems generate the net thrust for the quadrotor and the desired roll and pitch angles, which serve as the input to the rotational subsystem. Thus, we now aim to obtain the value of net thrust and the desired roll and pitch angles from the virtual control inputs presented in the following theorem.
\begin{theorem}\label{thm:thurst_desired_roll_pitch}
The net thrust and the desired roll and pitch angles (the actual control inputs) are 
    \begin{align}
     u_T=&~m \sqrt{\delta_x^2+\delta_y^2+\left(\delta_z+g\right)^2}, \label{eq:net_thrust}\\
     \phi_d=&~\sin^{-1}\left(\dfrac{m\left[\delta_x\sin\psi_d-\delta_y\cos\psi_d\right]}{u_T}\right), \label{eq:phid}\\
    \theta_d=&~\tan^{-1}\left(\dfrac{\delta_x\cos\psi_d+\delta_y\sin\psi_d}{\delta_z+g}\right).\label{eq:thetad}
\end{align}
\end{theorem}
\begin{proof}
By squaring and adding $\delta_x$, $\delta_y$, and $\delta_z+g$ from \eqref{eq:deltas}, one may obtain
\begin{align}
\nonumber   \delta_{x}^2 + \delta_{y}^2 + \left(\delta_{z}+g\right)^2 &= \dfrac{u_{T}^2}{m^2} \left(  
\cos^2\phi_d \sin^2\theta_d \cos^2\psi_d +  \sin^2\phi_d \sin^2\psi_d +2 \cos\phi_d\sin\theta_d\cos\psi_d\sin\phi_d\sin\psi_d +\cos^2\phi_d\cos^2\theta_d \right.\\
&\left.+ \cos^2\phi_d\sin^2\theta_d\sin^2\psi_d + \sin^2\phi_d\cos^2\psi_d - 2 \cos\phi_d\sin\theta_d\sin\psi_d\sin\phi_d\cos\psi_d  \right). \label{eq:u_T_calculation_1}
\end{align}
By collecting similar terms together and performing some algebraic simplifications, the expression in \eqref{eq:u_T_calculation_1} becomes
\begin{equation*}
 \delta_{x}^2 + \delta_{y}^2 + \left(\delta_{z}+g\right)^2 =  \dfrac{u_{T}^2}{m^2} \left[  
\cos^2\phi_d \sin^2\theta_d \left( \cos^2\psi_d +\sin^2\psi_d \right) +  \sin^2\phi_d \left( \sin^2\psi_d +\cos^2\psi_d\right) +\cos^2\phi_d\cos^2\theta_d \right],
\end{equation*}
which can be simplified to
\begin{equation}\label{eq:u_T_calculation_2}
 \delta_{x}^2 + \delta_{y}^2 + \left(\delta_{z}+g\right)^2 =  \dfrac{u_{T}^2}{m^2} \left[  
\cos^2\phi_d \sin^2\theta_d  +  \sin^2\phi_d  +\cos^2\phi_d\cos^2\theta_d \right]
\end{equation}
using the trigonometric relation $\cos^2\psi_d +\sin^2\psi_d = 1$. We may write \eqref{eq:u_T_calculation_2} in more convenient from as
\begin{equation*}
  \delta_{x}^2 + \delta_{y}^2 + \left(\delta_{z}+g\right)^2 =  \dfrac{u_{T}^2}{m^2} \left[  
\cos^2\phi_d \left(\sin^2\theta_d +\cos^2\theta_d\right)  +  \sin^2\phi_d \right] 
\end{equation*}
which is equivalent to
\begin{equation} \label{eq:u_T_calculation_3}
  \delta_{x}^2 + \delta_{y}^2 + \left(\delta_{z}+g\right)^2 =  \dfrac{u_{T}^2}{m^2} \left[  
\cos^2\phi_d  +  \sin^2\phi_d \right] = \dfrac{u_{T}^2}{m^2} .
\end{equation}
Therefore, on solving for $u_T$, one can obtain the net thrust from \eqref{eq:u_T_calculation_3} as
\begin{equation}
    u_T=m \sqrt{\delta_x^2+\delta_y^2+\left(\delta_z+g\right)^2}.
\end{equation}
Next, we endeavor to obtain the desired roll and pitch angles for a given yaw angle. Multiplying \eqref{eq:deltaydef} by $\dfrac{1}{\sin\psi_d}$ and subtracting it from the expression $\eqref{eq:deltaxdef}\times \dfrac{1}{\cos\psi_d}$ leads us to arrive at
\begin{equation}\label{eq:phid_calculation_1}
    \delta_{x} \times \dfrac{1}{\cos\psi_d} -  \delta_{y} \times \dfrac{1}{\sin\psi_d} = \frac{u_T}{m}\left( \cos{\phi_{d}} \sin{\theta_{d}} + \dfrac{\sin\phi_{d} \sin\psi_{d}}{\cos\psi_{d}} \right)-\frac{u_T}{m}\left( \cos\phi_{d} \sin\theta_{d}  - \dfrac{\sin\phi_{d} \cos\psi_{d}} {\sin\psi_{d}} \right).
\end{equation}
Upon further simplifications, one may obtain \eqref{eq:phid_calculation_1} as
\begin{equation*}
     \dfrac{\delta_{x}\sin\psi_d - \delta_{y}\cos\psi_d }{\sin\psi_d \cos\psi_d}  = \frac{u_T}{m}\left[ \sin\phi_{d} \left( \dfrac{ \sin^2\psi_{d} + \cos^2\psi_d}{\sin\psi_d\cos\psi_{d}}\right)\right],
\end{equation*}
which further reduces to
\begin{equation}\label{eq:phid_calculation_2}
     \delta_{x}\sin\psi_d - \delta_{y}\cos\psi_d   = \frac{u_T \sin\phi_{d}}{m}
\end{equation}
using the trigonometric relation $ \sin^2\psi_{d} + \cos^2\psi_d =1$ and performing some more algebraic simplifications. On solving \eqref{eq:phid_calculation_2} for $\phi_d$ yields
\begin{equation*}
\phi_d=\arcsin\left[\dfrac{m\left(\delta_x\sin\psi_d-\delta_y\cos\psi_d\right)}{u_T}\right].
\end{equation*}
Now, we obtain the desired roll angle from the virtual control inputs. From \eqref{eq:deltazdef}, we get
\begin{equation*}
    \delta_{z} + g = \dfrac{u_{T}}{m} \left( \cos\phi_d \cos\theta_d \right),
\end{equation*}
which implies
\begin{equation} \label{eq:thetad_calculation_1}
\cos\theta_d \cos\phi_d = \dfrac{m\left(\delta_{z} + g\right)}{u_{T} }.
\end{equation}
After multiplying \eqref{eq:deltaxdef} with $\cos\psi_d$ and \eqref{eq:deltaydef} with $\sin\psi_d$ and adding them up results in
\begin{equation*}
    \delta_x \cos\psi_d + \delta_{y}\sin\psi_d=\frac{u_T}{m}\left( \cos{\phi_{d}} \sin{\theta_{d}} \cos^2\psi_{d} + \sin\phi_{d} \sin\psi_{d} \cos\psi_d +\cos\phi_{d} \sin\theta_{d} \sin^2\psi_{d} - \sin\phi_{d} \cos\psi_{d} \sin\psi_{d} \right),
\end{equation*}
which, on some algebraic simplification, results in
\begin{equation}\label{eq:thetad_calculation_2}
    \delta_x \cos\psi_d + \delta_{y}\sin\psi_d=\frac{u_T}{m}\left[\cos{\phi_{d}} \sin{\theta_{d}} \left(\cos^2\psi_{d} + \sin^2\psi_{d}\right) \right].
\end{equation}
Using the trigonometric relation $\cos^2\psi_{d} + \sin^2\psi_{d} =1$, we obtain $\cos{\phi_{d}} \sin{\theta_{d}}$ from \eqref{eq:thetad_calculation_2} as
\begin{equation*}
   \cos{\phi_{d}} \sin{\theta_{d}} =\frac{m}{u_{T}}\left[\cos{\phi_{d}} \sin\delta_x \cos\psi_d + \delta_{y}\sin\psi_d  \right],
\end{equation*}
which on division by \eqref{eq:thetad_calculation_1} leads us to arrive at
\begin{equation}\label{eq:thetad_calculation_3}
    \tan{\theta_{d}} =\dfrac{\cos{\phi_{d}} \sin\delta_x \cos\psi_d + \delta_{y}\sin\psi_d}{\delta_z+g} \implies \theta_d = \arctan\left(\dfrac{\cos{\phi_{d}} \sin\delta_x \cos\psi_d + \delta_{y}\sin\psi_d}{\delta_z+g}\right).
\end{equation}
This completes the proof.
\end{proof}
\subsection{Attitude Subsystem}
\begin{figure}[!ht]
		\centering
		\includegraphics[width=0.85\linewidth]{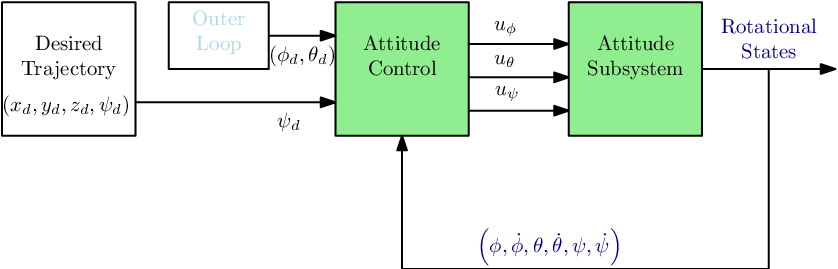}
		\caption{Illustration of the inner loop design.}
        \label{fig:inner_loop}
	\end{figure}

    The attitude subsystem forms the inner loop, which consists of the rotational dynamics of the quadrotor UAV. A schematic representation of the inner loop is shown in \Cref{fig:inner_loop}. The desired yaw trajectory is application-specific and is given by the high-level guidance system, while the desired roll and pitch trajectories are obtained by the position control loop, given in \eqref{eq:phid} and \eqref{eq:thetad}. This is primarily due to the fact that the roll and pitch angles are coupled with the motion of the UAV in $x$ and $y$ directions. Unlike the positioning subsystem, it is a fully actuated system since it has three degrees of freedom to control with three independent control inputs, namely, roll, pitch, and yaw moments. Toward this objective, we design the control inputs for the attitude subsystem that will ensure that the UAV tracks its desired orientation trajectory while satisfying the spatial constraints. 
    
    We define the orientation error vector $\mathbf{E}_\mathrm{O}$ as,  $\mathbf{E}_\mathrm{O}=\mathbf{\Theta}-\mathbf{\Theta}_\mathrm{d}=\Upsilon_k$ (element-wise), for $k=\phi$, $\theta$, and $\psi$, as defined previously. Subsequent time differentiation of $\mathbf{E}_\mathrm{O}$, results in 
	\begin{equation}\label{eq:Upsilonddot}
		\ddot{\Upsilon}_{k}=\ddot{\Theta}_{k}-\ddot{\Theta}_{d_k}=\mathcal{F}_k+\mathcal{G}_{k} u_{k}+h_{k}-\ddot{\Theta}_{d_k},
	\end{equation}
 where $\mathcal{F}_{k}$ and $\mathcal{G}_{k}$ are as defined in \eqref{eq:attitudedynamics}.
\begin{remark}
    We presume that a disturbance observer, such as the method outlined in \cite{doi:10.1016/j.automatica.2007.01.008}, can provide an estimate of the disturbance $h_{k}$, which we denote as $\hat{h}_k$. It was shown in \cite{doi:10.1016/j.automatica.2007.01.008} that $\hat{h}_{k}$ converges to $h_{k}$ within a finite time, and thus the error $\tilde{h}:=h_{k}-\hat{h}_{k}$ remains bounded. It may be observed from \eqref{eq:rotdyn1} that the uncertainties are a function of angular velocity and its rates. Therefore, it is reasonable to express them, for mathematical convenience, as $\tilde{h}_{k}=h\textsubscript{0}\mathcal{C}(\dot{\Theta}_{k})$, where $\mathcal{C}(\dot{\Theta}_{k})$ is a smooth function depending on the rate of change of Euler angles and $h\textsubscript{0}$ is an unknown constant.
\end{remark}
It can be observed from \eqref{eq:Upsilonddot} that the orientation error dynamics also has a relative degree of two with respect to the control input, $u_k$. The problem of tracking the orientation of a trajectory with constraints can be thought of as regulating the error, $\Upsilon_k$, to zero through a suitable choice of $u_k$. This is presented in the following theorem.
	\begin{theorem}\label{thm:attitude}
		Consider the UAV's attitude dynamics as in \eqref{eq:attitudedynamics} and  \eqref{eq:Upsilonddot}. If the UAV generates moments as
		\begin{align}
\nonumber u_{k}=&~\dfrac{1}{\mathcal{G}_{k}}\left[-\mathcal{F}_{k}+\ddot{\Theta}_{d_k}-\mathcal{Z}_{k}\Upsilon_{k}^2 \dot \Upsilon_{k} \left\{3s(\Upsilon_{k})\left(\overline{\Upsilon_{k}}^2-\underline{\Upsilon_{k}}^2\right)+\left(3\underline{\Upsilon_{k}}^2-5\Upsilon_{k}^2\right)\right\} -\hat{h}_{k}-\bar{h}_{k}\mathcal{C}(\dot{\Theta}_{k}) \right.\\
&\left.~ -\mathcal{N}_{k}\left[\dot \Upsilon_{k}+\mathcal{Z}_{k}\Upsilon_{k}^3\left\{s(\Upsilon_{k})\left(\overline{\Upsilon_{k}}^2-\underline{\Upsilon_{k}}^2\right)+\left(\underline{\Upsilon_{k}}^2-\Upsilon_{k}^2\right)\right\}\right] -\Upsilon_{k}\left\{\dfrac{1-s(\Upsilon_{k})}{\underline{\Upsilon_{k}}^2-\Upsilon_{k}^2}
+\dfrac{s(\Upsilon_{k})}{\overline{\Upsilon_{k}}^2-\Upsilon_{k}^2}\right\}\right],\label{eq:ui}
\end{align}
where $\mathcal{Z}_{k}$, $\mathcal{N}_{k}$ for ${k}=\phi,\theta,$ and $\psi$ are positive constants, $\hat{h}_{k}$ is the available estimate of the uncertainties,  with an update rule $\bar{h}_{k}$ chosen as
\begin{equation}\label{eq:updateterm}
 \dot{\bar{h}}_{k}=\left[\dot \Upsilon_{k}+\mathcal{Z}_{k}\Upsilon_{k}^3\left\{s(\Upsilon_{k})\left(\overline{\Upsilon_{k}}^2-\underline{\Upsilon_{k}}^2\right)+\left(\underline{\Upsilon_{k}}^2-\Upsilon_{k}^2\right)\right\}\right]\mathcal{C}(\dot{\Theta}_{k}),
\end{equation}
where $\mathcal{Z}_{k}>0$ is a constant and $s(\Upsilon_{k})$ is defined as
\begin{equation}\label{eq:sgammai}
    s(\Upsilon_{k})=\begin{cases}
        	1; &\mathrm{if}\,\,\Upsilon_{k}>0,\\
	0; &\mathrm{if}\,\,\Upsilon_{k}\leq 0,
    \end{cases}
\end{equation}
then the UAV's orientation error remains bounded within $\underline{\Upsilon_{k}} <\Upsilon_{k}<\overline{\Upsilon_{k}}$, where  $\underline{\Upsilon_{k}}$ is given by $\underline{\Upsilon_{k}}=\mathcal{Q}_{k}-\underline{\mathcal{B}_{\mathrm{0i}}}$ and  $\overline{\Upsilon_{k}}=\mathcal{Q}_{k}-\overline{\mathcal{B}_{\mathrm{0i}}}$. 
\end{theorem}
	\begin{proof}
		Consider another asymmetric barrier Lyapunov function candidate, $\mathcal{V}_{k}$ for ${k}=\phi,\theta$, and $\psi$ as
		\begin{equation}\label{eq:vi}
			\mathcal{V}_{k}=\dfrac{1-s(\Upsilon_{k})}{2}\ln\left[\dfrac{\underline{\Upsilon_{k}}^2}{\underline{\Upsilon_{k}}^2-\Upsilon^2} \right]+\dfrac{s(\Upsilon_{k})}{2}\ln\left[\dfrac{\overline{\Upsilon_{k}}^2}{\overline{\Upsilon_{k}}^2-\Upsilon^2} \right]+\dfrac{h_{\mathrm{e}_k}^2}{2},
		\end{equation}
		where $h_{\mathrm{e}_k} = h\textsubscript{0} - \bar{h}_{k}$. From hereafter, we drop the argument of $s(\Upsilon_{k})$, and $s$ must be understood as a function of $\Upsilon_{k}$. On differentiating \eqref{eq:vi} with respect to time, one may obtain
		\begin{equation}\label{eq:vidot1}
			\dot{\mathcal{V}}_{k}= \dfrac{1-s}{2}\left[\dfrac{\underline{\Upsilon_{k}}^2-\Upsilon_{k}^2}{\underline{\Upsilon_{k}}^2}\right]\left[\dfrac{-\left(\underline{\Upsilon_{k}}^2\right)^2}{\left(\underline{\Upsilon_{k}}^2-\Upsilon_{k}^2\right)^2}\right]\left(\dfrac{-2\Upsilon_{k}\dot \Upsilon_{k}}{\underline{\Upsilon_{k}}^2}\right)
			+
			\dfrac{s}{2}\left[\dfrac{\overline{\Upsilon_{k}}^2-\Upsilon_{k}^2}{\overline{\Upsilon_{k}}^2}\right]\left[\dfrac{-\left(\overline{\Upsilon_{k}}^2\right)^2}{\left(\overline{\Upsilon_{k}}^2-\Upsilon_{k}^2\right)^2}\right]\left(\dfrac{-2\Upsilon_{k}\dot \Upsilon_{k}}{\overline{\Upsilon_{k}}^2}\right)
			+h_{\mathrm{e}_{k}}\dot{h}_{\mathrm{e}_{k}}, 
		\end{equation}
		whose simplification yields
		\begin{equation}\label{eq:vidot2}
			\dot{\mathcal{V}}_{k}=\dfrac{1-s}{2}\left[\dfrac{2\Upsilon_{k}\dot \Upsilon_{k}}{\underline{\Upsilon_{k}}^2-\Upsilon_{k}^2}\right]
			+
			\dfrac{s}{2}\left[\dfrac{2\Upsilon_{k}\dot \Upsilon_{k}}{\overline{\Upsilon_{k}}^2-\Upsilon_{k}^2}\right]
			=\left[\dfrac{1-s}{\underline{\Upsilon_{k}}^2-\Upsilon_{k}^2}
			+
			\dfrac{s}{\overline{\Upsilon_{k}}^2-\Upsilon_{k}^2}\right]  \Upsilon_{k}\dot \Upsilon_{k}+h_{\mathrm{e}_{k}}\dot{h}_{\mathrm{e}_{k}}.
		\end{equation}
		Now, assume a variable $\lambda_{k}=\dot{\Upsilon}_{k}-\sigma_{k}$, where $\sigma_{k}$ is a stabilizing function that needs to be designed. Using the relation $\dot{ \Upsilon}_{k}=\lambda_{k}+\sigma_{k}$, one can write \eqref{eq:vidot2} as
		\begin{equation}\label{eq:vidot3}
			\dot{\mathcal{V}}_{k}=\left[\dfrac{1-s}{\underline{\Upsilon_{k}}^2-\Upsilon_{k}^2}
			+
			\dfrac{s}{\overline{\Upsilon_{k}}^2-\Upsilon_{k}^2}\right]  \Upsilon_{k}\left(\lambda_{k}+\sigma_{k}\right)+h_{\mathrm{e}_{k}}\dot{h}_{\mathrm{e}_{k}}.
		\end{equation}
		If we choose the stabilizing function $\sigma_{k}$, with $\mathcal{Z}_{k}>0$, such that
		\begin{align}\label{eq:sigmai}
			\nonumber   \sigma_{k}=&~  -\dfrac{\left(\underline{\Upsilon_{k}}^2-\Upsilon_{k}^2\right)\left(\overline{\Upsilon_{k}}^2-\Upsilon_{k}^2\right)}{\left[\left(1-s\right)\left(\overline{\Upsilon_{k}}^2-\Upsilon_{k}^2\right)+s\left(\underline{\Upsilon_{k}}^2-\Upsilon_{k}^2\right)\right]}\mathcal{Z}_{k}\Upsilon_{k}^3,\\
			=&~-\left[s\left(\overline{\Upsilon_{k}}^2-\Upsilon_{k}^2\right)+\left(1-s\right)\left(\underline{\Upsilon_{k}}^2-\Upsilon_{k}^2\right)\right]\mathcal{Z}_{k}\Upsilon_{k}^3
			=-\left[s\left(\overline{\Upsilon_{k}}^2-\underline{\Upsilon_{k}}^2\right)+\left(\underline{\Upsilon_{k}}^2-\Upsilon_{k}^2\right)\right]\mathcal{Z}_{k}\Upsilon_{k}^3,
		\end{align}
		then the derivative of the Lyapunov function candidate, given in \eqref{eq:vidot3}, becomes
		\begin{equation*}
			\dot{\mathcal{V}}_{k}=\Upsilon_{k}\left[\dfrac{1-s}{\underline{\Upsilon_{k}}^2-\Upsilon_{k}^2}
			+
			\dfrac{s}{\overline{\Upsilon_{k}}^2-\Upsilon_{k}^2}\right]  \left[\lambda_{k}-\dfrac{\left(\underline{\Upsilon_{k}}^2-\Upsilon_{k}^2\right)\left(\overline{\Upsilon_{k}}^2-\Upsilon_{k}^2\right)\mathcal{Z}_{k}\Upsilon_{k}^3}{\left\{\left(1-s\right)\left(\overline{\Upsilon_{k}}^2-\Upsilon_{k}^2\right)+s\left(\underline{\Upsilon_{k}}^2-\Upsilon_{k}^2\right)\right\}}\right]+h_{\mathrm{e}_{k}}\dot{h}_{\mathrm{e}_{k}}.
		\end{equation*}
		Equivalently
		\begin{equation}\label{eq:vidot4}
			\dot{\mathcal{V}}_{k}=-\mathcal{Z}_{k}\Upsilon_{k}^4\left[\dfrac{1-s}{\underline{\Upsilon_{k}}^2-\Upsilon_{k}^2}
			+\dfrac{s}{\overline{\Upsilon_{k}}^2-\Upsilon_{k}^2}\right]  \dfrac{\left(\underline{\Upsilon_{k}}^2-\Upsilon_{k}^2\right)\left(\overline{\Upsilon_{k}}^2-\Upsilon_{k}^2\right)}{\left[\left(1-s\right)\left(\overline{\Upsilon_{k}}^2-\Upsilon_{k}^2\right)+s\left(\underline{\Upsilon_{k}}^2-\Upsilon_{k}^2\right)\right]}
			+\Upsilon_{k}\lambda_i\left[\dfrac{1-s}{\underline{\Upsilon_{k}}^2-\Upsilon_{k}^2}
			+\dfrac{s}{\overline{\Upsilon_{k}}^2-\Upsilon_{k}^2}\right]+h_{\mathrm{e}_{k}}\dot {h}_{\mathrm{e}_{k}}.
		\end{equation}
		After some algebraic simplification, one may obtain \eqref{eq:vidot4} as
		\begin{equation}\label{eq:vidot5}
			\dot{\mathcal{V}}_{k}=-\mathcal{Z}_{k}\Upsilon_{k}^4+    \Upsilon_{k}\lambda_{k}\left[\dfrac{1-s}{\underline{\Upsilon_{k}}^2-\Upsilon_{k}^2}
			+\dfrac{s}{\overline{\Upsilon_{k}}^2-\Upsilon_{k}^2}\right]+h_{\mathrm{e}_{k}}\dot{h}_{\mathrm{e}_{k}}.
		\end{equation}
		Consider another Lyapunov candidate $\mathcal{D}_{k}$ as
		\begin{equation}\label{eq:viplusone1}
			\mathcal{D}_{k}=\mathcal{V}_{k}+\dfrac{1}{2}\lambda_{k}^2
			=
			\dfrac{1-s}{2}\ln\left[\dfrac{\underline{\Upsilon_{k}}^2}{\underline{\Upsilon_{k}}^2-\Upsilon_{k}^2}\right]+\dfrac{s}{2}\ln\left[\dfrac{\overline{\Upsilon_{k}}^2}{\overline{\Upsilon_{k}}^2-\Upsilon_{k}^2}\right]+\dfrac{h_{\mathrm{e}_{k}}^2}{2}+\dfrac{1}{2}\lambda_{k}^2,
		\end{equation}
		whose time derivative can be obtained using \eqref{eq:vidot5} as
		\begin{equation}\label{eq:viplusone2}
			\dot{\mathcal{D}}_{k}=\dot{\mathcal{V}}_{k}+\lambda_{k}\dot{\lambda}_{k}=-\mathcal{Z}_{k}\Upsilon_{k}^4+    \Upsilon_{k}\lambda_{k}\left[\dfrac{1-s}{\underline{\Upsilon_{k}}^2-\Upsilon_{k}^2}+
			+\dfrac{s}{\overline{\Upsilon_{k}}^2-\Upsilon_{k}^2}\right]+h_{\mathrm{e}_{k}}\dot{h}_{\mathrm{e}_{k}}+\lambda_{k}\left(\ddot \Upsilon_{k}-\dot \sigma_i\right),
		\end{equation}
		where $\dot{\sigma}_{k}$ can be obtained by differentiating  \eqref{eq:sigmai} with respect to time, as
		\begin{equation*}
	     \dot \sigma_{k}= -3\mathcal{Z}_{k}\Upsilon_{k}^2 \dot \Upsilon_{k} \left[s\left(\overline{\Upsilon_{k}}^2-\Upsilon_{k}^2\right)+\left(1-s\right)\left(\underline{\Upsilon_{k}}^2-\Upsilon_{k}^2\right)\right]-\mathcal{Z}_{k}\Upsilon_{k}^3\left[ s\left(-2\Upsilon_{k} \dot \Upsilon_{k}\right)+\left(1-s\right)\left( -2\Upsilon_{k} \dot \Upsilon_{k}\right)\right],
		\end{equation*}
		which, on some algebraic simplification, results in
		\begin{equation}\label{eq:sigmadot1}
			\dot{\sigma}_{k}= -3\mathcal{Z}_{k}\Upsilon_{k}^2 \dot \Upsilon_{k} \left[s\left(\overline{\Upsilon_{k}}^2-\Upsilon_{k}^2\right)+\left(1-s\right)\left(\underline{\Upsilon_{k}}^2-\Upsilon_{k}^2\right)\right]+2\mathcal{Z}_{k}\Upsilon_{k}^4 \dot \Upsilon_{k}.
		\end{equation}
		After rearranging the terms of \eqref{eq:sigmadot1}, we arrive at 
		\begin{equation}\label{eq:sigmadot}
		 \dot{\sigma}_{k}=-\mathcal{Z}_{k}\Upsilon_{k}^2 \dot \Upsilon_{k} \left[3s\left(\overline{\Upsilon_{k}}^2-\Upsilon_{k}^2\right)+3\left(1-s\right)\left(\underline{\Upsilon_{k}}^2-\Upsilon_{k}^2\right)\right]-2\Upsilon_{k}^2=-\mathcal{Z}_{k}\Upsilon_{k}^2 \dot \Upsilon_{k} \left[3s\left(\overline{\Upsilon_{k}}^2-\underline{\Upsilon_{k}}^2\right)+\left(3\underline{\Upsilon_{k}}^2-5\Upsilon_{k}^2\right)\right].
		\end{equation}
		Upon substituting the value of $\dot{\sigma}_{k}$ from \eqref{eq:sigmadot} and $\ddot{\Upsilon}_{k}$  from \eqref{eq:Upsilonddot} into \eqref{eq:viplusone2}, one may obtain
		\begin{align}
			\nonumber   \dot{\mathcal{D}}_{k}=&~
			\lambda_{k}\left[\mathcal{F}_{k}+\mathcal{G}_{k}u_{k}+h_{k}-\ddot{\Theta}_{d_k}+\mathcal{Z}_{k}\Upsilon_{k}^2 \dot \Upsilon_{k} \left\{3s\left(\overline{\Upsilon_{k}}^2-\underline{\Upsilon_{k}}^2\right)+\left(3\underline{\Upsilon_{k}}^2-5\Upsilon_{k}^2\right)\right\}\right]+h_{\mathrm{e}_{k}}\dot{h}_{\mathrm{e}_{k}}\\
			&~-\mathcal{Z}_{k}\Upsilon_{k}^4+    \Upsilon_{k}\lambda_{k}\left[\dfrac{1-s}{\underline{\Upsilon_{k}}^2-\Upsilon_{k}^2}
			+\dfrac{s}{\overline{\Upsilon_{k}}^2-\Upsilon_{k}^2}\right].\label{eq:viplusone3}
		\end{align}
		If we choose $u_{k}$ as per \eqref{eq:ui} and use the relation $\lambda_{k}=\dot{\Upsilon}_{k}-\sigma_{k}$, then this leads us to arrive at
		\begin{equation}\label{eq:viplusone4}
			\dot{\mathcal{D}}_{k}=-  \mathcal{Z}_{k}\Upsilon_{k}^4-\mathcal{N}_{k}\lambda_{k}^2+\lambda_{k} \tilde{h}_{k}-\lambda_{k}\bar{h}_{k}\mathcal{C}(\dot{\Theta}_{k})+h_{\mathrm{e}_{k}}\dot{h}_{\mathrm{e}_{k}}.
		\end{equation}
  Using the relations $\tilde{h}_{k}=h\textsubscript{0}\mathcal{C}(\dot{\Theta}_{k})$ and $\dot{h}_{\mathrm{e}_{k}}=\dot{h}\textsubscript{0}-\dot{\bar h}_{k}$ together with $\dot{h}\textsubscript{0}=0$, we get
		\begin{equation}\label{eq:viplusone5}
			\dot{\mathcal{D}}_{k}=- \mathcal{Z}_{k}\Upsilon_{k}^4-\mathcal{N}_{k}\lambda_k^2+\lambda_k h\textsubscript{0}\mathcal{C}(\dot{\Theta}_{k}) -\lambda_k\bar{h}_{k}\mathcal{C}(\dot{\Theta}_{k})- h_{\mathrm{e}_{k}}\dot{\bar{h}}_{k}.
   \end{equation}
   By substituting the value of $\dot{\bar{h}}_{k}$ from \eqref{eq:updateterm} into \eqref{eq:viplusone5} and using the relation $h_{\mathrm{e}_k} = h\textsubscript{0} - \bar{h}_{k}$, one may obtain 
   \begin{equation}
     \dot{\mathcal{D}}_{k}= -  \mathcal{Z}_{k}\Upsilon_{k}^4-\mathcal{N}\lambda_k^2+\lambda_k h_{\mathrm{e}_{k}}\mathcal{C}(\dot{\Theta}_{k})-h_{\mathrm{e}_{k}}\left[\dot \Upsilon_{k}+\mathcal{Z}_{k}\Upsilon_{k}^3\left\{s\left(\overline{\Upsilon_{k}}^2-\underline{\Upsilon_{k}}^2\right)+\left(\underline{\Upsilon_{k}}^2-\Upsilon_{k}^2\right)\right\}\right]\mathcal{C}(\dot{\Theta}_{k}),
   \end{equation}
   which, upon simplifying further, leads us to
   \begin{equation}\label{eq:vdotplusonefinal}
   \dot{\mathcal{D}}_{k}=    -  \mathcal{Z}_{k}\Upsilon_{k}^4-\mathcal{N}_{k}\lambda_{k}^2+\lambda_{k}h_{\mathrm{e}_{k}}\mathcal{C}(\dot{\Theta}_{k})-\lambda_{k} h_e \mathcal{C}(\dot{\Theta}_{k})
			= -  \mathcal{Z}_{k}\Upsilon_{k}^4-\mathcal{N}_{k}\lambda_{k}^2 \leq 0.
   \end{equation}
		Thus, the stability of the considered system, given by \eqref{eq:Upsilonddot}, is ensured in the Lyapunov sense. Using the results in \Cref{lem:lemma1}, it can be inferred from \eqref{eq:vdotplusonefinal} that the orientation tracking error will remain in the bound $\underline{\Upsilon_{k}} <\Upsilon_{k}(t)<\overline{\Upsilon_{k}}$ if its initial value is such that $\underline{\Upsilon_{k}} <\Upsilon_{k}(0)<\overline{\Upsilon_{k}}$. 
	\end{proof}
It can be deduced from \Cref{thm:attitude} that the quadrotor's attitude always remains in the given set despite the presence of partial information about its inertia matrix, provided its initial attitude also lies within the same set. The following theorem provides further insights into the convergence of attitude angles of the quadrotor.
 \begin{theorem}\label{thm:attitude_boundedness}
     Consider the closed-loop systems given in \eqref{eq:Upsilonddot} and \eqref{eq:ui}. Provided UAV's initial orientation is such that $\underline{\Upsilon_{k}}<\Upsilon_{k}(0)<\overline{\Upsilon_{k}}$ in line with \Cref{assum:qi}, we can guarantee that the UAV's attitude angle remains bounded for all time $t\geq 0$. Moreover, the attitude angle remains confined to the set $\mathcal{S}_{k} \in \left( -\mathcal{Q}_{k},\mathcal{Q}_{k}\right)$, which is defined as the region $\Lambda_{k}\coloneqq \left\{\Theta_{k}\;|\;\Theta_{k}\in \left(-\underline{\mathcal{O}_{k}}-\underline{\mathcal{X}_{0_k}},\overline{\mathcal{O}_{k}}+\overline{\mathcal{X}_{0_k}}\right)\right\}\subset \mathcal{S}_k$,where $\underline{\mathcal{O}_{k}}=\underline{\Upsilon_{k}}\sqrt{1-e^{-2\dot{\mathcal{D}}_{k}(0)}}$ and $\overline{\mathcal{O}_{k}}=\overline{\Upsilon_{k}}\sqrt{1-e^{-2\dot{\mathcal{D}}_{k}(0)}}$. 
 \end{theorem}
 	\begin{proof}
    It can be inferred from \Cref{thm:attitude} that the orientation error remains in the bound $\underline{\Upsilon_{k}}<\Upsilon_{k}(t)<\overline{\Upsilon_{k}}$ provided its initial value lies within it. Also, \eqref{eq:vdotplusonefinal} implies $\mathcal{D}_{k}(t)<\mathcal{D}_{k}(0)$. Using these two relations, one may write
    \begin{equation}\label{eq:viplus1dot01}
     \dot{\mathcal{D}}_{k}(0)\geq     
            \begin{cases}
              \dfrac{1}{2}\ln\left(\dfrac{\overline{\Upsilon_{k}}^2}{\overline{\Upsilon_{k}}^2-\Upsilon_{k}^2(t)}\right);& \Upsilon_{k}\in(0,\overline{\Upsilon_{k}}),\\
              \dfrac{1}{2} \ln\left(\dfrac{\underline{\Upsilon_{k}}^2}{\underline{\Upsilon_{k}}^2-\Upsilon_{k}^2(t)}\right);& \Upsilon_{k}\in (-\underline{\Upsilon_{k}},0].
          \end{cases}
      \end{equation}
By taking the exponential on both sides and after some algebraic simplification, the expression in \eqref{eq:viplus1dot01} can be simplified to
\begin{equation*}
      e^{2\dot{\mathcal{D}}_{k}(0)}\geq
     \begin{cases}
              \dfrac{\overline{\Upsilon_{k}}^2}{\overline{\Upsilon_{k}}^2-\Upsilon_{k}^2(t)};& \Upsilon_{k}\in(0,\overline{\Upsilon_{k}}),\\
            \dfrac{\underline{\Upsilon_{k}}^2}{\underline{\Upsilon_{k}}^2-\Upsilon_{k}^2(t)};& \Upsilon_{k}\in (-\underline{\Upsilon_{k}},0]  
          \end{cases}   
          \implies
               e^{2\dot{\mathcal{D}}_{k}(0)}\Upsilon_{k}^2(t)\leq
     \begin{cases}
     \overline{\Upsilon_{k}}^2\left(e^{2\dot{\mathcal{D}}_{k}(0)}-1\right);& \Upsilon_{k}\in(0,\overline{\Upsilon_{k}}),\\
    \underline{\Upsilon_{k}}^2\left(e^{2\dot{\mathcal{D}}_{k}(0)}-1\right)  ;& \Upsilon_{k}\in (-\underline{\Upsilon_{k}},0],
          \end{cases}
\end{equation*}
which on solving for $\Upsilon_{k}(t)$ results in
  \begin{equation}\label{eq:viplus1dot02}
     \Upsilon_{k}(t)\leq
     \begin{cases}
     \overline{\Upsilon_{k}}\sqrt{1-e^{-2\dot{\mathcal{D}}_{k}(0)}};& \Upsilon_{k}\in(0,\overline{\Upsilon_{k}}),\\
    \underline{\Upsilon_{k}}\sqrt{1-e^{-2\dot{\mathcal{D}}_{k}(0)}}  ;& \Upsilon_{k}\in (-\underline{\Upsilon_{k}},0].
          \end{cases}
  \end{equation}
Thus, it can be inferred from \eqref{eq:viplus1dot02} that the orientation error $\Upsilon(t)$ remains bounded by $-\underline{\Upsilon_{k}}\sqrt{1-e^{-2\dot{\mathcal{D}}_{k}(0)}}:=-\underline{\mathcal{O}_{k}}\leq \Upsilon_{k}(t)\leq \overline{\Upsilon_{k}}\sqrt{1-e^{-2\dot{\mathcal{D}}_{k}(0)}}:=\overline{\mathcal{O}_{k}}$ for all $t\geq 0$. Using the relation $\Theta_{k}(t)=\Upsilon_{k}(t)+\Theta_{d_k}(t)$ together with \Cref{assum:qi} implies that $-\mathcal{O}_{k}-\mathcal{X}_{0_k}\leq\Theta_{k}(t)\leq \mathcal{O}_{k}+\mathcal{X}_{0_k}$. Moreover, $\underline{\mathcal{O}_{k}}<\underline{\Upsilon_{k}}$ and $\overline{\mathcal{O}_{k}}<\overline{\Upsilon_{k}},$ implies  $\underline{\mathcal{O}_{k}}+\underline{\mathcal{X}_{\mathrm{0i}}}<\underline{\Upsilon_{k}}+\underline{\mathcal{X}_{\mathrm{0i}}}=\mathcal{Q}_{k}$ and $\overline{\mathcal{O}_{k}}+\overline{\mathcal{X}_{\mathrm{0i}}}<\overline{\Upsilon_{k}}+\underline{\mathcal{X}_{\mathrm{0i}}}=\mathcal{Q}_{k}$. Thus, the orientation of the quadrotor remains in set $\mathcal{S}_{k}$, that is $\mathcal{Q}_{k}(t) \in \mathcal{S}_{k}\, \forall\,t\geq 0$.
\end{proof}
\begin{corollary}
        The UAV's angular velocity remains bounded for all time $t\geq 0$ provided its initial orientation remains in the set given by $\underline{\Upsilon_{k}}<\Upsilon_{k}(0)<\overline{\Upsilon_{k}}$ in line with \Cref{assum:qi}.  Furthermore, the error $\Upsilon_k$ converges asymptotically to zero. As a consequence, the quadrotor converges to its desired orientation trajectory without violating its spatial constraints.
\end{corollary}
\begin{proof}
It can be inferred from \Cref{thm:attitude_boundedness} that the $\Theta_{k}$ remains within a bounded set $\mathcal{S}_{k}$. One can express the attitude angular rate, $\dot{\Theta}_{k}$, using relations $\lambda_{k}=\dot{\Upsilon}_{k}-\sigma_{k}$ and $\Upsilon_{k}=\Theta_{k}-\Theta_{d_k}$ as 
\begin{equation}\label{eq:dot_Theta_boundedness}
    \dot{\Theta}_{k}= \lambda_{k}+\sigma_{k}+\dot{\Theta}_{d_k}.
\end{equation}
From \eqref{eq:viplusone1}, we have $\dfrac{1}{2}\lambda_{k}^2(t)\leq \mathcal{D}_{k}(0)$, which implies $\lvert \lambda_{k}(t) \rvert \leq \sqrt{2\mathcal{D}_{k}(0)}$. One may also notice from \eqref{eq:sigmai} that $\sigma_{k}$ is only a function of bounded signal $\Theta_{k}$, and thus it remains bounded for all time $t \geq 0$, whereas the last term $\dot{\Theta}_{d_k}$ is also bounded according to \Cref{assum:qi}. As a consequence, the right-hand side of the expression $\dot{\Theta}_{k}= \lambda_{k}+\sigma_{k}+\dot{\Theta}_{d_k}$ is also bounded, ultimately inferring that the angular velocity of the UAV is bounded.

To show that attitude error converges to zero asymptotically, we differentiate \eqref{eq:wiplusonefinal} with respect to time along the state trajectories, which yields
\begin{equation}\label{eq:ddotvplus1}
\ddot{\mathcal{D}}_{k}=-4\mathcal{Z}_{k}\Upsilon_{k}^3\dot{\Upsilon}_{k}-2\mathcal{N}_{k}\lambda_k\dot{\lambda}_k.
\end{equation}
Note that the right-hand side of \eqref{eq:ddotvplus1} 
is a function of bounded signals $\Upsilon_{k}$, $\dot{\Upsilon}_{k}$, $\lambda_{k}$, and $\dot{\lambda}_{k}$, which implies $\ddot{\mathcal{D}}_{k}$. Therefore, $\dot{\mathcal{D}}_{k}$ is uniformly continuous. Using Barbalet's lemma, it can be inferred that $\gamma_{k}(t)\to 0$ as $t\to \infty$, and thus the concerned system, given in \eqref{eq:Upsilonddot}, is asymptotically stable. As a matter of fact, the UAV converges to its desired trajectory asymptotically.
 \end{proof}

\section{Simulations}\label{sec:simulation}
\begin{figure}[!htb]
\centering
\begin{subfigure}{0.45\linewidth}
	\centering
    \includegraphics[width=\linewidth]{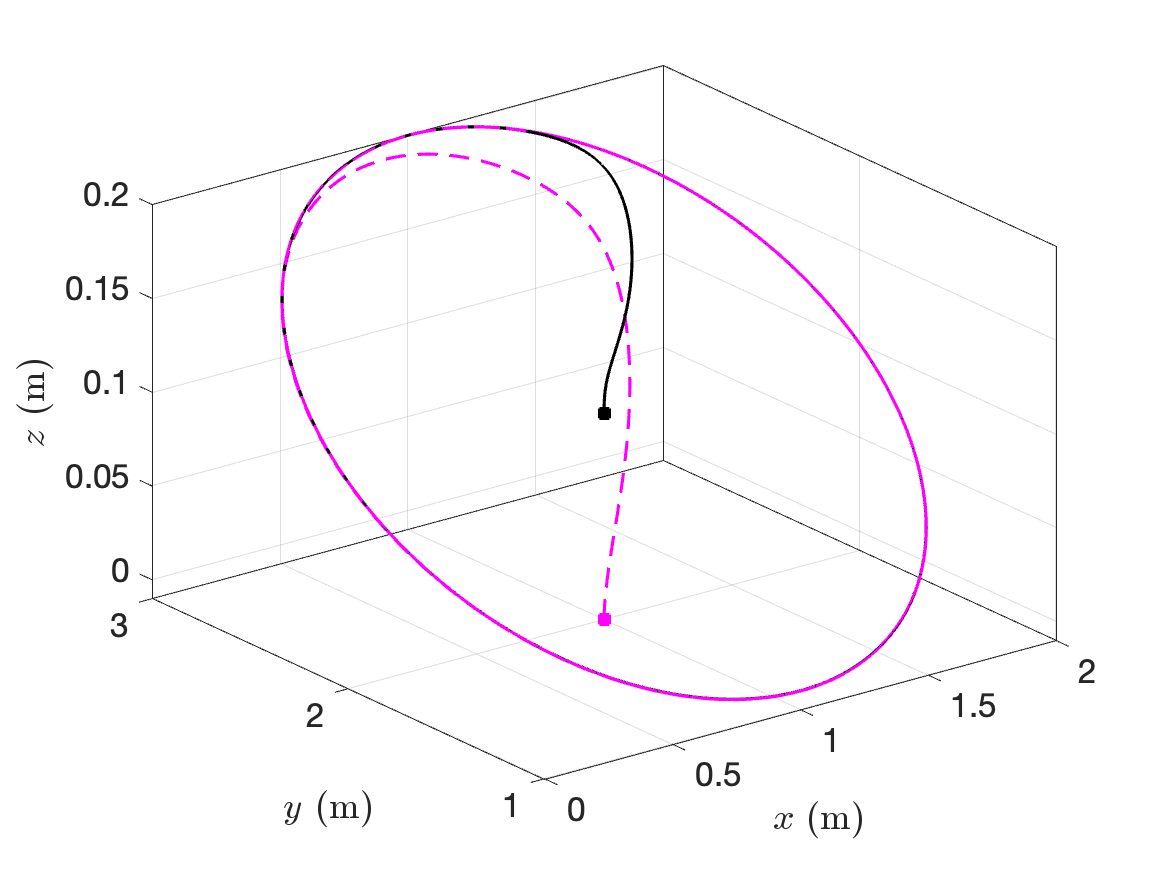}
     \caption{Trajectory.}
    \label{fig:circular_path}
\end{subfigure}%
\begin{subfigure}{0.45\linewidth}
    \centering
    \includegraphics[width=\linewidth]{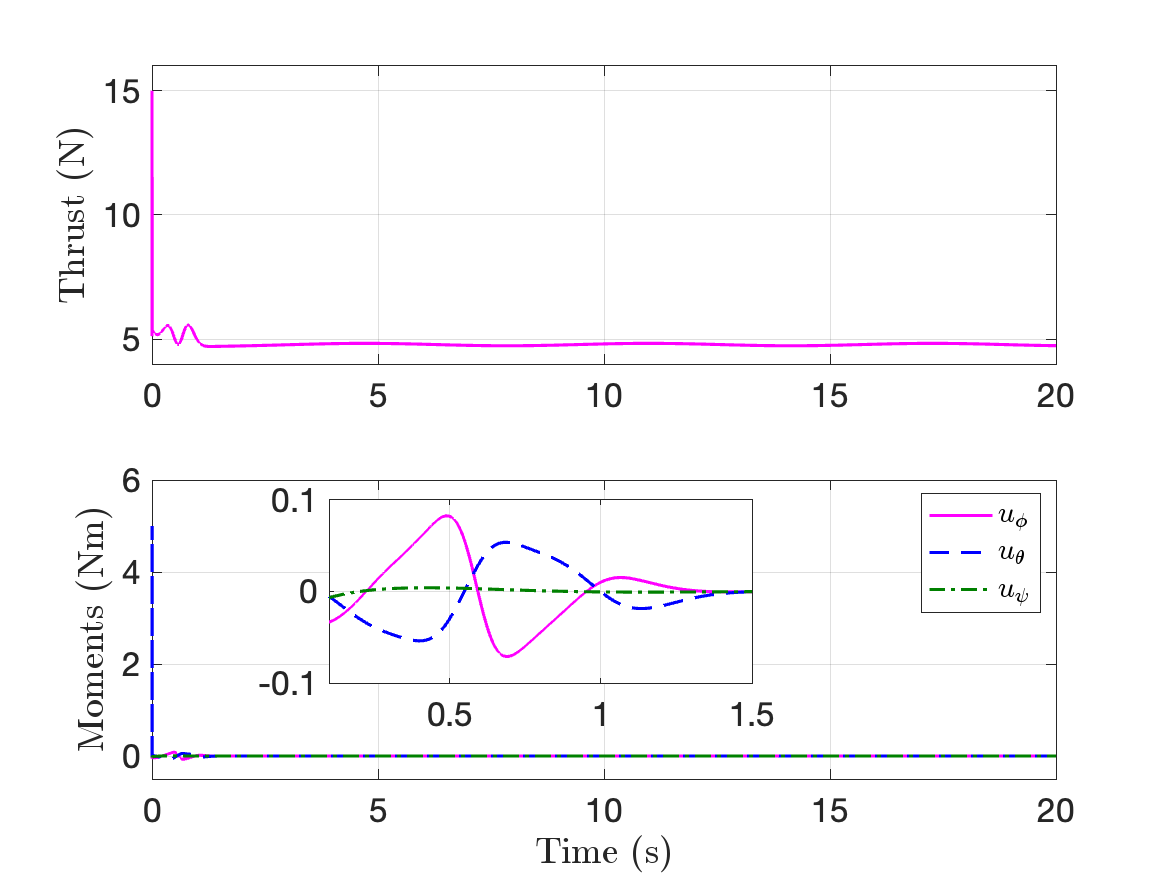}
    \caption{Control inputs.}
    \label{fig:circular_input}
\end{subfigure}
\begin{subfigure}{0.45\linewidth}
    \centering
    \includegraphics[width=\linewidth]{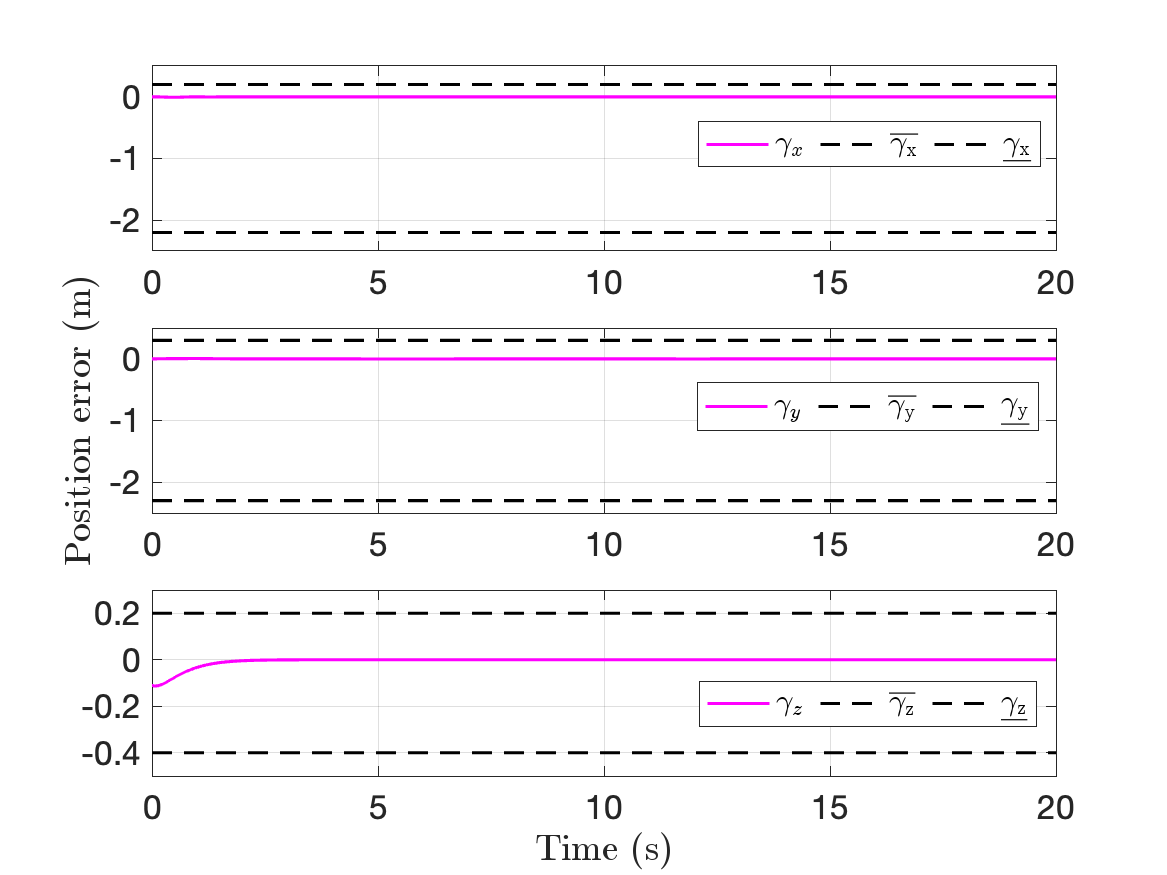}
    \caption{Position errors.}
    \label{fig:circular_gamma_bound}
\end{subfigure}%
\begin{subfigure}{0.45\linewidth}
    \centering
    \includegraphics[width=\linewidth]{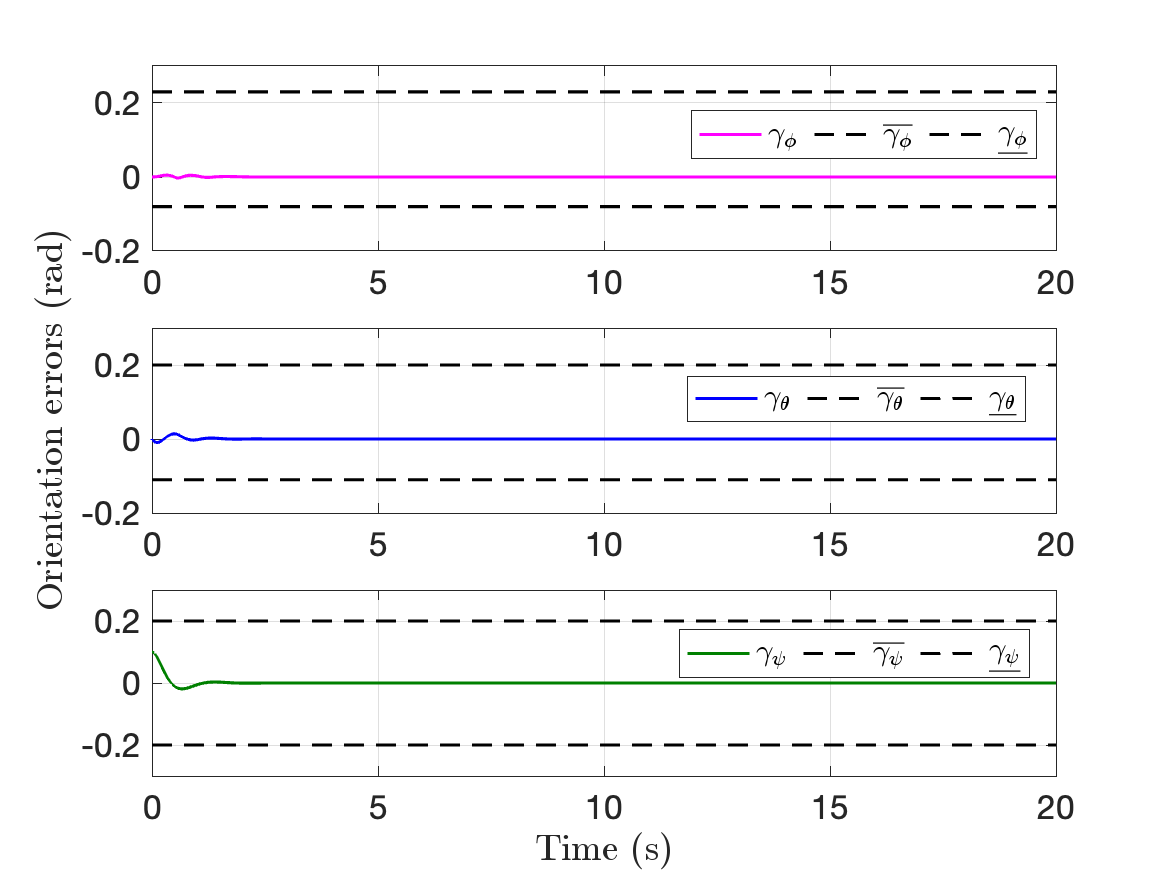}
    \caption{Orientation errors.}
    \label{fig:circular_upsilon_bound}
\end{subfigure}
\begin{subfigure}{0.45\linewidth}
    \centering
    \includegraphics[width=\linewidth]{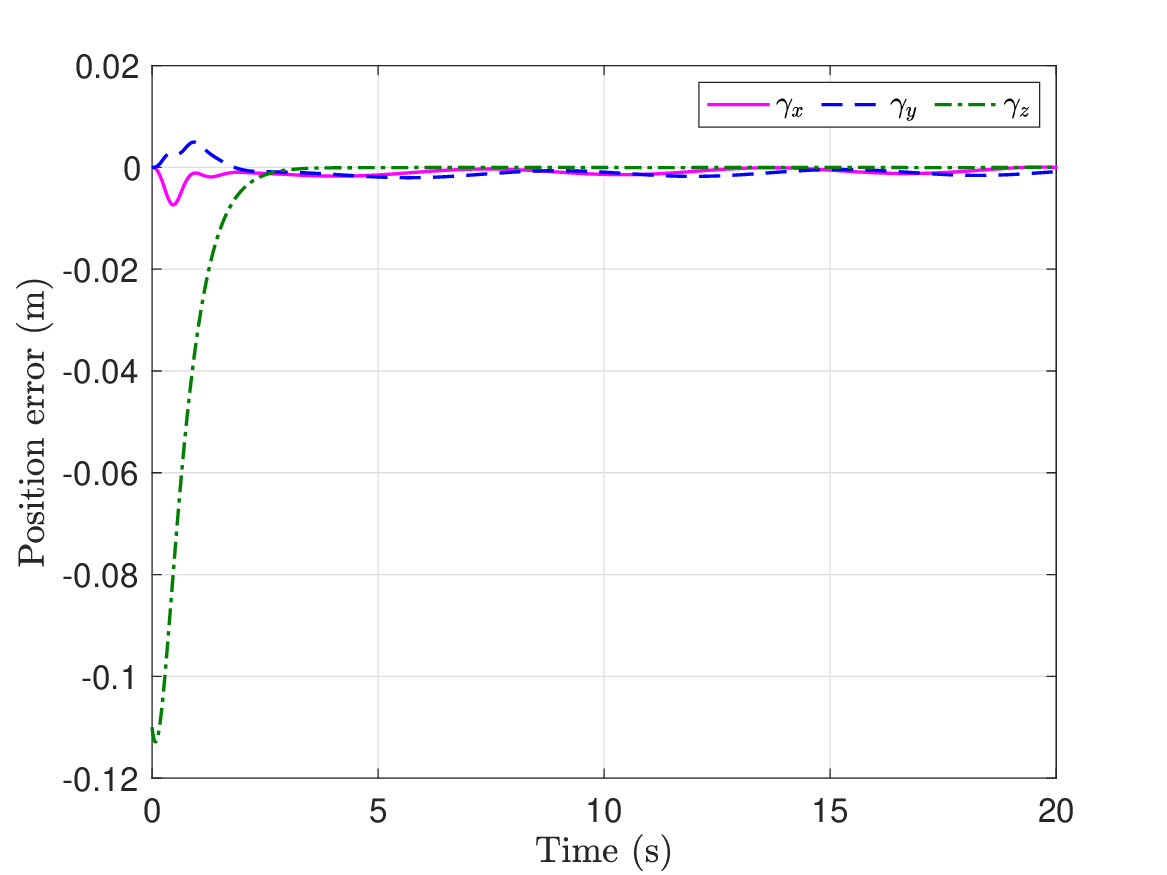}
    \caption{Zoomed position errors.}
    \label{fig:circular_gamma_wb}
\end{subfigure}%
\begin{subfigure}{0.45\linewidth}
    \centering
    \includegraphics[width=\linewidth]{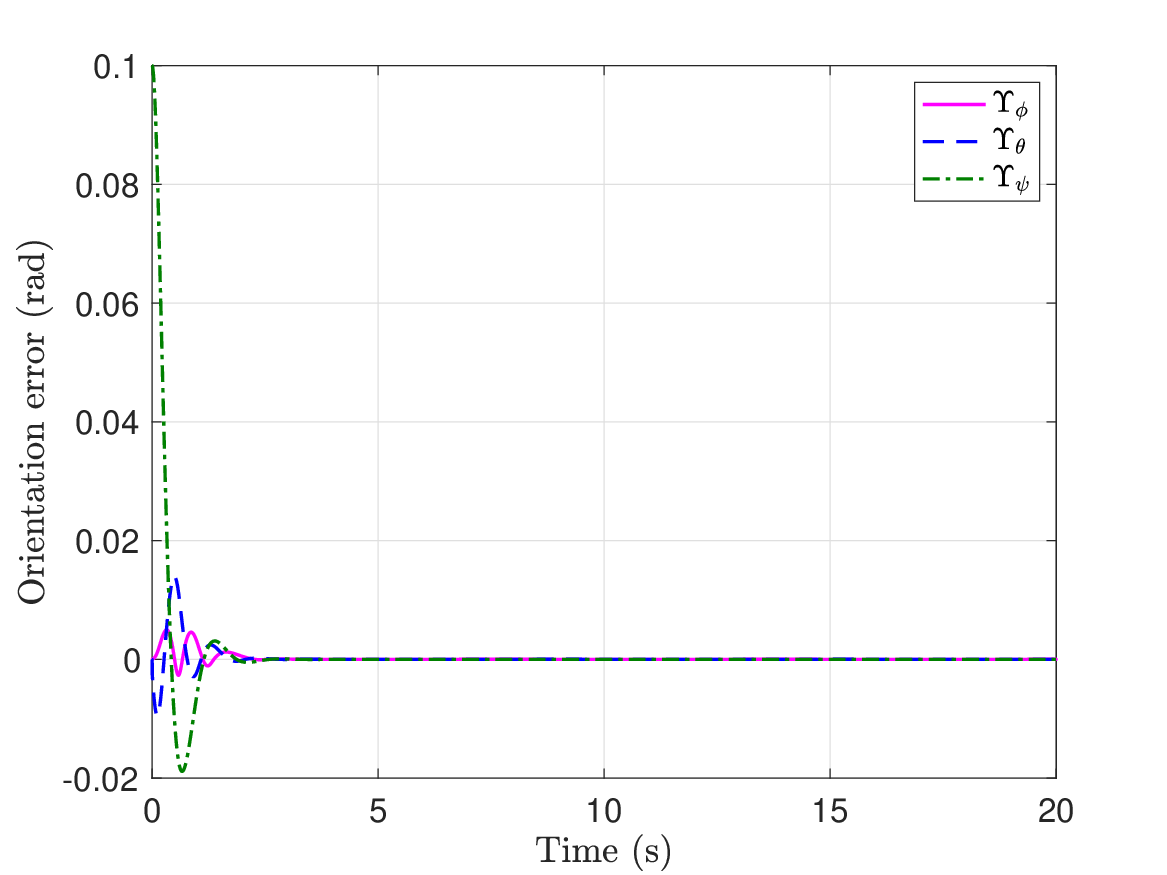}
    \caption{Zoomed orientation errors.}
    \label{fig:circular_upsilon_wb}
\end{subfigure}
\caption{UAV following an orbital path.}
\label{fig:circular_tracking}
\end{figure}
This section demonstrates the effectiveness of the proposed control strategy, as outlined in \Cref{thm:deltai,thm:attitude}, through a series of scenarios involving constrained motion. To ensure that the control inputs remain within the physical limitations of the actuators, we impose constraints on the thrust and moments, restricting their magnitudes to $15$ N and $\pm 3$ rad/s, respectively. The subsequent plots feature square markers indicating the initial position of the quadrotor (in magenta) and the starting point of the desired trajectory (in black). The design parameters for the translational subsystem are set to $\mathcal{K}_{\ell}=100$ and $\mathcal{M}_{\ell}=5$, for $\ell=x$, $y$, and $z$, while the attitude subsystem parameters are chosen as $\mathcal{Z}_{k}=100$ and $\mathcal{N}_{k}=5$, for $k=\phi,\theta$, and $\psi$, unless otherwise specified. To demonstrate the applicability of the proposed strategy further, we consider the UAV parameters of a commercial Pelican quadrotor provided by Ascending Technologies \cite{doi:10.1007/s11071-014-1425-y}. The UAV parameters are as follows: mass $m=0.485$ kg, moments of inertia $J_{xx}=3.4e-3$, $J_{\mathrm{yy}}=3.4e-3$, $J_{zz}=4.7e-3$, arm length $d=0.35$ m, rotor inertia $J_r=3.4e-5$, thrust coefficient $C_T=2.9842e-5$, and torque coefficient $C_Q=3.2320$.

\begin{figure}[!ht]
    \centering
    \begin{subfigure}{0.45\linewidth}
\centering
    \includegraphics[width=\linewidth]{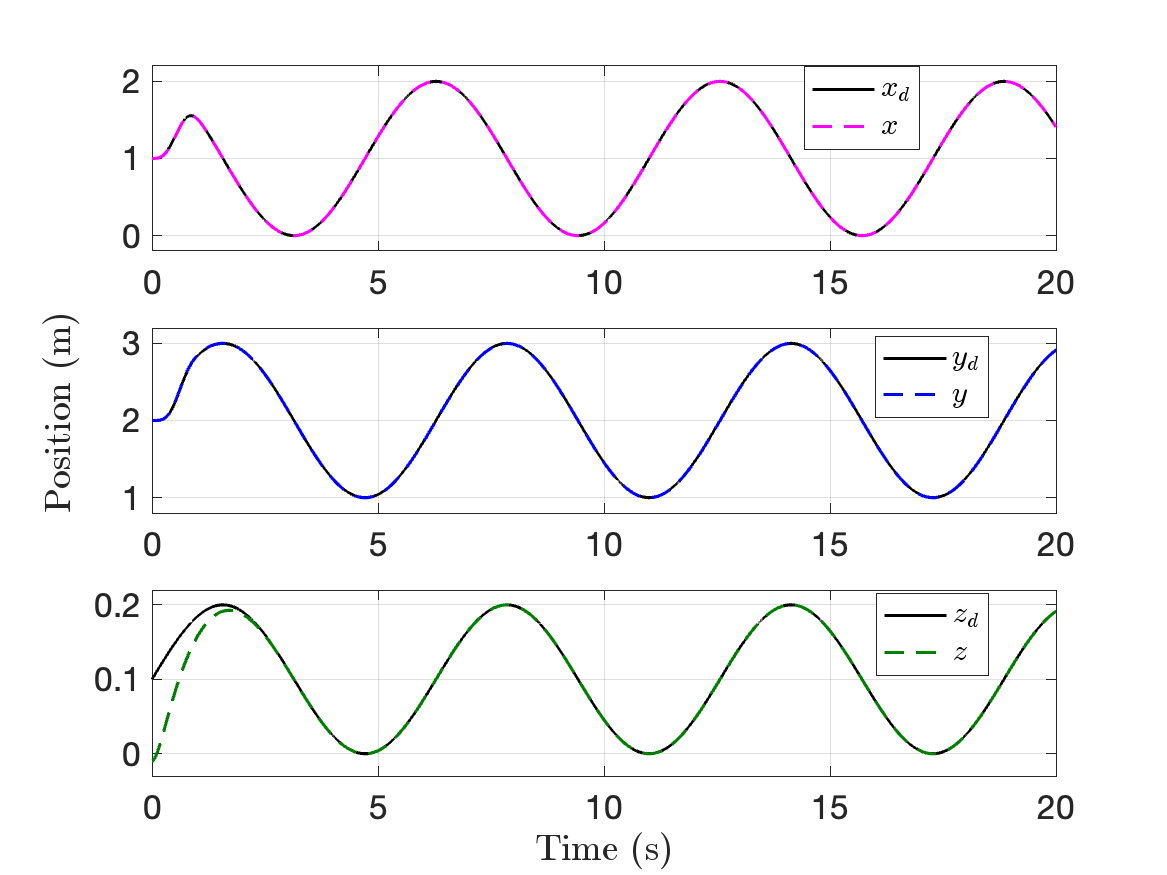}
    \caption{Position.}
    \label{fig:circular_position}
\end{subfigure}%
\begin{subfigure}{0.45\linewidth}
\centering
    \includegraphics[width=\linewidth]{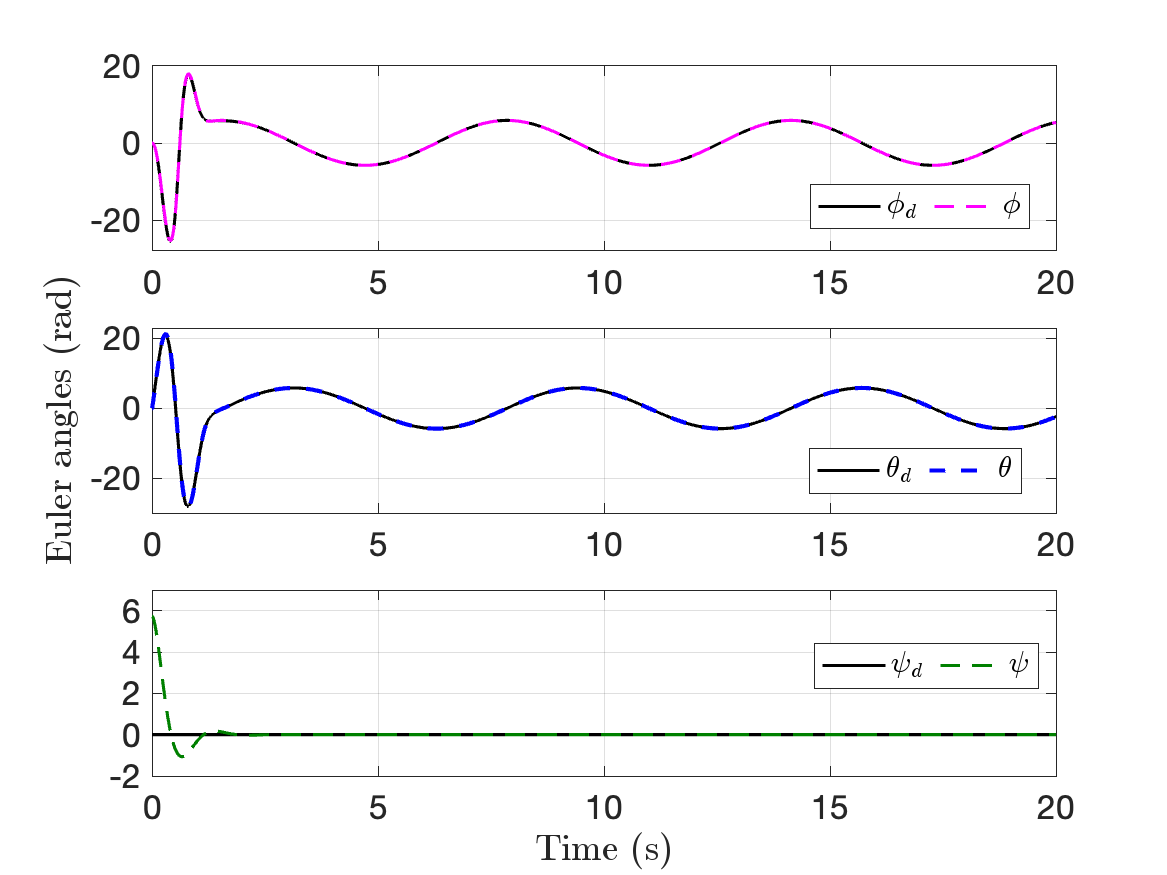}
    \caption{Orientation.}
    \label{fig:circular_orienatation}
\end{subfigure}
\begin{subfigure}{0.45\linewidth}
\centering
    \includegraphics[width=\linewidth]{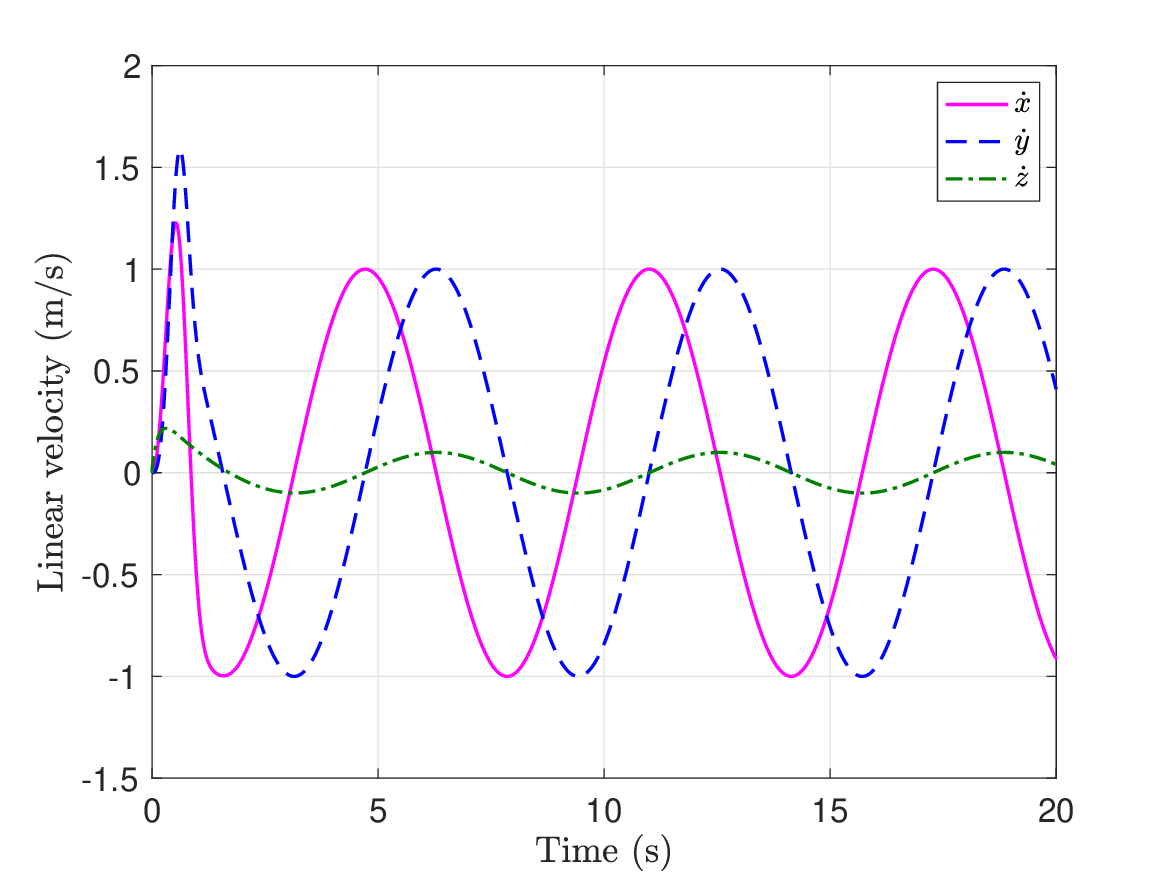}
    \caption{Linear velocity.}
    \label{fig:circular_linear_velocity}
\end{subfigure}%
\begin{subfigure}{0.45\linewidth}
\centering
    \includegraphics[width=\linewidth]{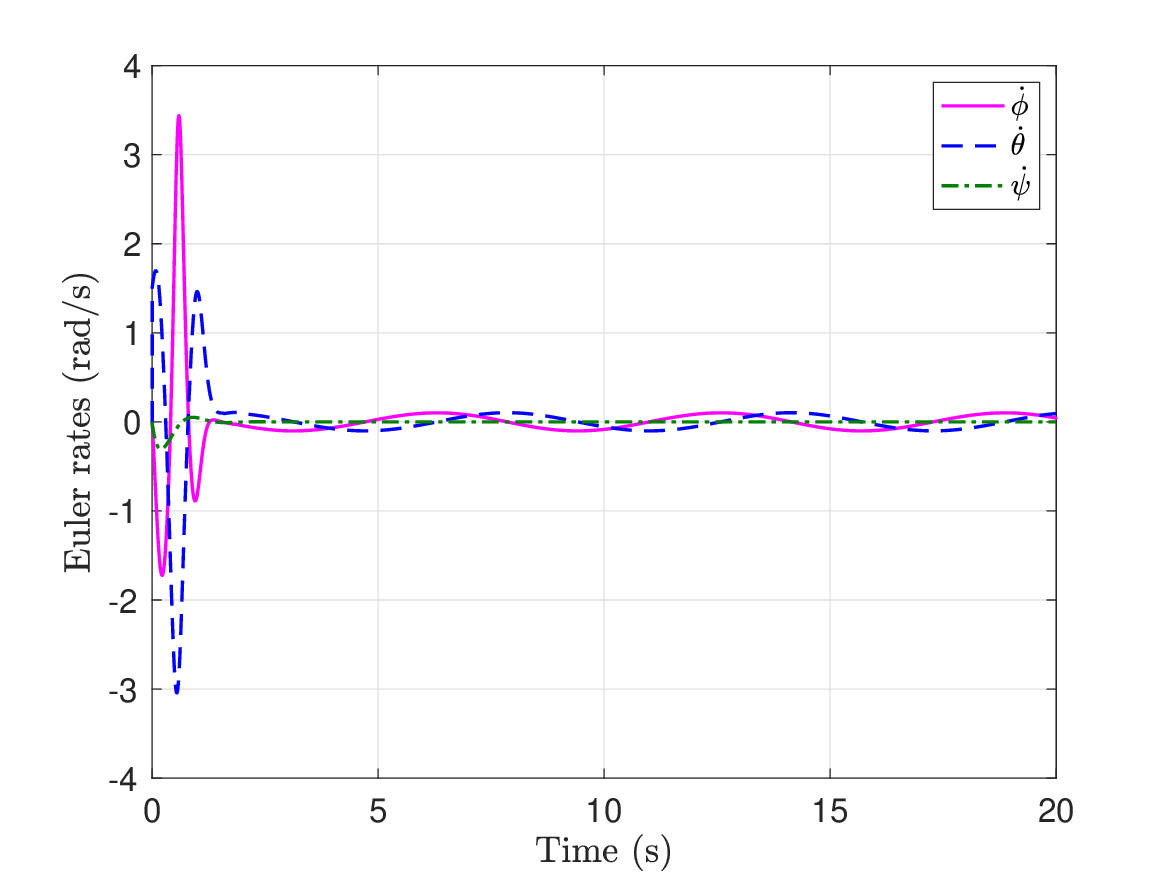}
    \caption{Euler rates.}
    \label{fig:circular_euler_rates}
\end{subfigure}
    \caption{UAV's position, orientation, and their rates when following an orbital path.}
    \label{fig:orbital_states}
\end{figure}
\begin{figure}[!hb]
	\centering
\begin{subfigure}{0.45\linewidth}
	\centering
    \includegraphics[width=\linewidth]{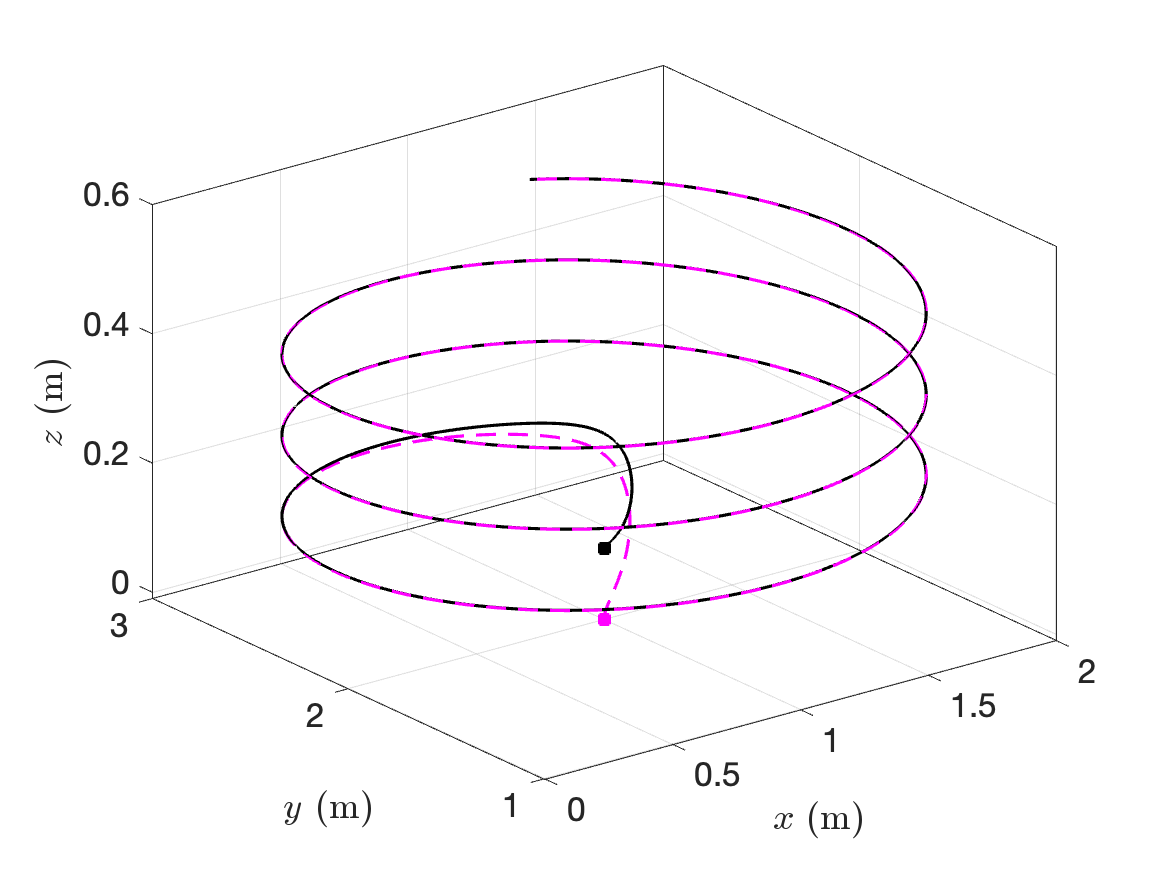}
     \caption{Trajectory.}
    \label{fig:helix_path}
\end{subfigure}%
\begin{subfigure}{0.45\linewidth}
    \centering
    \includegraphics[width=\linewidth]{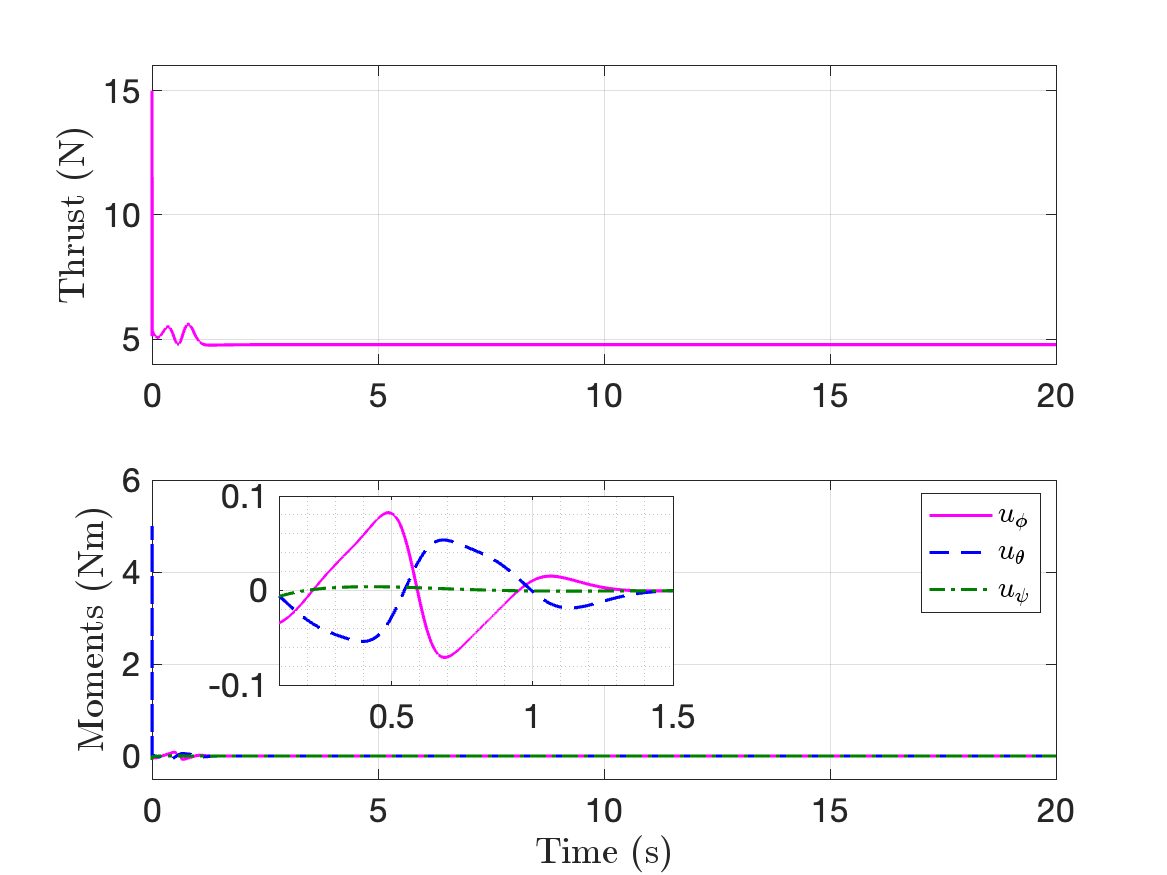}
    \caption{Control inputs.}
    \label{fig:helix_input}
\end{subfigure}
\begin{subfigure}{0.45\linewidth}
    \centering
    \includegraphics[width=\linewidth]{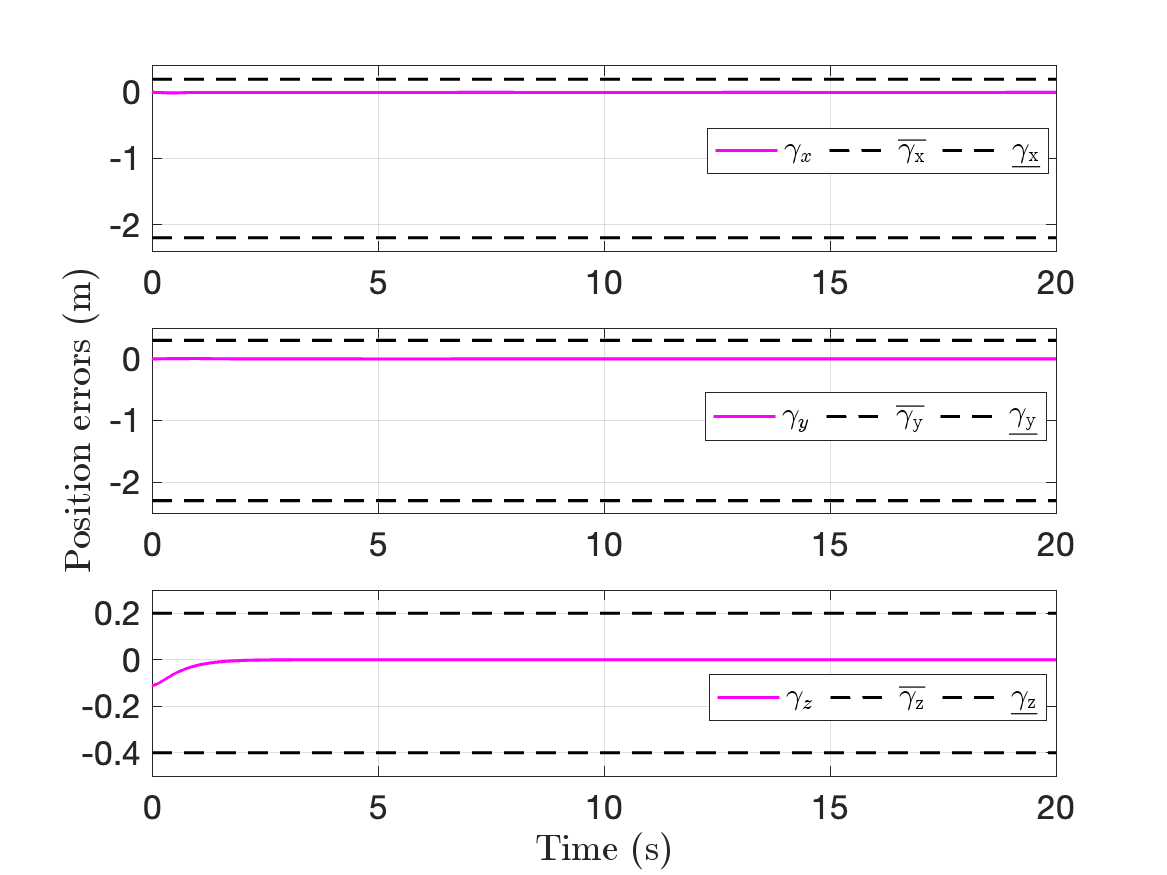}
    \caption{Position errors.}
    \label{fig:helix_gamma_bound}
\end{subfigure}%
\begin{subfigure}{0.45\linewidth}
    \centering
    \includegraphics[width=\linewidth]{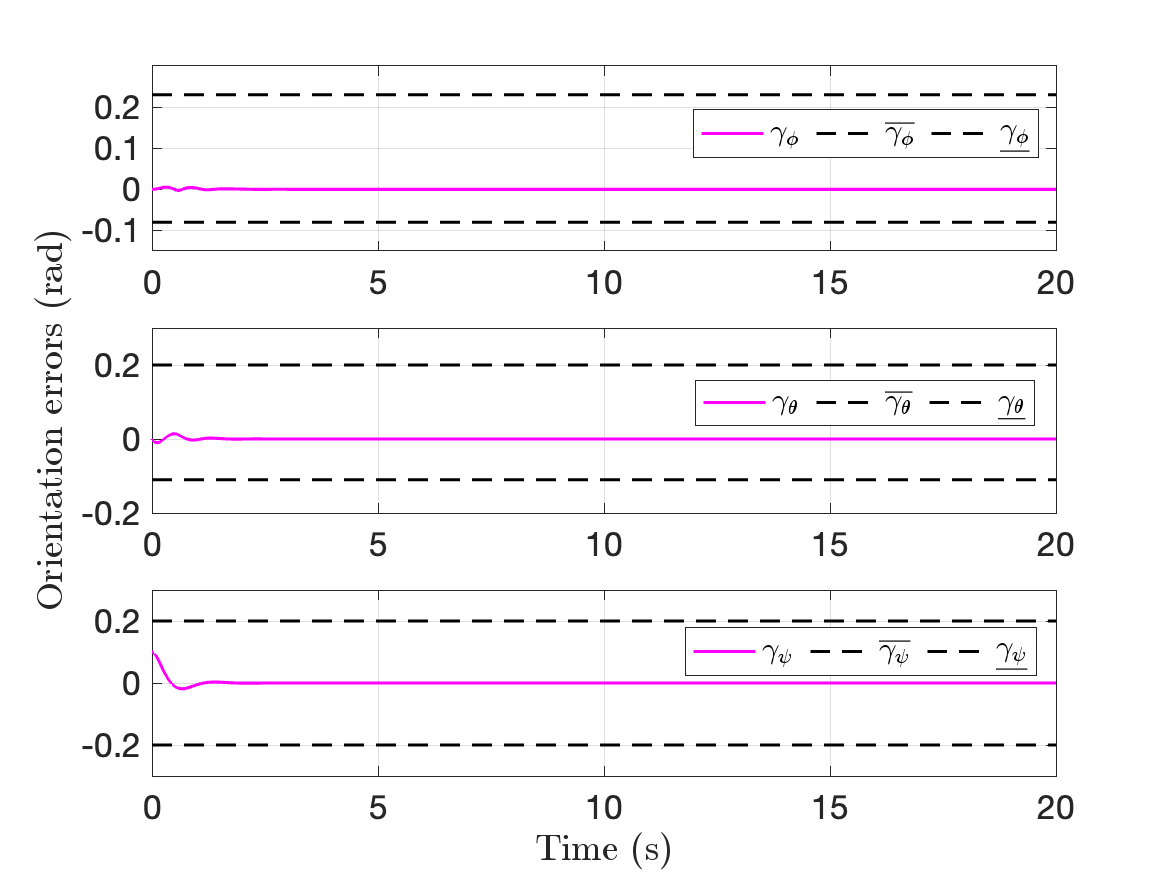}
    \caption{Orientation errors.}
    \label{fig:helix_upsilon_bound}
\end{subfigure}
\begin{subfigure}{0.45\linewidth}
    \centering
    \includegraphics[width=\linewidth]{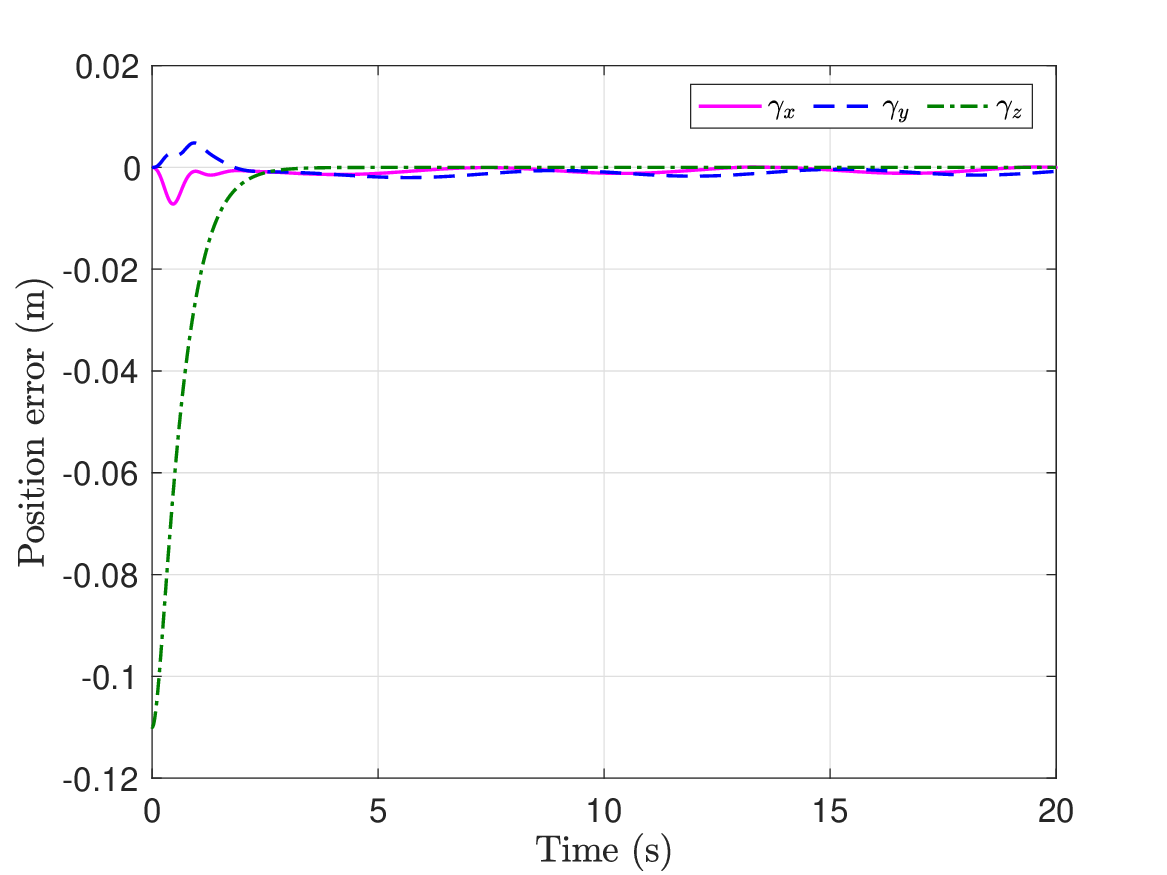}
    \caption{Zoomed position errors.}
    \label{fig:helix_gamma_wb}
\end{subfigure}%
\begin{subfigure}{0.45\linewidth}
    \centering
    \includegraphics[width=\linewidth]{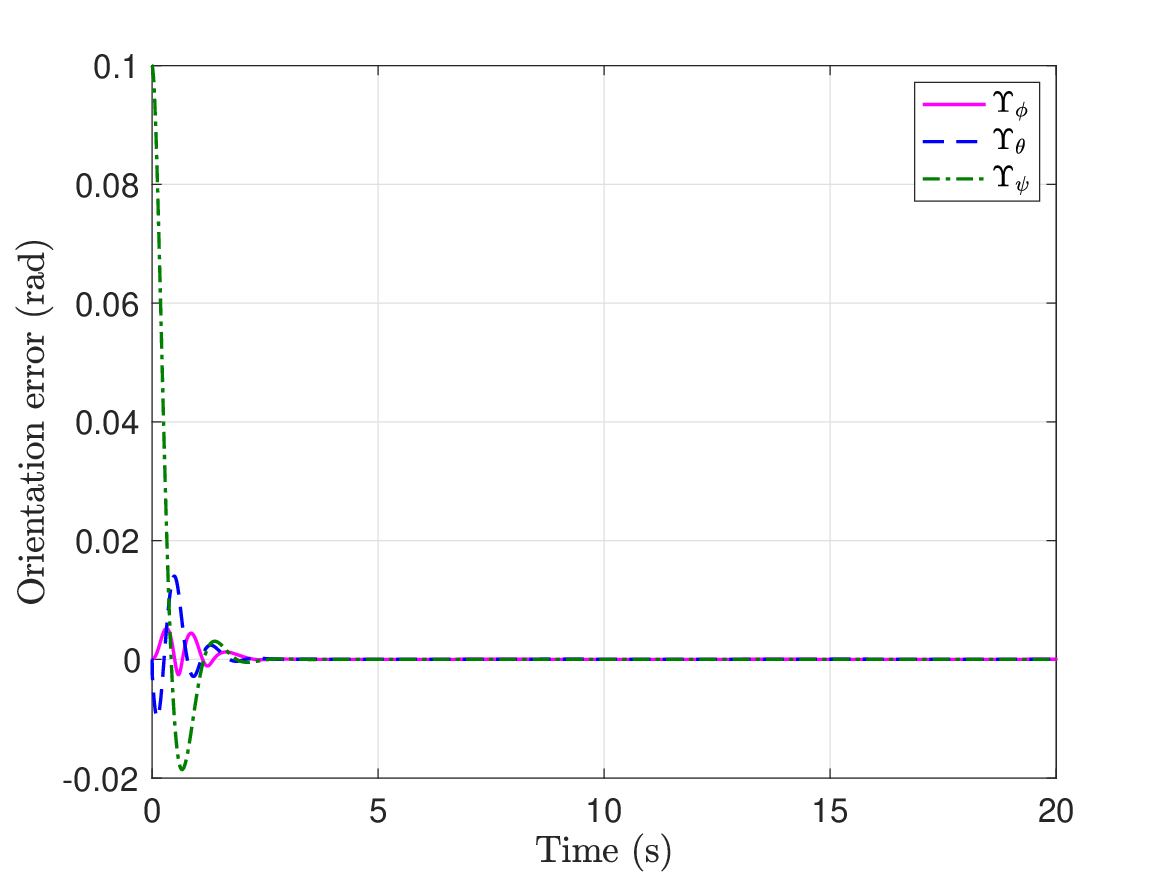}
    \caption{Zoomed orientation errors.}
    \label{fig:helix_upsilon_wb}
\end{subfigure}
\caption{UAV following a helical path.}
\label{fig:helix_tracking}
\end{figure}
We first demonstrate the performance of quadrotor under \eqref{eq:deltai} and \eqref{eq:ui} when it is required to follow an orbital-like path through \Cref{fig:circular_tracking,fig:orbital_states}. This kind of path is particularly important when the UAV is doing an inspection of steady objects such as bridges, vehicles, etc. The desired path is characterized by
\begin{align*}
    x_d(t)=&~ 1 + \left( 1- e^{-3t^3} \right) \cos t ~\text{m},\\
    y_d(t)=&~ 1 + \left( 1- e^{-5t^3} \right) \sin t ~\text{m},\\
    z_d(t)=&~ 0.1 + 0.1 \sin t ~ \text{m},\\
    \psi_d(t)=&~ 0 ~ \text{rad}.
\end{align*}
Now, the objective of the quadrotor is to track its time-varying desired trajectories such that $|x|<2.2$ m, $|y|<3.3$ m, $|z|<0.4$ m, $|\phi|<0.5$ rad, $|\theta|<0.6$ rad, and $|\psi|<0.2$ rad. Therefore, by \Cref{thm:deltai,thm:attitude} the lower and upper bounds of the tracking errors are $\underline{\gamma_x}=2.2$ m, $\overline{\gamma_x}=0.2$ m, $\underline{\gamma_y}=1.3$ m, $\overline{\gamma_x}=0.3$ m, $\underline{\gamma_z}=0.3$ m, $\overline{\gamma_z}=0.2$ m, $\underline{\Upsilon_\phi}=0.08$ rad, $\overline{\Upsilon_\phi}=0.23$ rad, $\underline{\Upsilon_\theta}=0.20$ rad, $\overline{\Upsilon_\theta}=0.11$ rad, and $\underline{\Upsilon_\psi}=0.20$ rad, $\overline{\Upsilon_\psi}=0.20$ rad. It can be observed from \Cref{fig:circular_path} that the quadrotor converges to its desired 3D time-varying trajectory and follows the same thereafter for all future times despite limited information about the inertia matrix. Also, the control demand is relatively more during transients (see \Cref{fig:circular_input}), but as soon as the error in the position and orientation vanishes, which is shown in \Cref{fig:circular_gamma_bound,fig:circular_upsilon_bound}, it reduces significantly. Note that such features are desirable to improve the flight time of quadrotors as they can carry only limited onboard power resources. It can also be observed from \Cref{fig:circular_input} that once the orientation tracking error vanishes, the moments required by the quadrotor becomes zero, which is one of the alluring features of the proposed strategy, since it will reduce the maneuverability of the quadrotor. The zoomed plots of the tracking errors are depicted in \Cref{fig:circular_gamma_wb,fig:circular_upsilon_wb}, from which one can observe that the tracking errors remain in their respective bounds, that is, $-\underline{\gamma_{\ell}}<\gamma_{\ell}<\overline{\gamma_{\ell}}$ for $\ell=x$, $y$, and $z$ and $-\underline{\Upsilon_{k}}<\Upsilon_k<\overline{\Upsilon_{k}}$ for $k=\phi$, $\theta$, and $\psi$. \Cref{fig:circular_position,fig:circular_orienatation} depicts the position and orientation trajectories of the quadrotor, and one may observe that it follows their desired trajectories with good precision. The linear velocity and Euler rates required by the quadrotor to follow an orbital-type path are shown \Cref{fig:circular_linear_velocity,fig:circular_euler_rates}. 

\begin{figure}[!ht]
    \centering
    \begin{subfigure}{0.45\linewidth}
\centering
    \includegraphics[width=\linewidth]{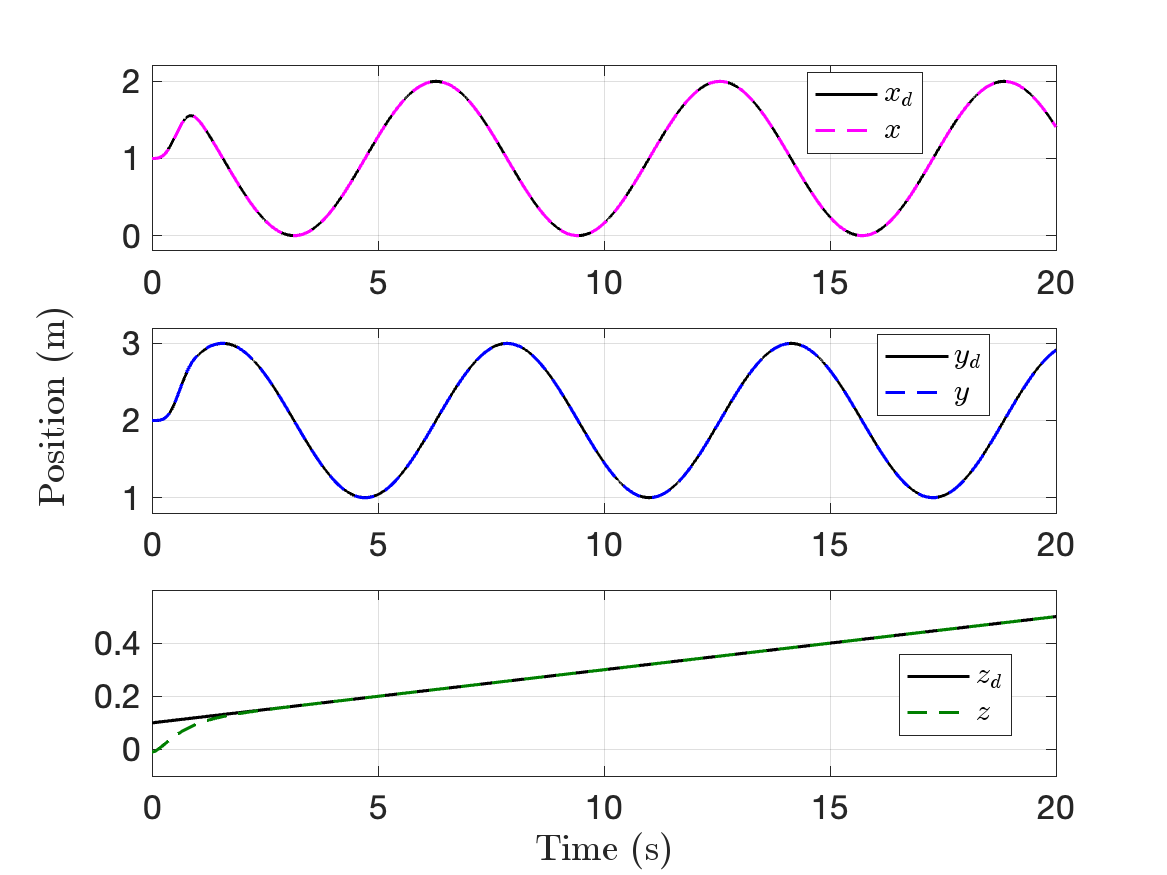}
    \caption{Position.}
    \label{fig:helix_position}
\end{subfigure}%
\begin{subfigure}{0.45\linewidth}
\centering
    \includegraphics[width=\linewidth]{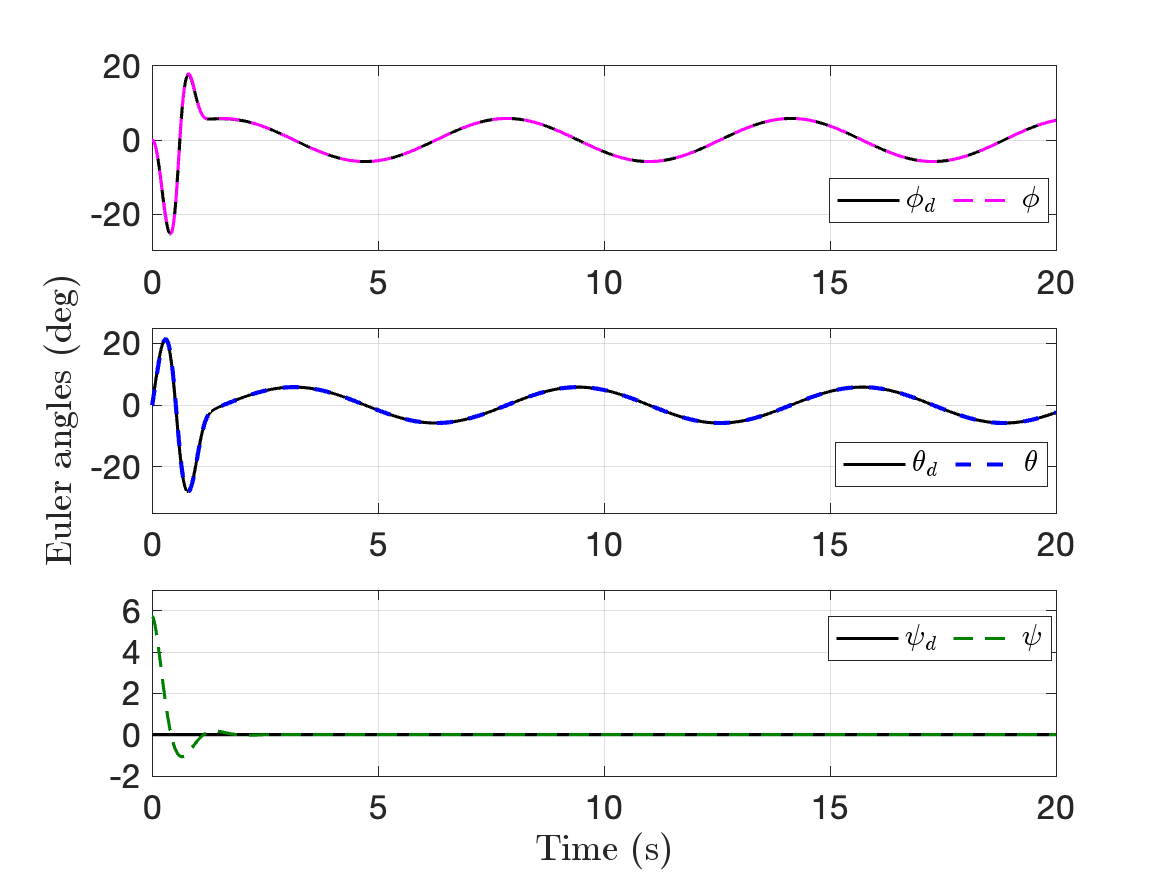}
    \caption{Orientation.}
    \label{fig:circle_orienatation}
\end{subfigure}
\begin{subfigure}{0.45\linewidth}
\centering
    \includegraphics[width=\linewidth]{circle_linear_velocity}
    \caption{Linear velocity.}
    \label{fig:helix_linear_velocity}
\end{subfigure}%
\begin{subfigure}{0.45\linewidth}
\centering
    \includegraphics[width=\linewidth]{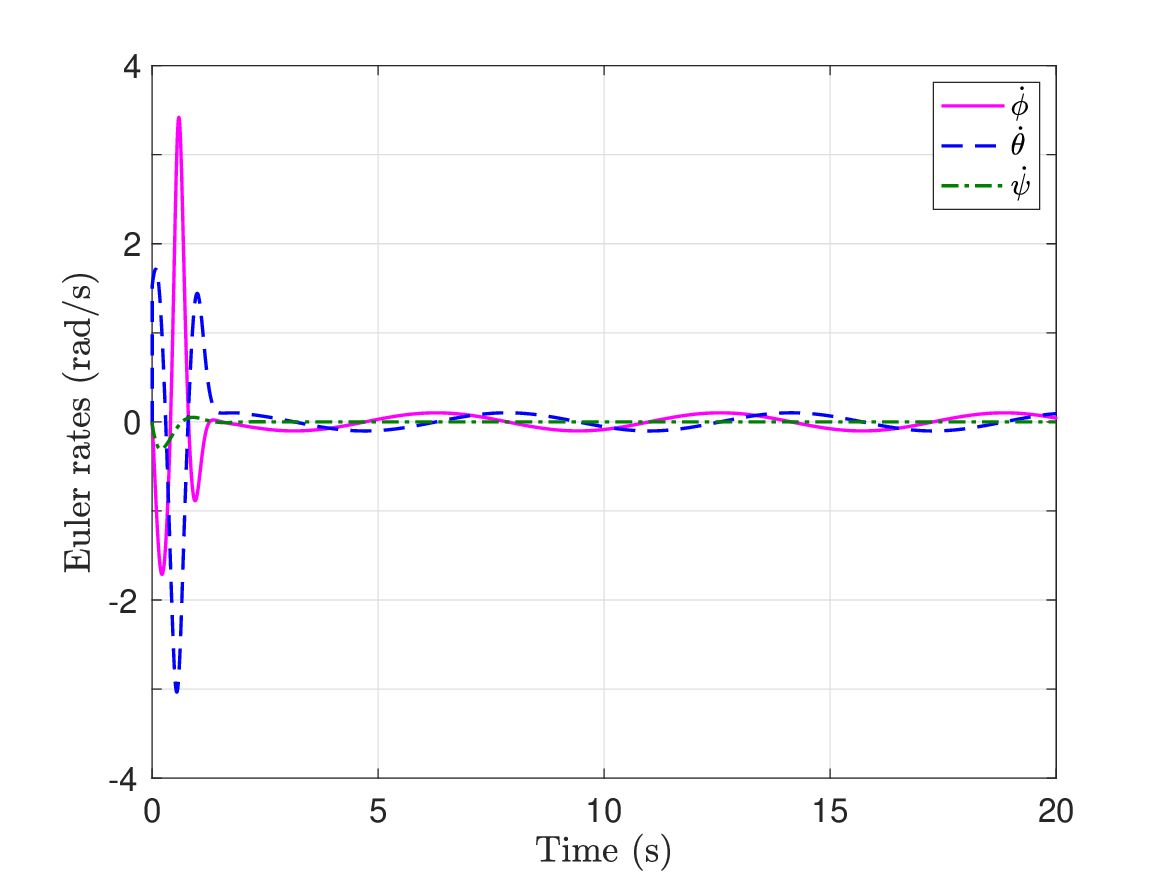}
    \caption{Euler rates.}
    \label{fig:helix_euler_rates}
\end{subfigure}
     \caption{UAV's position, orientation, and their rates when following a helical path.}
    \label{fig:helix_states}
\end{figure}

We now consider a scenario where the quadrotor needs to follow a 3D helix-like path, which is parameterized in time as
\begin{align*}
    x_d(t)=&~ 1 + \left( 1- e^{-3t^3} \right) \cos t ~\text{m},\\
    y_d(t)=&~ 1 + \left( 1- e^{-5t^3} \right) \sin t ~\text{m},\\
    z_d(t)=&~ 0.1 + t/50 ~ \text{m},\\
    \psi_d(t)=&~ 0 ~ \text{rad}.
\end{align*}
These types of paths are crucial for inspections of tall objects like chimneys, buildings, and trees. During these tasks, the UAV must orbit the structure, gradually increasing its height while ensuring the object of interest remains within the field of view of the camera or other sensors. Moreover, its position should also remain constrained to achieve maximal coverage with less maneuverability demands. In this case, the objective is to track its time-varying desired trajectories such that the position and orientation of the quadrotor remain in the set given by $|x|<2.2$ m, $|y|<3.3$ m, $|z|<0.7$ m, $|\phi|<0.5$ rad, $|\theta|<0.6$ rad, and $|\psi|<0.2$ rad. Therefore, in accordance with \Cref{thm:deltai,thm:attitude}, the lower and upper bounds of the tracking errors are chosen as $\underline{\gamma_x}=2.2$ m, $\overline{\gamma_x}=0.2$ m, $\underline{\gamma_y}=2.3$ m, $\overline{\gamma_y}=0.3$ m, $\underline{\gamma_z}=0.6$ m, $\overline{\gamma_z}=0.2$ m, $\underline{\Upsilon_\phi}=0.08$ rad, $\overline{\Upsilon_\phi}=0.23$ rad, $\underline{\Upsilon_\theta}=0.20$ rad, $\overline{\Upsilon_\theta}=0.11$ rad, and $\underline{\Upsilon_\psi}=0.20$ rad,  $\overline{\Upsilon_\psi}=0.20$ rad. Under the proposed control inputs \eqref{eq:deltai} and \eqref{eq:ui}, we show the performance of the quadrotor through \Cref{fig:helix_tracking,fig:helix_states}. One can observe from \Cref{fig:helix_path} that the quadrotor tracks its desired time-varying trajectory precisely by exhibiting similar trajectory-tracking behavior as for the orbital-type path. Note that once the quadrotor converges on to its desired trajectory, it remains on it irrespective of the curvature changes. Also, like the previous engagement scenario, the quadrotor tracks its desired trajectory without violating the imposed spatial constraints. 
\begin{figure}[!ht]
	\centering
\begin{subfigure}{0.45\linewidth}
	\centering
    \includegraphics[width=\linewidth]{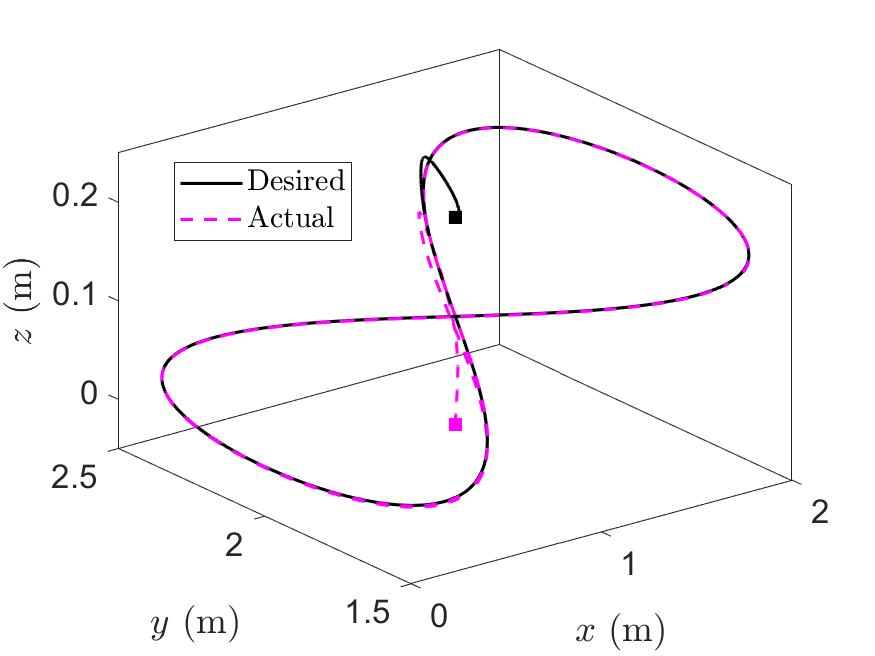}
     \caption{Trajectory.}
    \label{fig:bow_path}
\end{subfigure}%
\begin{subfigure}{0.45\linewidth}
    \centering
    \includegraphics[width=\linewidth]{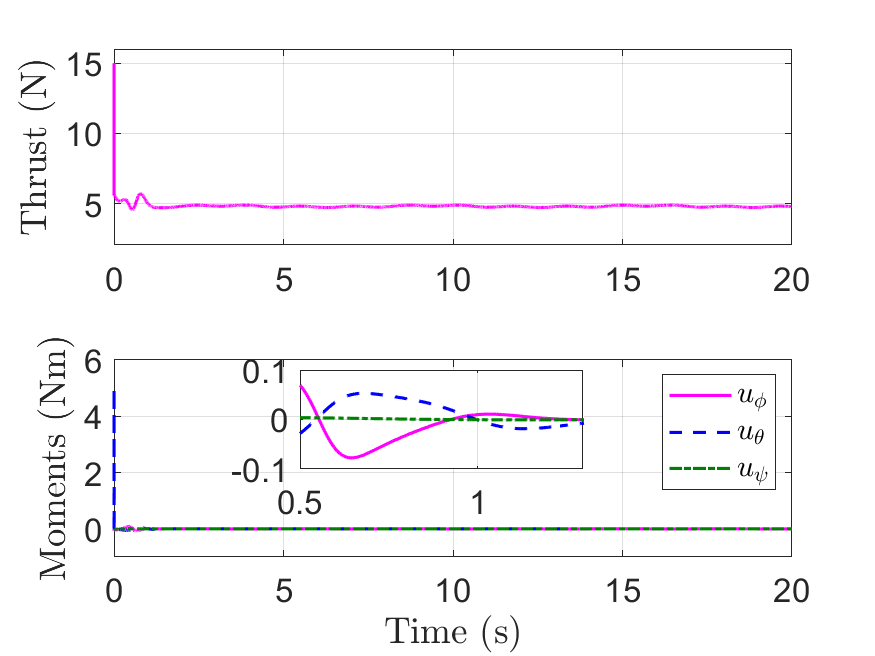}
    \caption{Control inputs.}
    \label{fig:bow_input}
\end{subfigure}
\begin{subfigure}{0.45\linewidth}
    \centering
    \includegraphics[width=\linewidth]{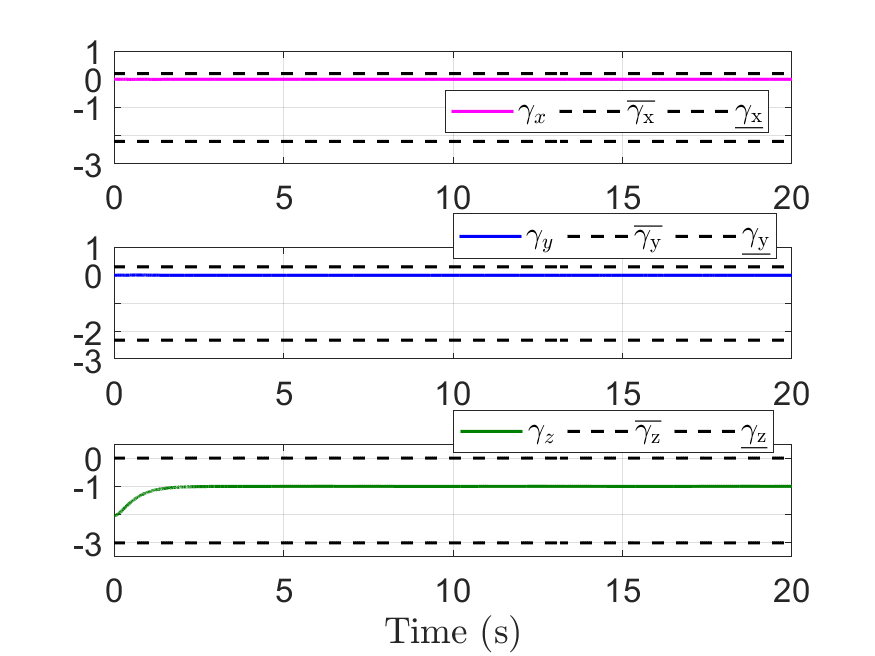}
    \caption{Position errors.}
    \label{fig:bow_gamma_bound}
\end{subfigure}%
\begin{subfigure}{0.45\linewidth}
    \centering
    \includegraphics[width=\linewidth]{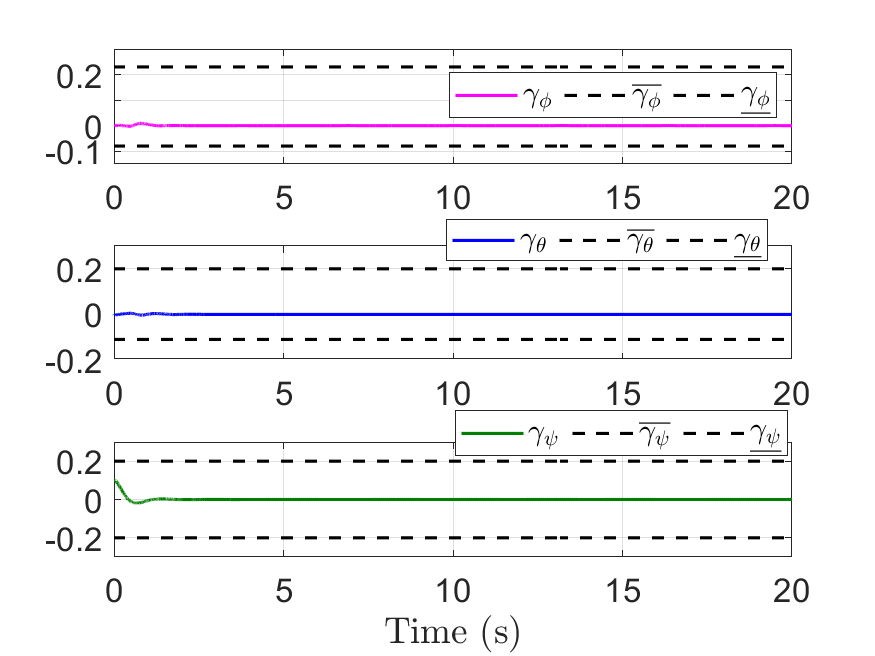}
    \caption{Orientation errors.}
    \label{fig:bow_upsilon_bound}
\end{subfigure}
\begin{subfigure}{0.45\linewidth}
    \centering
    \includegraphics[width=\linewidth]{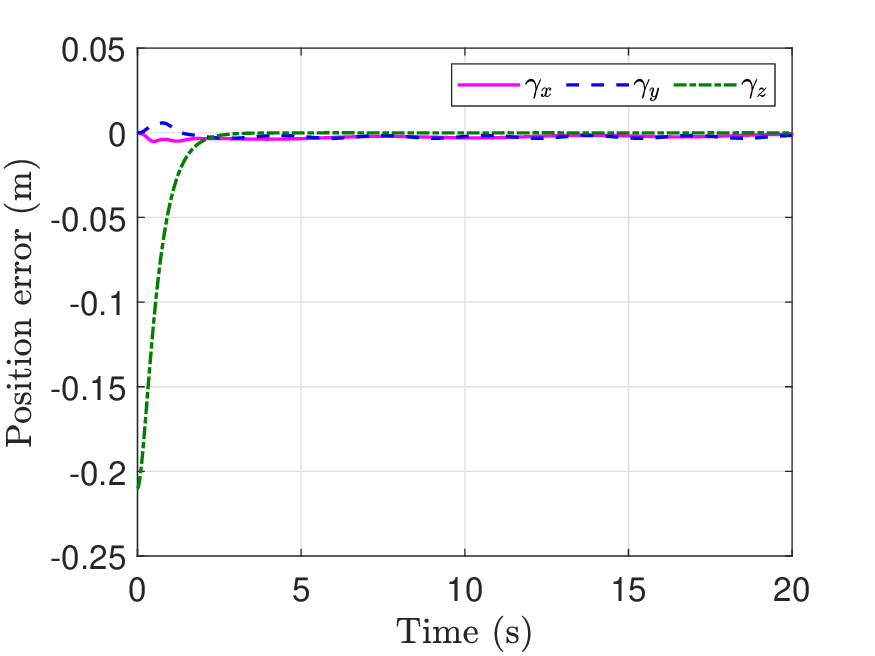}
    \caption{Zoomed position errors.}
    \label{fig:bow_gamma_wb}
\end{subfigure}%
\begin{subfigure}{0.45\linewidth}
    \centering
    \includegraphics[width=\linewidth]{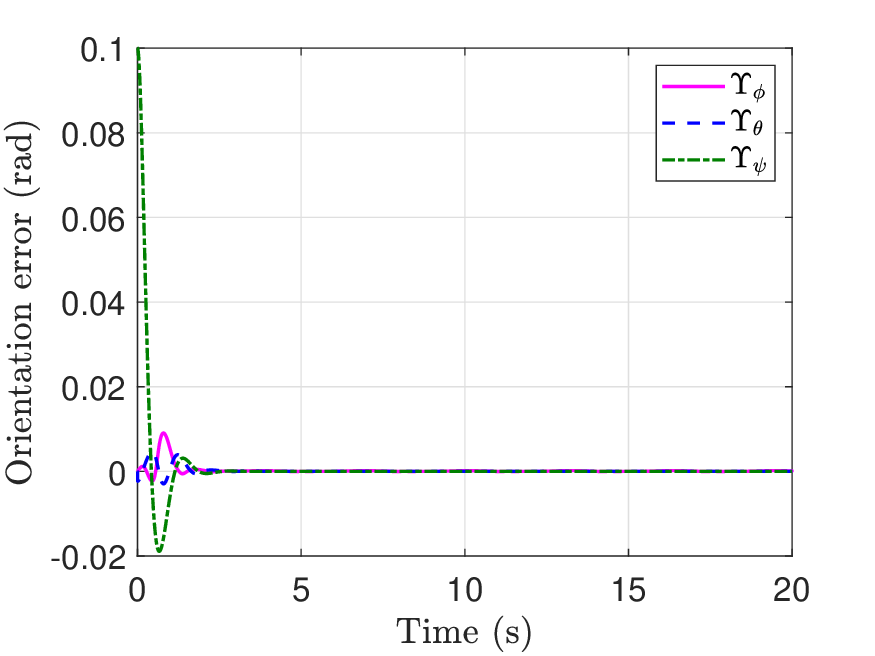}
    \caption{Zoomed orientation errors.}
    \label{fig:bow_upsilon_wb}
\end{subfigure}
\caption{UAV following a bow-shaped path.}
\label{fig:bow_tracking}
\end{figure}
\begin{figure}[!ht]
    \centering
    \begin{subfigure}{0.45\linewidth}
\centering
    \includegraphics[width=\linewidth]{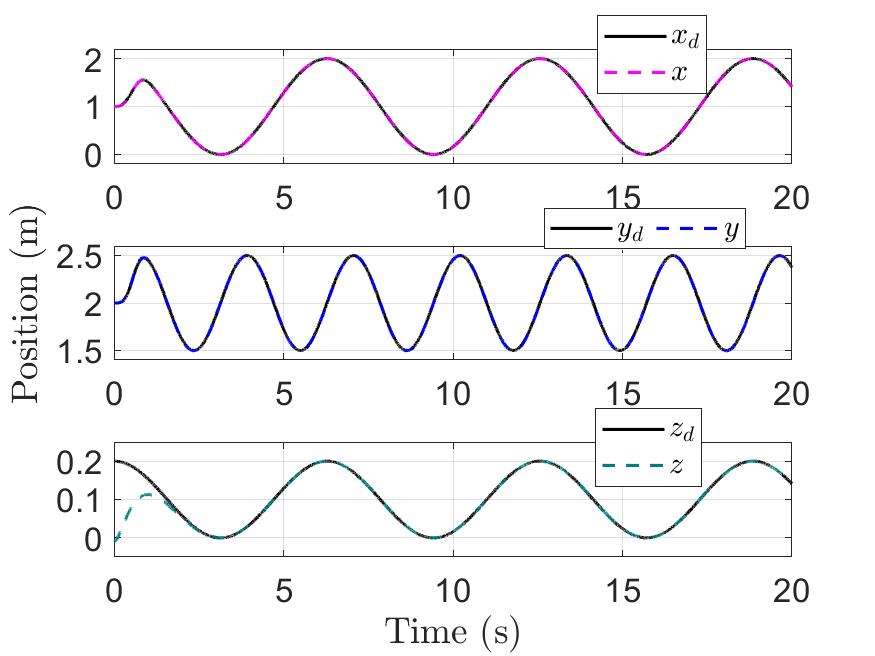}
    \caption{Position.}
    \label{fig:bow_position}
\end{subfigure}%
\begin{subfigure}{0.45\linewidth}
\centering
    \includegraphics[width=\linewidth]{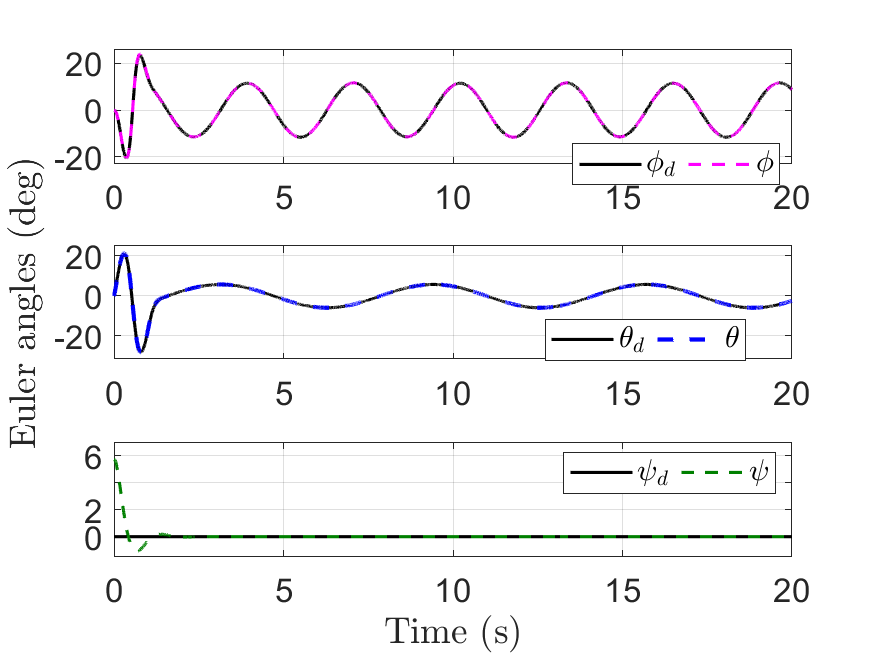}
    \caption{Orientation.}
    \label{fig:bow_orienatation}
\end{subfigure}
\begin{subfigure}{0.45\linewidth}
\centering
    \includegraphics[width=\linewidth]{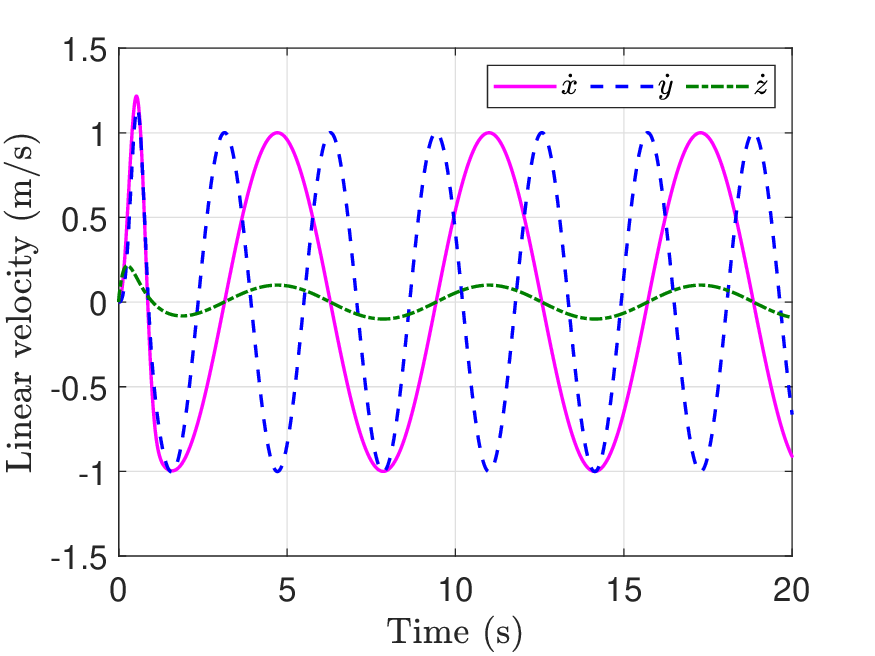}
    \caption{Linear velocity.}
    \label{fig:bow_linear_velocity}
\end{subfigure}%
\begin{subfigure}{0.45\linewidth}
\centering
    \includegraphics[width=\linewidth]{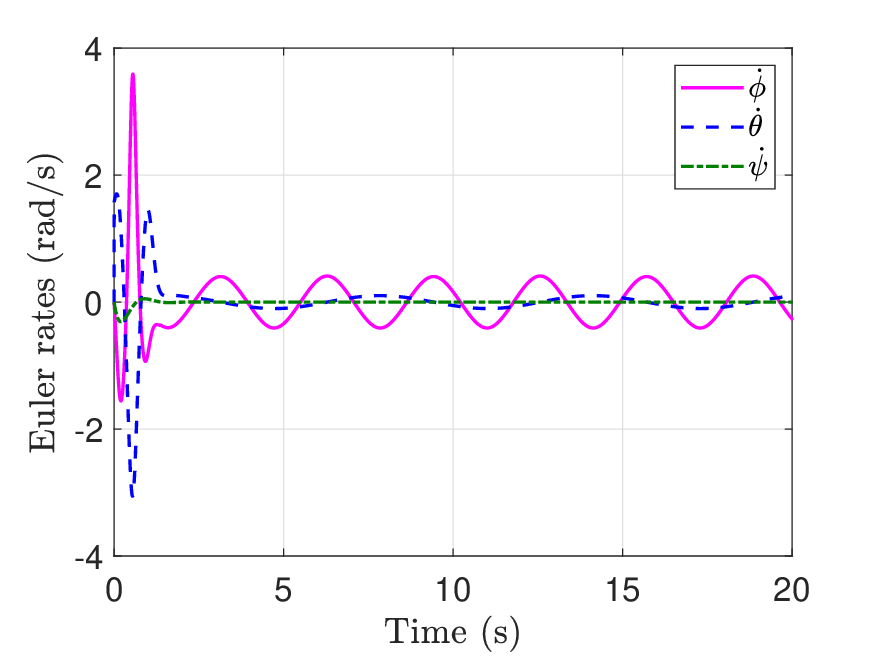}
    \caption{Euler rates.}
    \label{fig:bow_euler_rates}
\end{subfigure}
     \caption{UAV's position, orientation, and their rates when following a bow-shaped path.}
    \label{fig:bow_states}
\end{figure}

We now demonstrate the performance of the proposed control strategy when the UAV is required to follow a more complex path, such as a 3D bow shape curve. The path is characterized by
\begin{align*}
   x_d(t)=&~ 1 + \left( 1- e^{-3t^3} \right) \cos t ~\text{m},\\
    y_d(t)=&~ 1 + \left( 1- e^{-5t^3} \right) \sin t \cos t ~\text{m},\\
    z_d(t)=&~ 0.1 + 0.1 \cos{t} ~ \text{m},\\
    \psi_d(t)=&~ 0 ~ \text{rad}.  
\end{align*}
For the given engagement scenario, the objective is to track its time-varying desired trajectories while its position and orientation remain in the set  $\lvert x \rvert < 2.2$ m, $\lvert y \rvert < 2.8$ m, $\lvert z \rvert < 0.4$ m, $\lvert \phi \rvert < 2.2$ rad, $\lvert \theta \rvert < 0.6$ rad, and $\lvert \psi \rvert < 0.2$ rad. Thus, the lower and upper bounds of the tracking errors are $\underline{\gamma_x}=2.2$ m, $\overline{\gamma_x}=0.2$ m, $\underline{\gamma_y}=1.3$ m, $\overline{\gamma_y}=0.3$ m, $\underline{\gamma_z}=0.6$ m, $\overline{\gamma_z}=0.2$ m, $\underline{\Upsilon_\phi}=0.25$ rad, $\overline{\Upsilon_\phi}=0.20$ rad, $\underline{\Upsilon_\theta}=0.20$ rad, $\overline{\Upsilon_\theta}=0.11$ rad, and $\underline{\Upsilon_\psi}=0.20$ rad,  and $\overline{\Upsilon_\psi}=0.20$ rad in line with \Cref{thm:deltai,thm:attitude}. While keeping the other settings the same as before, we depict the performance of the proposed strategy through \Cref{fig:bow_tracking,fig:bow_states}. It can be observed that the UAV follows its desired path while exhibiting a similar behavior as the orbital path.

It is worth noting that the UAV converges to its desired trajectory (under certain mild assumptions) while adhering to the spatial constraints imposed on it regardless of the path geometry. This further attests to the claim that the proposed strategy remains invariant to the path geometry. Furthermore, inferences from \Cref{fig:orbital_states,fig:helix_states,fig:bow_states} indicate the boundedness of the closed-loop signals, as proved rigorously in the previous sections.


\section{Concluding remarks}\label{sec:conclusion}
This paper presents a novel nonlinear control strategy for 3D trajectory tracking of quadrotor unmanned aerial vehicles (UAVs), accommodating limited knowledge of the inertia matrix. The proposed approach ensures that the quadrotor tracks the desired trajectory while respecting spatial constraints on its motion, subject to mild assumptions on its initial configuration. Through rigorous theoretical analysis, we have established that the quadrotor's position, velocity, orientation, and Euler angle rates remain bounded, thereby guaranteeing a stable and reliable response. The proposed strategy exhibits robustness against path curvature changes and even accommodates changes in the desired path, exhibiting its versatility and adaptability. Future research directions may include real-time outdoor implementation and consideration of bounded inputs, which could further enhance the practical applicability of this work.

\bibliographystyle{ieeetr}
\bibliography{References/references} 
\end{document}